\documentclass[11pt, a4paper, pdftex, twoside, dvipsnames]{article}
\usepackage{amsmath,amssymb,amsthm,amsfonts}
\usepackage{bm}  

\everymath{\displaystyle}
\allowdisplaybreaks
\usepackage{mathtools}
\usepackage[makeroom]{cancel}
\usepackage{newtxtext,newtxmath}
\usepackage[title]{appendix}
\usepackage{epigraph}
\usepackage[labelfont=bf]{caption}
\usepackage{subcaption}
\usepackage{rotating}
\usepackage{float}
\usepackage{subcaption}
\usepackage{graphicx}  
\usepackage[section]{placeins}
\usepackage{comment}

\usepackage{threeparttable}
\usepackage{multirow}
\usepackage{multicol}
\usepackage{blkarray}
\usepackage{booktabs}
\usepackage{arydshln}
\usepackage{tabularx}
\usepackage{longtable, array}
\newcolumntype{P}[1]{>{\raggedright\arraybackslash}p{#1}}
\usepackage[x11names]{xcolor}
\usepackage[bookmarks=false, bookmarksnumbered=true,
colorlinks=true,setpagesize=false,
pdftitle={},pdfauthor={TakafumiUsui},pdfsubject={},
pdfkeywords={}, bookmarkstype=toc, urlcolor=black, citecolor=blue,
linkcolor=black, filecolor=black]{hyperref}
\usepackage{indentfirst}
\usepackage{enumitem}
\usepackage{cases}
\usepackage{url}
\usepackage{setspace}
\usepackage{ulem}
\usepackage{lscape}
\usepackage{here}
\usepackage{authblk}
\usepackage{framed}
\usepackage{adjustbox}
\usepackage{pdflscape}
\usepackage{listings}
\usepackage{cleveref}
\crefname{equation}{Eq.}{Eqs.}
\crefname{figure}{Figure}{Figures}
\crefname{table}{Table}{Tables}
\crefname{line}{Algorithm}{Algorithms}
\crefname{asmp}{Assumption}{Assumptions}
\crefname{section}{Section}{Sections}
\crefname{chapter}{Chapter}{Chapters}
\crefname{appsec}{Appendix}{Appendixes}

\renewcommand{\ref}{\cref}
\usepackage{subcaption}
\usepackage[a4paper,left=2cm,right=2cm,top=2.5cm,bottom=2.5cm]{geometry}
\usepackage{fancyhdr}
\fancypagestyle{firstpagestyle}{

\fancyhf{} 
\fancyfoot[C]{\thepage}
}
\fancypagestyle{CVfirstpagestyle}{

\fancyhf{} 
\fancyfoot[C]{\thepage}
}
\usepackage{natbib}
\title{The climate in climate economics\thanks{We thank Pratyuksh Bansal, Yongyang Cai, Thomas Lontzek, Alena Miftakhova, Karl Schmedders, Christian Tr\"ager, Takafumi Usui, Rick van der Ploeg, as well as seminar participants at the University of Lausanne, the University of Zurich, and various conferences and workshops for very useful conversations and comments. This work was supported by the Swiss National Science Foundation (SNF), under project ID  \lq\lq Can Economic Policy Mitigate Climate-Change?\rq\rq, and the Swiss Platform for Advanced Scientific Computing (PASC), under project ID \lq\lq Computing Equilibria in Heterogeneous Agent Macro Models on Contemporary HPC Platforms\rq\rq, for research support, and the Swiss National Supercomputing Center (CSCS) under project ID 995.}}
\author{Doris Folini\thanks{Institute for Atmospheric and Climate Science, ETHZ; Email:
    \href{mailto:doris.folini@env.ethz.ch}{doris.folini@env.ethz.ch}.}, \;
  Felix K\"ubler\thanks{Department for Banking and Finance, University of Z\"urich; Swiss Finance Institute (SFI); Email: \href{mailto:fkubler@gmail.com}{felix.kuebler@bf.uzh.ch}.}, \;
  Aleksandra Malova\thanks{Department of Economics, University of Lausanne; Email: \href{mailto:malova.alex@unil.ch}{aleksandra.malova@unil.ch}.}, \;
  Simon Scheidegger\thanks{Department of Economics, University of Lausanne; Enterprise for Society (E4S); Email: \href{mailto:simon.scheidegger@unil.ch}{simon.scheidegger@unil.ch}.}}
\date{\today}

\begin{document}
\maketitle

\begin{abstract}

To analyze climate change mitigation strategies, economists rely on simplified climate models -- so-called climate emulators -- that provide a realistic quantitative link between CO2 emissions and global warming at low computational costs. 
In this paper, we propose a generic and transparent calibration and evaluation strategy for these climate emulators that is based on freely and easily accessible state-of-the-art benchmark data from climate sciences. We demonstrate that the appropriate choice of the free model parameters can be of key relevance for the predicted social cost of carbon. The key idea we put forward is to calibrate the simplified climate models to benchmark data from comprehensive global climate models that took part in the Coupled Model Intercomparison Project, Phase 5 (CMIP5). In particular, we propose to use four different test cases that are considered pivotal in the climate science literature: two highly idealized tests to separately calibrate and evaluate the carbon cycle and temperature response, an idealized test to quantify the transient climate response, and a final test to evaluate the performance for scenarios close to those arising from economic models, and that include exogenous forcing.
As a concrete example, we re-calibrate the climate part of the widely used DICE-2016, fathoming the CMIP5 uncertainty range of model responses: the multi-model mean as well as extreme, but still permissible climate sensitivities and carbon cycle responses. We demonstrate that the functional form of the climate emulator of the DICE-2016 model is fit for purpose, despite its simplicity, but its carbon cycle and temperature equations are miscalibrated, leading to the conclusion that one may want to be skeptical about predictions derived from DICE-2016. 
We examine the importance of the calibration for the social cost of carbon in the context of a partial equilibrium setting where interest rates are exogenous, as well as the simple general equilibrium setting from DICE-2016. We find that the model uncertainty from different consistent calibrations of the climate system can change the social cost of carbon by a factor of four if one assumes a quadratic damage function.
When calibrated to the multi-model mean, our model predicts similar values for the social cost of carbon as the original DICE-2016, but with a strongly reduced sensitivity to the discount rate and about one degree less long-term warming.
The social cost of carbon in DICE-2016 is oversensitive to the discount rate, leading to extreme comparative statics responses to changes in preferences.

  \end{abstract}



\newpage

\section{Introduction}
\label{sec:intro}

Anthropogenic climate change and the associated economic damages constitute a substantial negative externality from CO2 emissions. Economic policy can mitigate this externality and potentially lead to significant welfare gains across all economic agents (see, e.g.,~\cite{Nordhaus2018,kotlikoff2021making,kotlikoff-2021_UQ,KKPS}). In order to determine the optimal mitigation strategies, economists need to develop quantitative models that produce a realistic link between CO2 emissions and global warming and that are informed by research in climate science as presented in the Intergovernmental Panel on Climate Change (IPCC)\footnote{For more details on IPCC, see, e.g.,~\url{https://www.ipcc.ch}.} reports, that is, the \lq\lq state-of-the-art\rq\rq \ in climate science. 

The backbone of the IPCC reports are the Coupled Model Intercomparison Projects (CMIP5 and CMIP6; see~\citet{taylor-et-al:12, eyring-et-al:16}), which bundle the output from a collection of global climate models (GCMs) that run on pre-defined future scenarios, notably future greenhouse gas (GHG) concentrations. The GCMs come essentially in two flavors: Earth System Models (ESMs), which cover biogeochemical processes like the carbon cycle and can take carbon emissions as input, and coupled Atmosphere-Ocean Global Climate Models (AOGCMs), which lack biogeochemical processes and can take only GHG concentrations as input. 
One fundamental challenge is that the computational costs for ESMs (and for AOGCMs) are so significant that they are not suitable for studying the two-way feedback between the earth system and human behavior.\footnote{See, e.g.,~\citet{danabasoglu-et-al:20}, who state that the Community Earth System Model Version 2 (CESM2), one of the most popular ESMs, needs more than 3000 CPU hours to simulate one year of the global climate.} Therefore, economic models focusing on this feedback have to rely on a much-simplified representation of the Earth system component. Simplified and computationally cheap-to-evaluate climate models come at different levels of complexity (and computational costs) and under different names: Climate emulators (CEs), energy balance models (EBMs), simple climate models (SCMs), or ESMs of intermediate complexity (EMICs). We avoid using this differentiation throughout this paper and, somewhat loosely, just use CE or the climate model.

There is a proliferation of different CEs used in the climate-change economics literature. The most popular emulators can be found in the Dynamic Integrated Model of Climate and the Economy (DICE)~\citep{Nordhaus2018}, in PAGE~\citep{hope2013critical}, as well as the one in~\cite{Golosov2014}. Unfortunately, as~\cite{Calel2017} point out, \textit{\lq\lq in failing to maintain clear links to the physical science literature, the climate components of these models have become opaque to the scientific community.\rq\rq} Even worse, there is no consensus in the (economics) literature about what makes a \textit{\lq\lq good\rq\rq} climate model, that is, which data is one trying to match when choosing the model's functional forms and parameters.

We close this gap by developing a comprehensive and transparent suite of tests that aims to answer the following key questions: i) When is a particular CE used in economics fit for purpose?, and ii) Which calibration strategy should one, in general, apply so that the CE is in line with state-of-the-art climate science? We advocate the use of four test cases that are standard in climate science, notably in the context of CMIP~\citep{eyring-et-al:16, keller-et-al:18}. Two of the said tests will be used for the calibration of the CE, whereas another two tests are used for evaluating the CE. For CEs as used by economists, these four tests can be regarded as comprehensive. However, the said test cases are typically not employed by economists to evaluate their emulators. We showcase our proposed strategy, including economic effects at an example from the seminal DICE model family: we re-calibrate one specific version -- DICE-2016~\citep{Nordhaus2018}\footnote{Below, we use the abbreviation DICE-2016 and DICE interchangeably.} -- according to our battery of tests to match the output of climate science models, thereby obtaining CDICE, short for \lq\lq \textbf{C}alibrated \textbf{DICE}\rq\rq. In addition, we illustrate the economic importance of a proper calibration of the CE.
 
A key functionality of any CE is to translate anthropogenic emissions, as computed by the economic model, into a global mean temperature change. The task is typically split into two parts (see, e.g., DICE~\citep{Nordhaus2018}): a \lq\lq carbon cycle\rq\rq  \ which translates anthropogenic emissions in the wake of human economic activity into changes in the atmospheric CO2 concentration, and a temperature model which translates changes in atmospheric CO2 concentrations into global mean temperature changes. 
We propose to first calibrate both parts of the CE independently, using one test case for each calibration. Each of the two test cases comes, from a climate science perspective, with a plausible range of outcomes. To capture this range, we suggest for each test case one intermediate and two extreme calibration targets, to be detailed below. This leaves us with a total of nine differently calibrated CEs, three calibrations for the carbon cycle times three calibrations for the temperature part. Finally, we advocate evaluating the performance of these differently calibrated CEs against two additional climate science benchmarks, also detailed below, that are closer to real-world applications. The four tests and the associated CMIP5-based benchmark data are well consolidated and explored in the literature.

The first test we use targets the calibration of the carbon cycle. It uses a highly idealized setup; an instantaneous emission of 100 GtC to the present-day atmosphere is followed in time. The CE is calibrated such that the (steadily decreasing) fraction of emitted carbon remaining in the atmosphere is in line with benchmark data adopted from~\citet{joos2013carbon}. We choose three specific calibration targets - one intermediate and two extremes - to mirror the range of carbon cycle responses considered plausible from a climate science point of view. As an intermediate target, we use the average response of all models in~\citet{joos2013carbon}; their multi-model mean (MMM).\footnote{The multi-model mean is a debated, but regularly used benchmark quantity in climate sciences (see, e.g.,~\cite{mesmer-pattern-2020,Mariethoz_2021}, and references therein).} As extreme calibration targets, we use two specific models from the study by~\citet{joos2013carbon}, leaving either a very large (cf. their model ``MESMO'') or small (cf. their model ``LOVECLIM'') fraction of emitted carbon in the atmosphere in the long run~\citep[cf.][]{joos2013carbon}. In the following, to make the different calibration targets transparent where necessary, we add the name of the corresponding target model to the general name CDICE.  

The second test we use targets the calibration of the temperature equations. Being highly idealized, it follows the evolution of the global mean temperature after an instantaneous quadrupling of atmospheric CO2 concentrations with respect to pre-industrial values. The CE is calibrated to meet benchmark data anchored in CMIP5 and adopted from~\citet{Geoffroy2013}. The range of warming displayed by the different CMIP5 models\footnote{This range in CMIP5 does not become narrower when observational data are used as an additional constraint. The newer, next-generation CMIP6 data is being established just now and is thus not yet fit for the purpose of this paper. Preliminary analyses suggest a larger range that, in contrast to CMIP5, gets narrower if additional observation-based constraints are used~\citep{Tokarska-et-al:20}.} is considered plausible from a climate science point of view, strong or weak warming are plausible limiting cases~\citep{knutti-et-al:17}. We choose three specific calibration targets again to capture this range: the MMM, as well as two models that show a rather strong (model HadGEM2-ES, Hadley Centre Global Environmental Model 2) or weak (model GISS-E2-R, Goddard Institute for Space Studies Model E) warming in the long run within CMIP5.

Any full-fledged quantitative economic treatment of climate change must take this  ``model uncertainty'' seriously. Even in a simple representative agent framework, the attitude of the representative agent towards risk and uncertainty becomes a crucial determinant for optimal carbon policy (see, e.g.,~\cite{barnett2020pricing,barnett2022climate}). To assess the effect of climate model uncertainty on economic models, the so-called equilibrium climate sensitivity (ECS) parameter is often varied within the range observed in CMIP5 models (see, e.g.,~\cite{Hassler2018a}, or~\cite{Nordhaus2018a}). The ECS measures the long-term increase in global average temperature expected to occur after an instantaneous doubling of the atmospheric CO2 concentration; different projected warmings imply a different ECS~\citep{Geoffroy2013}. While being endogenously determined in AOGCMs, the ECS is an exogenous parameter determining long-run warming in CEs. The approach of just changing the ECS parameter to sample the uncertainty range is often a useful shortcut, but it misses part of the relevant temperature dynamics. We will show below that it makes a difference whether one changes only the ECS of the CE or, as we do, really calibrates the CE against CMIP5 model data whose ECS (high for HadGEM2-ES, low for GISS-E2-R) can then be associated with upper and lower bounds on the true costs of climate change and also provide an average scenario (MMM) that can, perhaps, be viewed as \lq \lq most likely\rq \rq. 

The third test evaluates the CE's transient climate response (TCR), using the same setup as in climate science: starting in 1850 at pre-industrial levels, the atmospheric CO2 concentration gradually increases at one percent per year until doubling after 70 years (quadrupling after 140 years). The TCR is defined as the warming after 70 years. It is not calibrated for but, rather, is an emergent property of both the CE and CMIP5 models - or any AOGCM or ESM in general. This test allows comparing the TCR of the CE with values from CMIP5. In fact, the TCR is a standard test in climate sciences and a compulsory test for any GCM to take part in CMIP, see~\citet{eyring-et-al:16}. As the test takes CO2 concentrations as input, not emissions, it addresses only the temperature equations of the CE. 

The fourth test case is closer to what integrated assessment models (IAMs) are expected to cope with in real applications: the CMIP5 simulations for the historical period and future representative concentration pathways (RCPs) for three different forcing levels (RCP26, RCP45, RCP60, RCP85).\footnote{The scenarios describe the results of different socioeconomic narratives that produce particular concentration profiles of greenhouse gases, aerosols, and other climatically relevant forcing agents over the 21st century. The RCP85 scenario, for instance, reflects a \lq\lq no policy\rq \rq \ narrative, in which total anthropogenic forcing reaches approximately $8.5 W/m^2$ in the year 2100. Conversely, the RCP26 scenario involves aggressive decarbonization, causing radiative forcing to peak at approximately $3 W/m^2$ around 2050 and to decline to approximately $2.6 W/m^2$ at the end of the 21st century.} Input data are either prescribed atmospheric CO2 concentrations or CO2 emissions.\footnote{\citet{meinshausen-et-al:11}, http://www.pik-potsdam.de/\textasciitilde mmalte/rcps.} In the latter case, the entire climate model is tested; in the former case, only its temperature equations. Within the CE, an assumption has to be made on non-CO2 forcings, which form part of the CMIP5 data. The CE climate is evaluated against data from the CMIP5 archive.


To exemplify our calibration strategy, we apply it to the climate part of the widely used DICE-2016 model by~\cite{Nordhaus2018}. As~\citet{Dietz2020} point out, the most commonly used CEs in economics, including Nordhaus' widely used DICE-2016 model, seem seriously flawed in that they cannot reproduce the evolution of atmospheric CO2 and temperature for the basic test case of an instantaneous carbon pulse to the present day atmosphere~\citep{joos2013carbon}. Based on this one test case, which mixes the response of the carbon cycle and the temperature equations, they conclude that \textit{\lq\lq economic models of climate change are out of line with state of the art in climate science\rq\rq}, and they recommend that \textit{\rq\rq the climate modules in economic models be replaced.\rq\rq} 
While we confirm the finding by~\citet{Dietz2020} that the carbon cycle of DICE-2016 leaves too much CO2 in the atmosphere and its temperature response to a sudden increase in CO2 concentration is too slow, our study also demonstrates that their conclusion that the climate model must be replaced is incorrect. In particular, we show that the functional form of the CE in DICE-2016 is fit for purpose and can be re-calibrated to match results from our test cases with respect to the multi-model mean as well as with regard to CMIP5  extreme cases. For the carbon cycle, no perfect calibration is possible as the functional form in DICE and the benchmark data from~\cite{joos2013carbon} are different. By contrast, the temperature equations in DICE are formally identical to those in~\citet{Geoffroy2013}, thus the calibration target can be matched exactly.\footnote{The original formulation of DICE hides this similarity somewhat, as it absorbs the time step of the climate part in the coefficients of the equations. We offer a more transparent formulation where the time step is explicitly exposed, which allows us to directly use coefficients from~\citet{Geoffroy2013} that are already calibrated to match the CMIP5 results.} Our study goes beyond~\citet{Dietz2020} in that we present a comprehensive set of tests to ascertain transparently whether a CE is fit for purpose and, if not, we also present means to re-calibrate a CE properly on the condition that its functional form is fit for purpose.

To examine the impact of different climate calibrations on the optimal price of carbon, we first consider a very simple partial equilibrium setup where we take the path of emissions, the growth rate, and the interest rate as exogenously given. As it is standard in economic models of climate change (see, e.g.~\cite{Golosov2014,Nordhaus_1979}), we assume that an aggregate damage function determines what fraction of output is destroyed as a function of the temperature. In these (deterministic) models, the optimal carbon tax equals the social cost of carbon (SCC), which can be computed as the marginal cost of carbon emissions, that is, as the sum of all future damages, discounted at the market interest rate, that results from an infinitesimal extra emission of CO2 into the atmosphere. The ratio of growth rate and interest rate then constitutes the correct discount factor for these calculations and, as it is well documented in the literature, one of the most important determinants of the SCC.
We show that for a given discount factor and emissions scenario, the model uncertainty from climate science can change the SCC by a factor of four if one assumes a quadratic damage function; one of the most common functional forms in the literature. This finding is (roughly) independent of discounting and emissions, but depends crucially on the functional form of the damage function. Simply adjusting the ECS in the equations of the climate emulator leads to significantly different estimates for the SCC than using our full-blown calibrations of extreme climate models. In this simple partial equilibrium setting, the incorrect calibration of the CE in DICE-2016 leads to an SCC that is oversensitive to the discount rate. This high sensitivity has important consequences for a general equilibrium setting, where the interest rates are endogenous.

In order to investigate the economic consequences of our different calibrations in a general equilibrium framework, we use the economic model from DICE-2016 together with our newly calibrated versions of ``CDICE'' to compute the SCC and the optimal mitigation. We find that in our re-calibrated model, when updated with respect to the MMM, the SCC and the optimal mitigation are very similar to the values found in DICE-2016 if we take the exact same economic parameters.\footnote{A significant difference is that the long-run temperature in DICE-2016 is substantially higher, that is, about one degree Celsius, higher than in our calibration. This finding holds both for the business as usual (BAU) scenario without any mitigation and the optimally mitigated scenario.}
We point out that for the BAU scenario, the temperature increases in DICE-2016 are below those predicted by our extreme calibration, that is, CDICE-HadGEM2-ES. This finding implies that the DICE-2016 predictions are within the range of what is considered plausible in climate science. However, being in the plausible range is merely caused by the fact that two different flaws in the original calibration of the temperature equations and the carbon cycle in DICE-2016 partially offset each other. In particular, the calibration gives the wrong results for the optimal economic response to climate change. In the optimal mitigation scenario, carbon taxes turn out to be too low in the sense that the temperature response of DICE-2016 now falls outside of the CMIP5 range. As a result, the model with an optimal carbon policy predicts a higher temperature than CDICE-HadGEM2-ES.

As pointed out above, the SCC in the DICE-2016 calibration is oversensitive to the discount rate. The role of the rate of time-preference of the social planner / representative agent for the SCC and optimal carbon taxes has been extensively documented and discussed in the literature (see, e.g.,~\cite{barrage2018careful,Hansel2018}). While this parameter is obviously important for the SCC, we show that a large part of the quantitative results are driven by an incorrect calibration of the climate module in DICE-2016. Moreover, in a model with time-varying growth rates (as DICE-2016), it is not only the rate of time preference, but also the curvature of cardinal utility that determines the discount rates. We show that this implies that the DICE-2016 climate calibration does not produce reliable results for important comparative statics questions.


The remainder of the paper is organized as follows.
In section~\ref{see isc:liter}, we briefly connect our paper to the related literature. In section~\ref{sec:methods}, we introduce our four test cases in detail. Section~\ref{sec:climate} explains how to re-calibrate the parameters in the climate part of DICE to produce forecasts that are consistent with our test cases. Section \ref{sec:peSCC} explores the implications of different CE calibrations for the SCC in a simple partial equilibrium model. Subsequently, in section~\ref{sec:simulation}, we use the re-calibrated DICE model, that is, CDICE, to study the role of the climate calibration in the economic model of DICE-2016. Section~\ref{sec:conclusion} concludes. In addition, we provide for the convenience of the reader a complete specification of the different variants of the DICE model (some of which we use in our numerical experiments below) in Appendix~\ref{sec:generic}, including the DICE versions from 2007, 2016, as well as the re-calibration that we suggest in this paper, that is, CDICE.


\section{Literature}
\label{see isc:liter}


Starting with Nordhaus' seminal work on climate change~\citep{Nordhaus_1979}--a massive field of research was spawned, including Nordhaus' own development of the DICE model (see, e.g.,~\citet{Nordhaus2012,Nordhaus2013,Nordhaus2018}).
For example,~\citet{Anderson2014} as well as~\citet{Miftakhova2018} conducted uncertainty quantification, whereas~\citet{Cai2019,Lemoine2014}, and~\cite{lontzek2015stochastic} added stochastic tipping points on top of the DICE-2007 model.~\citet{Hwang2017} performed learning about equilibrium climate sensitivity with fat tails, whereas~\citet{Popp2004} studied endogenous growth in the DICE framework.~\citet{WouterBotzen2012} and~\cite{Michaelis2020} studied DICE with an alternative damage function. The fact that DICE was criticized for a flawed climate model (in particular by~\cite{Dietz2020}) puts the quantitative predictions of these contributions into a questionable position and motivates our study. We show below that the climate model in DICE-2016 can be re-calibrated to perform well in our tests.

\cite{Golosov2014} develop a simple CE that is at present widely used as an alternative to the climate module of DICE (see, e.g.,~\cite{Hassler2018,Hassler2019,kotlikoff-2021_UQ}). Due to its simplicity, the model is computationally much more tractable than DICE. However, the simplicity of this model implies that certain stylized facts of temperature changes in response to CO2 in the atmosphere cannot be captured adequately.\footnote{For example, since in \cite{Golosov2014} there is only one temperature equation that links CO2 in the atmosphere directly to the current temperature, the temperature must decrease as soon as there are net-zero emissions and CO2 concentration in the atmosphere decreases. We will show in section \ref{SCC:CDICE} that the timing of warming plays an important role for the SCC.} This model, therefore, does not lend itself to our calibration strategy.\footnote{Interestingly, ~\cite{Dietz2020} argue that although the model in~\cite{Golosov2014} is very simple, it matches some important stylized facts that can be derived from CMIP5 data better than the original DICE-2016 calibration.}

More directly related to our paper, there have been several important contributions about the calibration of CEs.~\cite{miftakhova2020statistical} propose a general emulation method for constructing low-dimensional approximations of complex dynamic climate models and develop an emulator for MAGICC to approximate the impact of emissions on global temperature.~\cite{cai2019climate} calibrate a spatial climate system to match four RCP scenarios and historical data.~\citet{Traeger2014} and~\citet{Cai2012,Cai2019} re-calibrate the DICE CE from a ten-year time step (as in the DICE-2007 formulation) or five-year time step (as in the DICE-2016 formulation) to higher frequency data. In the climate-science literature,~\citet{Calel2017} discuss the physics of different climate models, including DICE.~\citet{Thompson2018} conducted an intercomparison study of models, including DICE to assess how they have been designed from a physics point of view.~\citet{Geoffroy2013} show how two-layer energy-balance models (formally identical to the temperature equations in DICE) can be used to match the results in CMIP5.~\cite{joos2013carbon} quantify responses to emission pulses in a carbon cycle intercomparison project. Our analysis relies heavily on these two latter papers. In particular, we use~\citet{Geoffroy2013}'s calibration of the temperature equation and take our carbon-cycle benchmarks from~\cite{joos2013carbon}. Our paper is closest to~\cite{Dietz2020}, and we compare our contributions in more detail in the following sections.

\section{A comprehensive framework to calibrate the climate in IAMs}
\label{sec:methods}

Naively, one would think that existing historical data suffices for calibrating and evaluating CEs used in IAMs. However, it turns out that, broadly speaking, the signal-to-noise ratio in that sort of data is simply too low for a reliable calibration of CEs with their focus on forcing from CO2.  Put differently, in observed records of annual global mean temperature, the anthropogenic signal due to CO2 emissions comes in combination with a range of other relevant effects. The change in global mean temperature over the last 150 years is not only due to changes in CO2. It also bears substantial imprints from other greenhouse gases, from aerosols of anthropogenic or natural origins, including volcanoes, and also from land use changes~\citep{gambhir-et-al:17, mengis2020non}. Reliably disentangling the different contributions is challenging. To illustrate, \cite{millar-friedlingstein:18} estimate that from 1880 to 2015, CO2 contributed about 0.7 degrees to global warming, whereas non-CO2 agents contributed about 0.3 degrees, with 5\% to 95\% uncertainty ranging from 0.5 to 1.5 degrees and -0.2 to +0.4 degrees, respectively (cf. their Figure 4). Hence, we need to base our suite of tests on historical observations, but also on the so-called third pillar of science: computational experiments. 

As detailed in the introduction, the CMIP experiments are a logical choice for this purpose. The range of models participating in CMIP then raises the issue that for any specific CMIP experiment (or CE test), one faces not a single benchmark, but a range of benchmarks. Given that the full range is considered plausible by climate scientists, it is important that any CE captures this model diversity in some way. In this light, a first and primary benchmark is the MMM,\footnote{For a recent application of the MMM, the multi-model-mean, in climate science, see, e.g.,~\cite{mesmer-pattern-2020}.} the average of all cases reported in CMIP5. The MMM is complemented by extreme cases bracketing the CMIP5 range. Two extreme cases that bracket the plausible temperature response to CO2 in the atmosphere in CMIP5 are the HadGEM2-ES model (strongly warming, ECS of 4.55 K) and the GISS-E2-R  model (weakly warming, ECS of 2.15 K)~\citep{Geoffroy2013}. Two extreme cases that bracket the amount of emitted carbon remaining in the atmosphere are MESMO and LOVECLIM, with about 55\% and 30\%, respectively, of the 100 GtC test pulse mass remaining in the atmosphere after 100 years, according to Figure 1a from~\citet{joos2013carbon}.

\subsection{Choice of test cases}
\label{sec:choice_of_bm}
%
The choice of the four test cases outlined in the introduction, that is, two for calibration, and two for evaluation of the CE, was guided by the following considerations. First, all test cases should be widely used in the relevant climate science literature and transparently documented. Thus, they can easily be reproduced and compared with published results. Second, some of the tests should be highly idealized to allow for independent calibration of the carbon cycle and the temperature response, respectively. Third, we want some test cases to be close to the ultimately envisaged {\it real-world applications}, with gradual changes in CO2 emissions, exogenous forcing, and associated temperature response. The proximity to real applications makes these tests kind of a gold standard, but also makes them more difficult to interpret than the highly idealized tests.

Taken together, the four test cases we propose to use may, loosely speaking, be regarded as \textit{necessary and sufficient} to answer whether an emulator is fit for purpose. They allow calibration and comprehensive evaluation of a CE like the one in DICE, from its individual components to real application cases. To this end, it is crucial that highly idealized settings, which are frequently used as benchmarks in the literature, are augmented with closer to reality tests. We advocate that the same four tests may be used to examine the performance of the climate parts of other IAMs featuring a carbon cycle and a set of temperature equations. Any climate model that fits the CMIP5 data for the four test cases will likely provide a good fit for all emissions paths arising in economic modeling.     

\subsection{Technical setup of test cases}
\label{sec:description_of_bm}

The first of four experiments in our test suite is used to calibrate the carbon cycle. Its highly idealized setup follows the temporal evolution of an instantaneous release of 100 Gt carbon (GtC) to the present-day atmosphere. The said 100 GtC roughly corresponds to ten years of present-day CO2 emissions from fossil fuels~\citep{lequere-et-al:18}. The test requires two steps, which we outline here. For more detailed explanations of the test and its benchmark data, we refer to~\citet{joos2013carbon}. In a first step, the CE is used to simulate the evolution of the climate from its 1850 equilibrium to 2015, using as input historical carbon emissions from 1850 to 2015. The simulation is then continued from 2015 onward, with annual carbon emissions being iteratively determined to keep the atmospheric CO2 concentration constant at the value of the year 2015. These emissions are then used in a second simulation that starts in 2015, but now with an additional, instantaneous pulse of 100 GtC to the atmosphere. Changes in atmospheric CO2 are then taken as the difference between this second and the first simulation. The test is used to calibrate the time scales and rates at which the atmospheric CO2 concentration decays in the wake of instantaneous carbon release to the atmosphere.\footnote{Depending on the concrete formulation of the CE's carbon cycle, the decay of atmospheric CO2 may result in increasing CO2 concentrations in other carbon reservoirs, notably ocean or land.}

The second test case in our proposed framework is used to calibrate the amplitude and time scales of a given emulator's temperature response, that is, how fast and how strongly the global mean temperature warms in response to a sudden quadrupling of CO2.\footnote{Note that while the test described here applies to any emulator, it will in particular target Equations~\eqref{eq:temp1} and~\eqref{eq:temp2} of the DICE-2016 model (cf. section~\ref{sec:climate} below).} Starting from pre-industrial equilibrium conditions, that is, in the year 1850, with atmospheric CO2 at 285 parts per million (ppm), the atmospheric CO2 concentration is instantaneously quadrupled, and the change in temperature with time is examined. The related benchmark data is anchored in CMIP5 and adopted from~\citet{Geoffroy2013}. 

The third test case and its benchmark data come directly from CMIP5: starting in 1850 at pre-industrial levels, the atmospheric CO2 concentration gradually increases from 285 ppm CO2 at one percent per year until quadrupling after 140 years. It is used to evaluate the TCR of the CE in comparison with the CMPI5 values of TCR. The test is still relatively simple, as it involves the temperature part of the CE solely, but not the carbon cycle. That is, it uses CO2 concentrations as input, not emissions. It involves forcing only from CO2. The change in CO2 concentration is not an instantaneous step function, but is gradual in time. The test thus examines a situation where the temperature response time scales overlap with an additional time scale associated with the gradually increasing forcing from CO2. 

\begin{figure}[t!]
  \centering
  \includegraphics[scale=0.32]{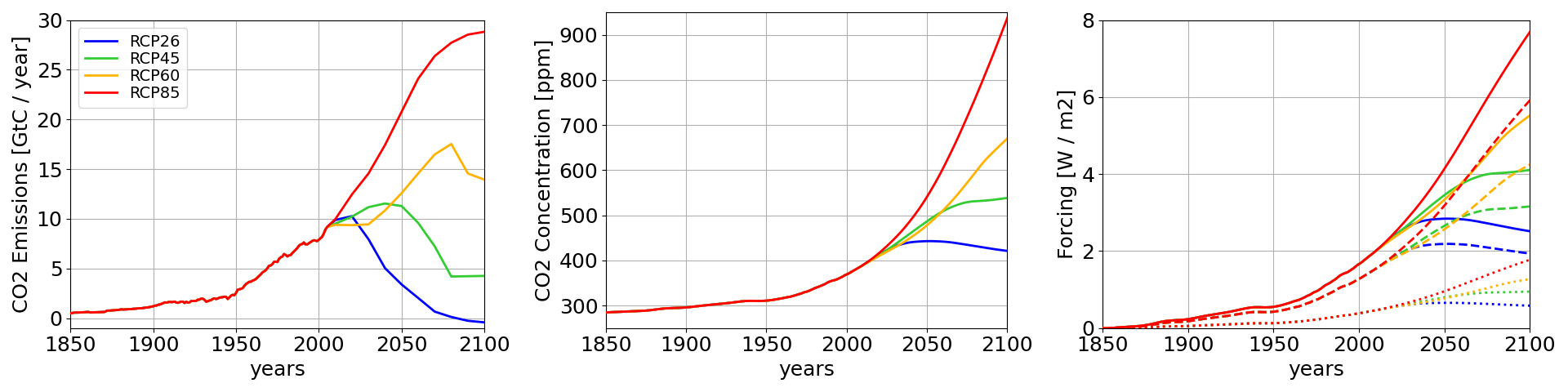}
  \captionof{figure}{Illustration of CMIP5 historical and future evolution (RCP26, RCP45, RCP60, and RCP85 scenarios, shown in blue, green, yellow, and red), from 1850 to 2100, of prescribed CO2 emissions (left, in GtC per year), alternatively prescribed CO2 concentrations (middle, in ppm CO2), and the forcing derived from prescribed CO2 concentrations (right, in W/m$^{2}$). The total forcing $F_{t}$ (solid lines) is decomposed into forcing from CO2 ($F^{\text{CO2}}_{t}$, dashed lines) and non-CO2 forcing assumed as $F^{\text{EX}}_{t} = 0.3 \cdot F^{\text{CO2}}_{t}$ (dotted lines).}
  \label{fig:CMIP_EMI_CONC_FORC}
  \end{figure}

The fourth test case is close to what DICE or some other IAM is expected to cope with in real applications: the CMIP5 simulations for the historical period and future representative concentration pathways (RCPs) for four different forcing levels of 2.6 W/m$^2$, 4.5 W/m$^2$, 6.0 W/m$^2$, and 8.5 W/m$^2$ by 2100 (RCP26, RCP45, RCP60, RCP85). The input data consists of either prescribed atmospheric CO2 concentrations or CO2 emissions~\citep[][http://www.pik-potsdam.de/\textasciitilde mmalte/rcps/]{meinshausen-et-al:11}, as illustrated in  the left and middle panel of Figure~\ref{fig:CMIP_EMI_CONC_FORC}. The former case tests only the temperature response to gradually changing CO2 concentrations. The latter case tests the full climate model of the CE, including the carbon cycle. In this test case, it is necessary to make an assumption on the non-CO2, exogenous radiative forcing $F^{\text{EX}}_{t}$, which in the actual CMIP simulations is modeled in great detail. In Appendix~\ref{sec:fex}, we show how this choice affects quantitative results of the full model. Here, we assume $F^{\text{EX}}_{t} = 0.3 \cdot F^{\text{CO2}}_{t}$, that is, proportional to the forcing from CO2 alone  ($F^{\text{CO2}}_{t}$) and resulting in a total radiative forcing, $F_{t} = F^{\text{CO2}}_{t} + F^{\text{EX}}_{t}$, which is 30\% larger than from CO2 alone (cf. section~\ref{sec:meth_temperature} below). This is in line with estimates for the radiative forcing in 2011 with respect to pre-industrial times~\citep[see, e.g.,][]{ClimateChange2014SynthesisReportIPCC2014, gambhir-et-al:17} and with estimates for the different RCPs in 2100: 26\% (RCP26), 33\% (RCP45), 32\% (RCP60), 35\% (RCP85), where the total forcing comes from the scenario and the CO2 forcing is computed from concentrations given in~\citet{meinshausen-et-al:11}, a base concentration of 285 ppm CO2, and $F_{\text{2XCO2}} = 3.68$ W/m$^2$. An illustration of the temporal evolution of the different forcings is given in Figure~\ref{fig:CMIP_EMI_CONC_FORC}, right panel. Note that estimates for the forcing from non-CO2 GHGs by~\citet{macdougall-et-al:13} (their Figure 6) show more temporal structure than $F^{\text{EX}}_{t}$. Additionally, the assumption of strict proportionality between CO2 and non-CO2 forcings is subject to debate in climate sciences, especially in the context of strong mitigation scenarios~\citep[see, e.g.,][]{mengis2020non}. Benchmark data for the evolution of atmospheric temperature is taken directly from the CMIP5 archive. The proximity to real applications makes this test difficult to interpret, yet it is a key test that CEs in economic models should pass. 
%
%

\subsection{Metric for performance evaluation}
\label{sec:metric}
%
%
The choice of a metric to assess the calibration and subsequent evaluation of a CE is not straightforward. The basic reason lies in the design of the CE, in its reduced functional form, and the limited number of free parameters as compared to ESMs, which capture only some aspects of the benchmark cases used for calibration and evaluation. This also implies that the choice of metric may depend on the concrete form of the CE, the simplicity or complexity of its carbon cycle, and temperature equations. If the CE explicitly covers further elements, like, for example, methane, then not only the metric but also the benchmarks for calibration and evaluation may need adaptation. 

Focusing on simple CEs, like the one in DICE, one prominent difference with respect to ESMs concerns the response time scales upon perturbation: an ESM features a whole range of response time scales, whereas the CE only a few, for example, two in the case of DICE-2016. Furthermore, an ESM explicitly covers a range of forcings; the CE explicitly covers only CO2 via its carbon cycle, while any other forcings are included in a more or less ad hoc fashion. The temperature may affect the carbon cycle in an ESM, but not in the simple CE considered here. A comprehensive discussion of why to favor any particular metric in view of these issues is beyond the scope of the present paper. Therefore, we limit ourselves to a short statement of what we did and why. 

We calibrate the carbon cycle of a CE via the response to an instantaneous 100 GtC pulse to the atmosphere. The comparison is performed against benchmark data from~\citet{joos2013carbon}, using the maximum norm over a time range of 100 years after the pulse. This choice of metric favors solutions, that is, calibrations that remain close to the benchmark data at all times over a time scale of interest in an economic context. The calibration of the temperature equations does not require any metric because their functional form is identical to the benchmark data from~\citet{Geoffroy2013}. Therefore, we can just adopt parameter values from the benchmark data.

We evaluate a calibrated CE first with regard to TCR, and we expose the CE to the corresponding test case from climate science, where TCR is defined as the change in global mean temperature after 70 years (and after 140 years) when the atmospheric CO2 concentration steadily increases from 1850 onward at a rate of one percent per year. We measure the performance of the CE with respect to CMIP5 models by comparing the TCR. Finally, we evaluate the performance of the calibrated CE with regard to climate projections and historical simulation data from CMIP5. Here, we explicitly renounce using a specific metric for mainly two reasons: the CMIP5 models span a range of plausible climates, and they explicitly include non-CO2 forcings that depend on the projection (RCP26, RCP45, or RCP85) and are included in the CE only in an ad hoc fashion. We consider the CE fit for purpose if a visual inspection shows that its climate in terms of atmospheric CO2 concentration and the global mean temperature falls within the range of CMIP5 models.

\section{CDICE - re-calibrating the climate of DICE}
\label{sec:climate}
%
%

So far, we put forward our suite of four test cases: one test to calibrate the carbon cycle of the CE, one test case to calibrate the temperature equation of the CE, and two test cases to examine emerging properties of the calibrated CE, notably its performance under gradually changing conditions in terms of CO2 and non-CO2 forcings. We claim that, taken together, these tests are necessary and sufficient to transparently evaluate and, if needed, re-calibrate the CE of an IAM, which is used in an economic context. 

In this section, we illustrate our claim with the example of the climate module of the seminal DICE-2016 model. We choose DICE-2016 as an example because it is widely used in the IAM modeling community and because it is known to be flawed~\citep{Dietz2020}. We re-examine these flaws in the light of our four test cases and present an improved version, CDICE, that has the same functional form as DICE-2016 but is fully re-calibrated. We thereby go beyond the work of~\citet{Dietz2020} in three essential points: i) while~\citet{Dietz2020} use one test, the 100 GtC pulse, to evaluate the CE, we stress that it is crucial to separately examine each part of the CE, carbon cycle, and temperature equations; ii) we provide corresponding test setups and show how they can be used to calibrate a CE, and not only to examine it; iii) we demonstrate that even a simple CE like the one in DICE is fit for purpose if properly calibrated in this way.

This section is structured as follows: First, we lay down in section~\ref{sec:meth_clim} the fundamental building blocks of DICE's CE, its carbon cycle and temperature equations, along with the commonly used calibration of DICE-2016. Next, we demonstrate in sections~\ref{sec:test_100GtC},~\ref{sec:test_4xco2},~\ref{sec:test_1pctco2}, and~\ref{sec:test_cmip5_rcp} how our battery of tests can be used to systematically test and re-calibrate DICE-2016. Finally, a critical synthesis of our learnings is provided in section~\ref{sec:test_synthesis}. A summary of all CDICE model parameters is given in Table~\ref{table:model_calibration} in Appendix~\ref{sec:models_summary}.
%
%
\subsection{Model equations}
\label{sec:meth_clim}
%
%
The climate model in DICE-2016 translates carbon emissions from the model's economic side into atmospheric CO2 concentrations (i.e., the carbon cycle; cf. section~\ref{sec:meth_carbcycle} below) and, subsequently, into a global mean temperature change (i.e., the temperature equations; cf. section~\ref{sec:meth_temperature} below), which then feeds back into the part of the model that represents the economic side of the IAM. Thus, the carbon cycle affects temperature, but not vice versa.

The carbon cycle is a linear three-box model, where the three carbon reservoirs represent the atmosphere (AT), the upper ocean (UO), and the lower ocean (LO), with the respective carbon masses $M = (M^{\text{AT}}, M^{\text{UO}}, M^{\text{LO}})$.  Carbon can be exchanged between the atmosphere and upper ocean and between the upper and lower ocean, but not directly between the atmosphere and lower ocean. The global mean temperature is modeled via a system of two ordinary differential equations that couple two heat reservoirs, that is, the atmosphere, including the upper ocean, and the lower ocean, $T = (T^{\text{AT}}, T^{\text{LO}})$. The climate is anchored at an assumed pre-industrial equilibrium state, $M_{\text{EQ}}$ and $T_{\text{EQ}}$. Climate change is quantified as deviations from this equilibrium state. In an economic context, time integration in DICE-2016 typically does not start at pre-industrial times, but at present-day, for example, in the year 2015. A corresponding initial state $M_{\text{INI}}$ and $T_{\text{INI}}$ must be determined so as to be in line with carbon emissions since pre-industrial times and the equations of the CE (cf. section~\ref{sec:meth_init}). 
%
%
\subsubsection{Carbon cycle}
\label{sec:meth_carbcycle}
%
%
The carbon cycle model~\citep[see, e.g.,][]{Keeling1973} formally reads
\begin{align}
    M_{t+1} = (I + \Delta_t \cdot B ) \cdot M_{t} + \Delta_t \cdot  E_{t},
\label{eq:cc_mass}
\end{align}
with $I$ being the identity matrix, $\Delta t$ being the time step in years, and $M_{t} = (M^{\text{AT}}_t, M^{\text{UO}}_t, M^{\text{LO}}_t)$ representing the carbon mass at time $t$ in the three reservoirs. The carbon emissions per year to the atmosphere, as well as to the ocean\footnote{This is an interesting option in the context of Carbon Dioxide Removal (CDR) techniques~\citep[see, e.g.,][]{rickels-et-al:18}.}, are specified via $E_{t}$. The time-constant matrix $B$, 
\begin{align}
B = \left(\begin{array}{ccc} 
b_{11} & b_{21} & b_{31} \\
b_{12} & b_{22} & b_{32} \\
b_{13} & b_{23} & b_{33} 
\end{array}\right),
\label{eq:matrix_B}
\end{align}
describes the mass transfer among reservoirs and has units \lq \lq mass fraction per time step\rq \rq. Assuming that mass conservation holds, that is,
\begin{equation}
    \sum_{i} b_{ji} = 0 \text{ for } j = 1,2,3,
\end{equation}
and that there is no direct mass transfer between AT and LO (implying that $b_{13} = b_{31} = 0$), leaves four free parameters in matrix $B$ that are used to calibrate the carbon cycle. In DICE, the said parameters are chosen as the transfer coefficients from AT to UO ($b_{12}$) and from UO to LO ($b_{23}$), as well as the equilibrium carbon mass ratios at pre-industrial times between the reservoirs, $r_{1} = M^{\text{AT}}_{\text{EQ}} / M^{\text{UO}}_{\text{EQ}}$ and $r_{2} = M^{\text{UO}}_{\text{EQ}} / M^{\text{LO}}_{\text{EQ}}$.
The remaining matrix entries $b_{ij}$ are then given by $b_{11} = -b_{12}$; $b_{21} = b_{12} \cdot r_{1}$; $b_{22} = -b_{21}-b_{23}$; $b_{32} = b_{23} \cdot r_{2}$; $b_{33} = -b_{32}$. 
Parameter values for CDICE are given in Table~\ref{table:FreeParamsClimDICE2}. In DICE-2016, the numerical values for the free parameters are given by $M_{\text{EQ}} = (588, 360, 1720)$, $r_{1} = 1.633$, $r_{2} = 0.209$, as well as $b_{12}=0.12$ and $b_{23}=0.007$, with a time step of $\Delta t = 5$ years being hard-wired (absorbed) into the coefficients $b_{ij}$. To alleviate this limitation of being bound to a time step of fixed size, the $b_{ij}$ in our formulation are specified in units of one year and are then explicitly multiplied with a time-step. The latter thus can be freely chosen, for example, as the time-step prescribed by the respective economic model.
%
%
\subsubsection{Temperature}
\label{sec:meth_temperature}
%

The two-layer energy balance model in DICE-2016 reads as
\begin{align}
    T^{\text{AT}}_{t+1} &= T^{\text{AT}}_{t} + \Delta_t \cdot c_{1} \left(F_{t} - \lambda T^{\text{AT}}_{t} - c_{3}\left(T^{\text{AT}}_{t} - T^{\text{OC}}_{t}\right)\right), \label{eq:temp1}\\
    T^{\text{OC}}_{t+1} &= T^{\text{OC}}_{t} + \Delta_t \cdot c_{4} \left(T^{\text{AT}}_{t} - T^{\text{OC}}_{t}\right),
\label{eq:temp2}
\end{align}
and thus formally corresponds to the one described in~\citet{Geoffroy2013}. $T^{\text{AT}}_{t}$ and $T^{\text{OC}}_{t}$ denote the temperature change with respect to pre-industrial times of the upper layer (atmosphere and upper ocean) as well as the lower layer (deep ocean), respectively, at time step $t$. The free parameters $c_{1}$, $c_{3}$, $c_{4}$, and $\lambda$ in ~\eqref{eq:temp1} and~\eqref{eq:temp2} are used for calibration. From a physics perspective, they may be interpreted as a heat exchange coefficient between the upper and lower layer, $\gamma  = c_{3}$, effective heat capacities of the upper and lower layer, $C = 1/c_{1}$ and $C_{0} = \gamma / c_{4}$, and the radiative feedback parameter $\lambda = F_{\text{2XCO2}}/T_{\text{2XCO2}}$, that is, the ratio of the forcing from a doubling of CO2 to the associated temperature change, that is, the ECS. Finally, $F_{t}$ denotes the total radiative forcing from CO2, $F^{\text{CO2}}_{t}$, and other exogenous factors, $F^{\text{EX}}_{t}$, such as GHGs other than CO2 and also aerosols:
\begin{align}
    F_{t} &= F_{\text{2XCO2}} \frac{\log(M^{\text{AT}}_{t}/M^{\text{AT}}_{\text{EQ}})}{\log(2)} + F^{\text{EX}}_{t}.
\label{eq:forcing}
\end{align}
In, DICE-2016, $F^{\text{EX}}_{t}$ is assumed to change linearly with time from 0.5 in the year 2015 to 1.0 in 2100. Parameter values for CDICE are given in Table~\ref{table:FreeParamsClimDICE}.
Parameter values in DICE-2016 are $c_{1}=0.1005$, $c_{3}=0.088$, $c_{4}=0.025$, and $\lambda = 3.6813/3.1 = 1.1875$, with a time step of $\Delta_t = 5$ years formally hard wired (absorbed) into $c_{1}$ and $c_{4}$. To lift this issue, we provide, here again, a generic formulation of the temperature equations such that the time-step can be chosen freely in discrete values of one year, that is, one year, two years, and so forth.
%
%
\subsubsection{Initialization at present day}
\label{sec:meth_init}
%
%
In the context of economic models, the time integration typically does not start at pre-industrial times but rather at the present day. Corresponding initial values for the carbon and temperature reservoirs used in CDICE for 2015 are  $M_{\text{INI}} = (851, 628, 1323)$ and $T_{\text{INI}} = (1.10, 0.27)$ for the MMM case (see section~\ref{sec:test_cmip5_rcp} for more details regarding the MMM and the alternative calibrations). Values for extreme calibrations of CDICE are also given in Table~\ref{table:IniVals2015ClimDICE2}

These initial values should meet three criteria: the atmospheric carbon mass and temperature change should be consistent with observed changes from pre-industrial to present times, and all values should be internally consistent in that they can be obtained by integrating the CE in time from its assumed pre-industrial equilibrium to the present. Regarding the first criterion, the 851 Gt atmospheric carbon corresponds to 400 ppm CO2, in line with measured concentrations in 2015.\footnote{https://www.co2levels.org.} The 1.1 K change in global mean temperature since pre-industrial times corresponds to the 2015 centered 11-year average (years 2010 to 2020) observation-based value.\footnote{http://berkeleyearth.org/global-temperature-report-for-2020.} We assert the third criterion by integrating the CE from its 1850 equilibrium state to the present day. We use the historical carbon emissions from CMIP5. We slightly adjust the non-CO2 forcing, by default, at 30\% of the CO2 forcing (see section~\ref{sec:description_of_bm}) to meet the warming target of 1.1 K by 2015. Such an adjustment is legitimate given that from a climate science perspective, this forcing comes with substantial uncertainty~\citep[see, e.g.,][]{mengis2020non}. From the internally consistent time series of $M_{t}$ and $T_{t}$ obtained in this way, we pick that time (year) with the desired atmospheric CO2 concentration, for example 400 ppm, as initial conditions, $M_{\text{INI}}$ and $T_{\text{INI}}$. 

The original initialization values used in DICE-2016 are $M_{\text{INI}} = (851, 460, 1740)$ and $T_{\text{INI}} = (0.85, 0.0068)$. The 0.85 K change in global mean temperature tends to be on the low side when compared to observation-based  estimates.\footnote{http://berkeleyearth.org/global-temperature-report-for-2020.}

\begin{table}[t!]
\centering
\begin{tabular}{ c c c c } 
 \toprule
 Model & $b_{12}$ & $b_{23}$ & $M_{\text{EQ}}$ \\ 
 \midrule
 DICE-2016, 5yr  &  0.12  & 0.007  & (588, 360, 1720)  \\ 
 CDICE           &  0.054 & 0.0082 & (607, 489, 1281)  \\ 
 \midrule
 CDICE-MESMO     &  0.059 & 0.0080 & (607, 305,  865) \\
 CDICE-LOVECLIM  &  0.067 & 0.0095 & (607, 600, 1385) \\
  \bottomrule
\end{tabular}
\caption{Values of free parameters in the carbon cycle, original DICE-2016 (with $\Delta_t = 5$ years) and our proposed, different calibrations of CDICE.}
\label{table:FreeParamsClimDICE2}
\end{table}
\begin{table}[t!]
\centering
\begin{tabular}{ c c c c c c c } 
 \toprule
 Model & $c_{1}$ & $c_{3}$ & $c_{4}$ & ECS & F$_{\text{2XCO2}}$ & $\lambda$ \\ 
 \midrule
 DICE-2016, 5yr  & 0.1005 & 0.088 & 0.025   & 3.1  & 3.6813 & 1.19 \\ 
 CDICE           & 0.137  & 0.73  & 0.00689 & 3.25 & 3.45   & 1.06 \\ 
 \midrule
 CDICE-HadGEM2-ES   & 0.154  & 0.55  & 0.00671 & 4.55 & 2.95  & 0.65  \\ 
 CDICE-GISS-E2-R      & 0.213  & 1.16  & 0.00921 & 2.15 & 3.65 & 1.70   \\
  \bottomrule
\end{tabular}
\caption{Values of the free parameters of the temperature equations for different versions of DICE: original DICE-2016 with $\Delta_t = 5$ years
and re-calibrated (CDICE). Also given are versions mirroring CMIP5 models with extreme ECS from~\citet{Geoffroy2013}, which are used for comparison purposes. Note that $\lambda = \text{F}_{\text{2XCO2}} / \text{ECS}$ is a derived quantity.}
\label{table:FreeParamsClimDICE}
\end{table}
%
%
%
%
\subsection{Atmospheric CO2 after 100 GtC pulse to the atmosphere}
\label{sec:test_100GtC}
%
%

\begin{figure}
  \centering
  \includegraphics[scale=0.34]{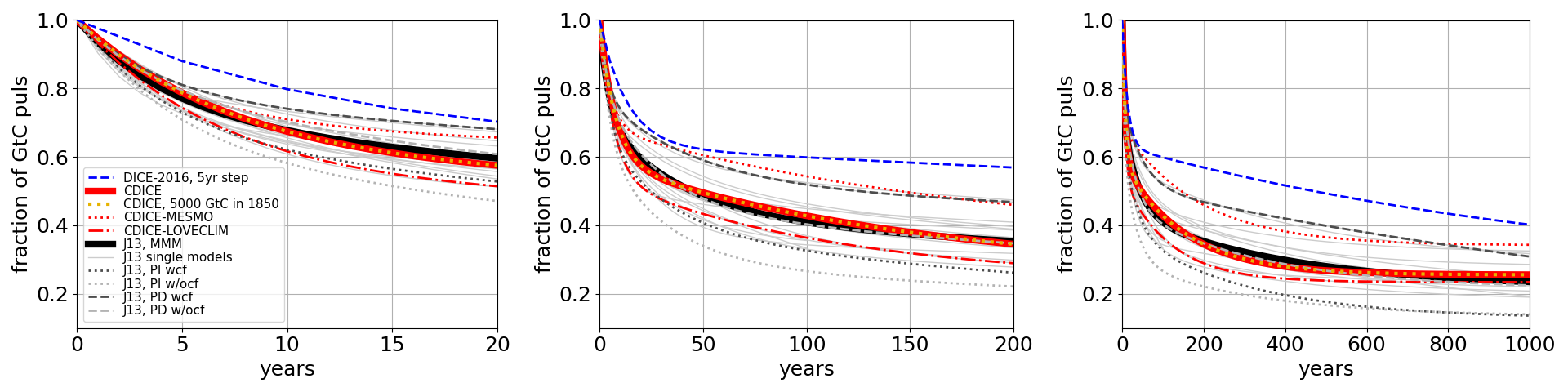}
  \captionof{figure}{Fraction of an instantaneous 100 GtC pulse remaining in the atmosphere (y-axis) as a function of time (x-axis, three different time scales from left to right). Shown are benchmark data adapted from~\citet{joos2013carbon} (black and gray lines, J13) and three calibrations against this data: CDICE (red solid), which is calibrated against the multi-model-mean (MMM, black solid), and two extreme calibrations, CDICE-MESMO (red dotted) and CDICE-LOVECLIM (red dash-dotted), which essentially capture the range of benchmark models (thin gray lines). DICE-2016 (blue dashed) removes too little CO2 from the atmosphere compared to the benchmark data. Absolute values of pulse height and equilibrium masses are irrelevant in DICE's carbon cycle model (CDICE, 5000 GtC on 1850 equilibrium masses, orange dotted), which is in contrast to findings in J13 (dashed and dotted gray lines, see their paper for details).}
  \label{fig:CCBM_100GtC_Puls}
\end{figure}

To evaluate and re-calibrate the carbon cycle of DICE-2016, we use the idealized test case of an instantaneous 100 GtC pulse to the present-day atmosphere, as described in section~\ref{sec:description_of_bm} and illustrated in Figure~\ref{fig:CCBM_100GtC_Puls}. 
Compared to benchmark data from~\citet{joos2013carbon} (black and gray lines), the known deficiency of DICE-2016 (blue dashed; see, e.g.,~\citet{Dietz2020}) are apparent: DICE-2016 overestimates the fraction of the pulse remaining in the atmosphere. Retaining the functional form of the carbon cycle, but re-calibrating the model against the MMM benchmark (black solid) using the maximum norm over the first 100 years after the pulse (see section~\ref{sec:metric}), the new best-fit version, CDICE (red solid), performs clearly better. Full agreement between CDICE and the MMM benchmark data cannot be reached because of the different functional forms of the carbon cycle.\footnote{The benchmark data in~\citet{joos2013carbon} has three characteristic decay time scales, whereas DICE with its three carbon reservoirs has only two.} Extreme calibrations against MESMO and LOVECLIM (CDICE-MESMO, red dotted, and CDICE-LOVECLIM, dash-dotted) are also shown. They are representative of scientifically plausible, but extreme carbon cycles that remove CO2 very efficiently (or inefficiently) from the atmosphere, and which we use later on in our economic analysis.
   
Figure~\ref{fig:CCBM_100GtC_Puls} also shows the functional limitation of the carbon cycle model in CDICE (or DICE in general). The fraction of the carbon pulse remaining in the atmosphere is sensitive to neither the pulse height nor the equilibrium state. The 100 GtC test case (red solid line) shows the same decay as a 5000 GtC pulse to pre-industrial equilibrium (orange dashed line). This is in contrast with informed expectations~\citep[see, e.g.,][]{joos2013carbon}: the pulse should decay more slowly if it is larger, if it is applied to present-day instead of pre-industrial conditions, or if climate feedback processes are taken into account (dashed and dotted gray lines, PI and PD indicating pre-industrial and present-day, respectively, adapted from~\citet{joos2013carbon}). 

\begin{figure}
  \centering
  \includegraphics[scale=0.34]{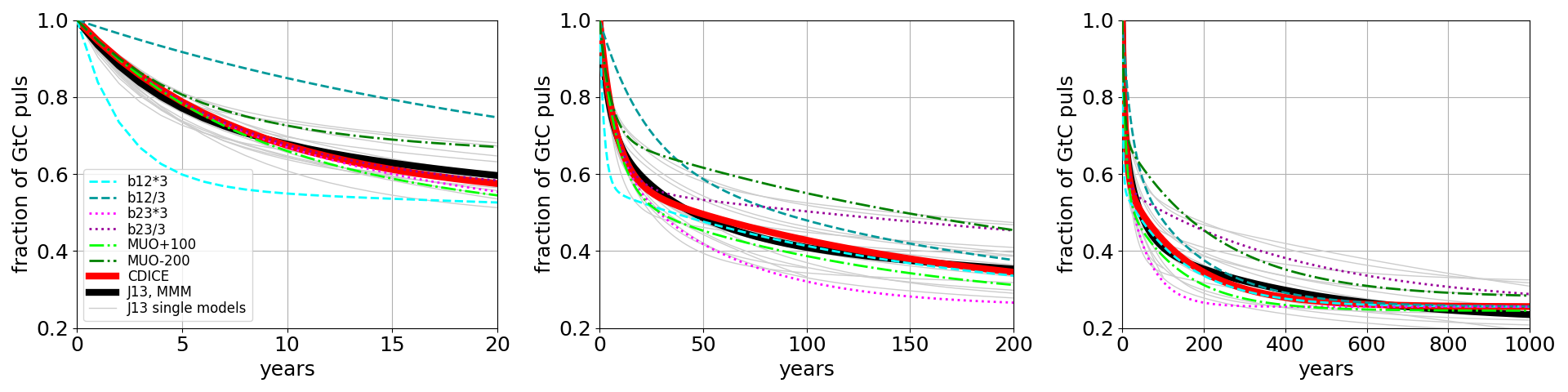}
   \captionof{figure}{Parameter sensitivities (thin colored lines) of the carbon cycle, anchored at CDICE (red), illustrated at the example of an instantaneous 100 GtC pulse to the atmosphere. Shown is the fraction of the pulse remaining in the atmosphere (y-axis) as a function of time (x-axis, three different time scales from left to right). Benchmark data from~\citet{joos2013carbon} is shown in black and gray.}
  \label{fig:Mod_CCBM_100GtC_Puls}
\end{figure}

The test case used here is further suited to examine sensitivities to the parameter values of the carbon cycle (colored lines) in Figure~\ref{fig:Mod_CCBM_100GtC_Puls}, anchored at CDICE (see Table~\ref{table:FreeParamsClimDICE}). As expected, the transfer coefficient $b_{12}$ between the atmosphere and the upper ocean has the largest effect on short time scales. On intermediate time scales, up to about 50 years, assumptions about the equilibrium masses in the upper and lower ocean start to matter. The effect of the transfer coefficient $b_{23}$ starts to have a clear effect only later on. The long-term evolution is equally dependent on $b_{23}$ and the equilibrium masses in the upper and lower ocean.

In this context, the eigenvalues (EV) of matrix $B$ are of interest, as they relate to decay or half-life time scales ($\tau = \Delta_t \cdot ln(0.5)/ln(EV)$) of an instantaneous perturbation (carbon emission). One EV equals $1$ and corresponds to the equilibrium solution. The other two EVs are $(1 + g \pm h)/2$, with $g = 1 - b_{12}\cdot(1+r_{1}) - b_{23}\cdot(1+r_{2})$ and $h = ( (1-g)^{2} - 4f)^{1/2}$, where $f = b_{12}\cdot b_{23} \cdot (1 + r_{2} ( 1+ r_{1}))$. Associated numerical values for DICE-2016 are 0.6796 and 0.9959, corresponding to half-life times of 9 and 851 years. For our CDICE calibration, we arrive at EVs of 0.8742 and 0.9933, corresponding to half-life times of 5 and 102 years. This highlights the difference between our carbon-cycle calibration and that in DICE-2016.

\begin{table}[t!]
\centering
\begin{tabular}{ c c c } 
 \toprule
 Model &  $M_{\text{INI}}$ & $T_{\text{INI}}$ \\ 
 \midrule
 DICE-2016, 5yr  &  (851, 460, 1740) & (0.85, 0.0068) \\ 
 CDICE           &  (851, 628, 1323) & (1.10, 0.27)   \\ 
 \midrule
 CDICE-HadGEM2-ES           &  (851, 628, 1323) & (1.10, 0.27)\\
 CDICE-GISS-E2-R            &  (851, 628, 1323) & (1.10, 0.27)\\
 CDICE-MESMO                &  (851, 403,  894) & (1.10, 0.27)\\
 CDICE-LOVECLIM             &  (850, 770, 1444) & (1.10, 0.27)\\
 CDICE-HadGEM2-ES-MESMO     &  (851, 403,  894) & (1.10, 0.27)\\
 CDICE-HadGEM2-ES-LOVECLIM  &  (850, 770, 1444) & (1.10, 0.27)\\
 CDICE-GISS-E2-R-MESMO      &  (851, 403,  894) & (1.10, 0.27)\\
 CDICE-GISS-E2-R-LOVECLIM   &  (850, 770, 1444) & (1.10, 0.27)\\
  \bottomrule
\end{tabular}
\caption{Initial conditions for the year 2015 as used in DICE-2016 and in CDICE, as well as for extreme calibrations of CDICE with respect to the carbon cycle and/or temperature. The atmospheric CO2 concentration is always ~400 ppm (~851 GtC). All other values result from integrating the differently calibrated CEs from 1850 to the present day using the CMIP5 setup as described in the fourth test case.}
\label{table:IniVals2015ClimDICE2}
\end{table}

\subsection{Temperature evolution upon quadrupling of CO2 concentration}
\label{sec:test_4xco2}
%
%
Our second test case, namely the temperature response to an instantaneous quadrupling of the atmospheric CO2 concentration (cf. section~\ref{sec:description_of_bm}), allows for an evaluation and re-calibration of the temperature equations of the CE under consideration. In principle, a metric would again be needed for this purpose (see section~\ref{sec:metric}). However, as CDICE (or DICE in general) and the benchmark data from~\citet{Geoffroy2013} share the same functional form (cf. section~\ref{sec:meth_temperature}), a re-calibration is trivial: we simply adopt the parameter values already fitted to the CMIP5 data by~\citet{Geoffroy2013}. Specifically, we obtain CDICE using their fit to the multi-model-mean and two extreme models, that is, CDICE-HadGEM2-ES and CDICE-GISS-E2-R. As explained above, these extreme cases play an important role in our economic analysis below (cf. sections~\ref{sec:peSCC} and~\ref{sec:simulation}). Table~\ref{table:FreeParamsClimDICE} summarizes the relevant coefficients for these alternative calibrations.

\begin{figure}
  \centering
  \includegraphics[scale=0.33]{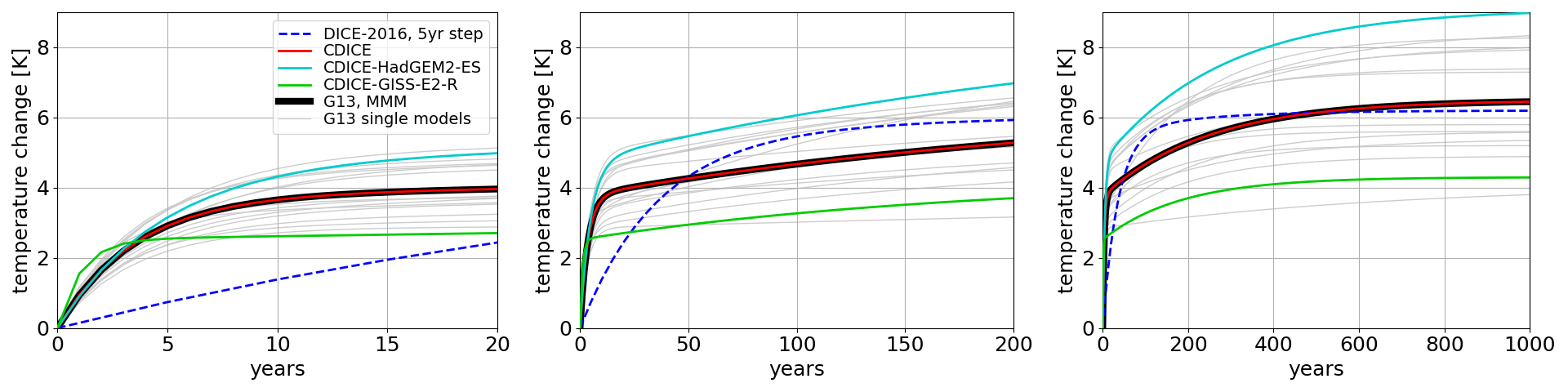}
  \captionof{figure}{Temperature response to instantaneous quadrupling of atmospheric CO2 with respect to pre-industrial values on time scales of 20 years (left), 200 years (middle), and 1000 years (right). Parameters in CDICE (red solid) can be chosen such as to exactly reproduce the calibration target, the CMIP5 multi-model mean (black solid, G13-MMM). Also shown are two extreme ECS calibrations (solid green and light blue, CDICE-HadGEM2-ES and CDICE-GISS-E2-R). In DICE-2016 (dashed blue, 5 year time step), temperature increase is too slow compared to any CMIP5 model (thin gray lines, G13, from~\citet{Geoffroy2013}). }
  \label{fig:TempBM_4xCO2_on_PI}
  \end{figure}

The test confirms previous findings that DICE-2016 warms too slowly compared to any CMIP5 model (dashed blue versus solid gray lines in Figure~\ref{fig:TempBM_4xCO2_on_PI}). All the CMIP5 models (solid gray lines) agree that warming occurs rapidly. Meanwhile, their level of warming differs by roughly a factor of two, in line with their differing ECS, given in Table~\ref{table:FreeParamsClimDICE}. The overly slow warming in DICE-2016 finds confirmation via formal expressions for the characteristic fast and slow response time scales associated with Equations~\eqref{eq:temp1} and~\eqref{eq:temp2}~\citep[see][their Table 1]{Geoffroy2013}, which yield numerical values of 4 and 249 years for the CMIP5 multi-model mean in~\citet{Geoffroy2013}, but 40 and 219 years for DICE-2016.

\begin{figure}
  \centering
  \includegraphics[scale=0.34]{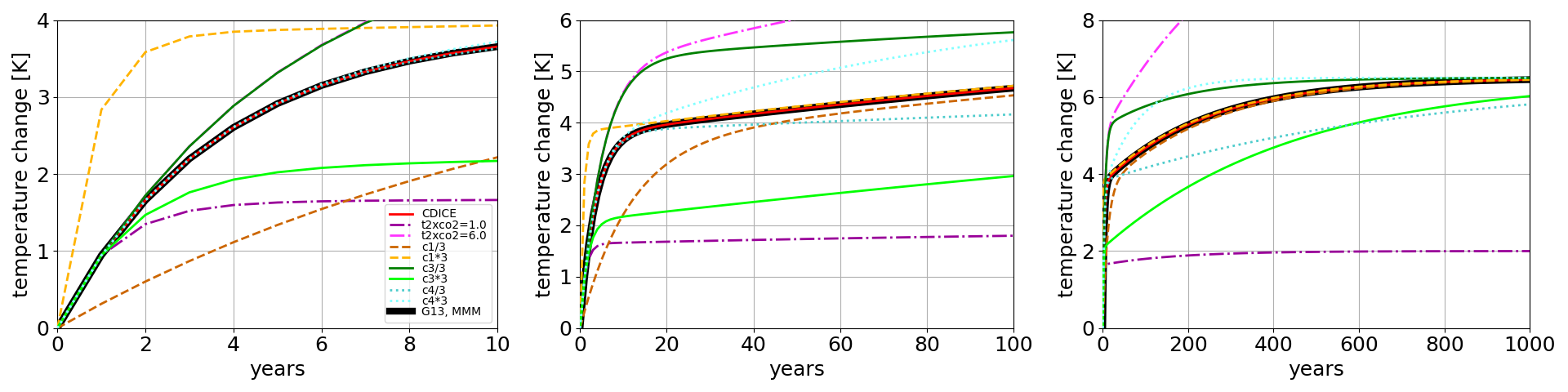}
  \captionof{figure}{Parameter sensitivity of the temperature equations in DICE, illustrated at the example of an instantaneous quadrupling of the CO2 concentration with respect to pre-industrial values. Shown is warming (y-axis) as a function of time (x-axis) for different time scales (from left to right). Parameters are varied one by one with respect to the CMIP5 multi-model mean from \citet{Geoffroy2013} (black solid, G13-MMM) and its adaptation in CDICE (red solid).}
  \label{fig:Mod_TempBM_4xCO2_on_PI}
\end{figure}

The test discussed here is further suited to shed some light on the role of the different parameter values in the temperature equations of CDICE, as illustrated in Figure~\ref{fig:Mod_TempBM_4xCO2_on_PI}. How quickly the temperature responds to an increase in atmospheric CO2 is mainly governed by the heat capacity of the upper ocean via the parameter $c_{1}$ (left panel, yellow and brown dashed lines), which, in turn, does not affect the mid-and long-term absolute warming (beyond roughly 20 years, middle and right panel). As expected, the very long-term warming is set by the ECS (right panel, pink and violet dash-dotted lines). The track toward long term equilibrium across intermediate time scales and levels of warming depends on the energy exchange between the upper and lower ocean, via the choice of $c_{3}$ (solid green lines, after roughly four years), and on the heat capacity of the deep ocean, entering in the form of $c_{4}$ (cyan dotted lines, after roughly 20 years).

In summary, the test case demonstrates that the temperature equations in DICE-2016 can be re-calibrated to reach general agreement with corresponding results from CMIP5. Specific CMIP5 models or the multi-model mean as given in~\citet{Geoffroy2013} can be recovered precisely by appropriate parameter choice.

%
\subsection{Temperature evolution as atmospheric CO2 increases at 1\% per year}
\label{sec:test_1pctco2}
%
%

The purpose of the third test is to evaluate a CE's TCR. In contrast, the first two test cases primarily aim at calibrating a CE.
The TCR is of interest as a measure of how the (model) climate responds to steadily changing conditions. From the point of view of a CE, one may also say that the TCR examines the CE's performance upon a series of perturbations, not a single one as in the tests used for calibration of the CE. Although still highly idealized, this test is thus closer to reality. Consequently, TCR is often considered a better measure to evaluate the performance of a climate model than ECS, which assumes the climate system to have equilibrated. A reasonable TCR is a necessary condition for any CE but, as we shall see, not a sufficient one.

\begin{figure}
  \centering
  \includegraphics[scale=0.50]{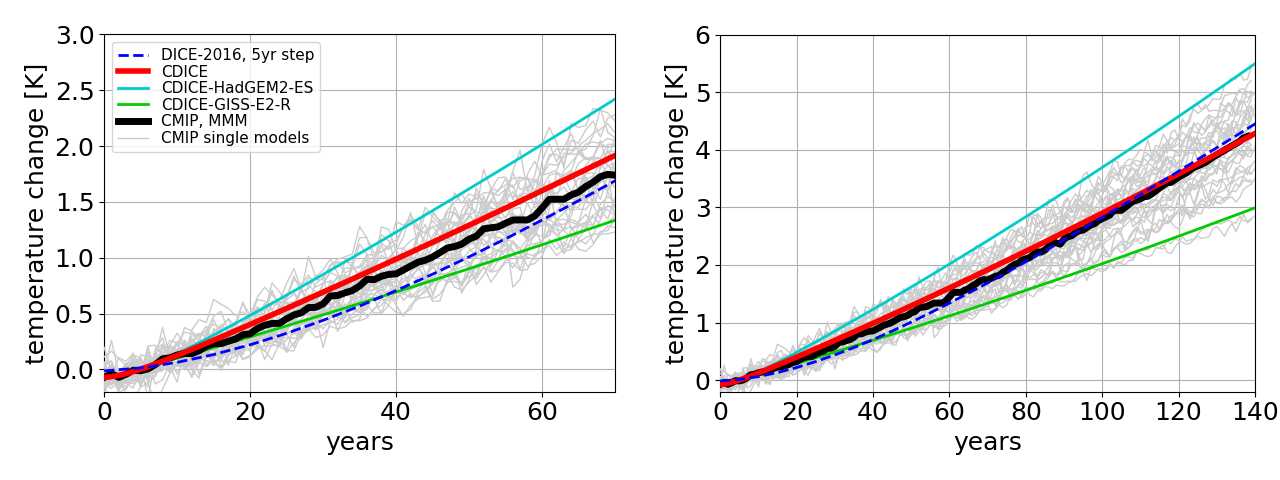}
  \captionof{figure}{Temperature response (y-axis) to transient, one percent per year, increase of the CO2 concentration as a function of years (x-axis) from 1850 onward. The CO2 concentration doubles within 70 years (left panel), and associated warming levels correspond, by definition, to the transient climate response (TCR) of the model. CO2 quadruples within 140 years (right panel), and TCR roughly doubles. DICE-2016 (dashed blue), our new version CDICE (solid red), as well as its extreme scaling variants (solid cyan and green), are all compatible with CMIP5 models and their TCR (individual models and multi-model mean, solid gray and black, respectively).}
  \label{fig:CMIPBM_1pctCO2}
\end{figure}

Of the test results shown in Figure~\ref{fig:CMIPBM_1pctCO2}, two aspects deserve highlighting. First, all variants of DICE (shown in different color coding) are compatible with the CMIP5 data (black and gray lines). The temperature evolution of DICE-2016 (blue dashed lines) now lies within the range of CMIP5 models (solid gray lines). However, the warming is still slightly slower than the CMIP5 multi-model mean (black solid) during the first few decades. CDICE (solid red) warms slightly faster than the CMIP5 multi-model mean. Variants of CDICE with extreme ECS (green and cyan lines, see Table~\ref{table:FreeParamsClimDICE} for parameters) indeed bracket the individual CMIP5 models. Second, the transient warming of the models remains clearly below their ECS, even after the quadrupling of CO2 after 140 years. The TCR of the models is between 1.3 K and 2.3 K across CMIP5 models (gray lines in Figure~\ref{fig:CMIPBM_1pctCO2}). Associated values for ECS are roughly 50\% to 100\% larger, ranging between 2.05 K and  4.55 K (see Table~\ref{table:FreeParamsClimDICE}). Upon the quadrupling of CO2 after 140 years, ECS still exceeds the transient temperature change by about 40\% to 70\%. 

Contrasting the above findings with those from Sect.~\ref{sec:test_4xco2} demonstrates that the model performance is highly sensitive to the concrete test. In particular, the striking under-performance of DICE-2016 in the idealized CO2 quadrupling test case is hardly detectable if CO2 increases gradually at a rate of one percent per year instead. The decent performance in terms of TCR may seem to come as a relief given the ample amount of existing literature using DICE-2016. This impression is, however, deceptive, as we will further elaborate on it in section~\ref{sec:simulation}. 
%
%

\subsection{CMIP5 historical and RCP evolution as simulated by DICE}
\label{sec:test_cmip5_rcp}
%

\begin{figure}
  \centering
  \includegraphics[scale=0.24]{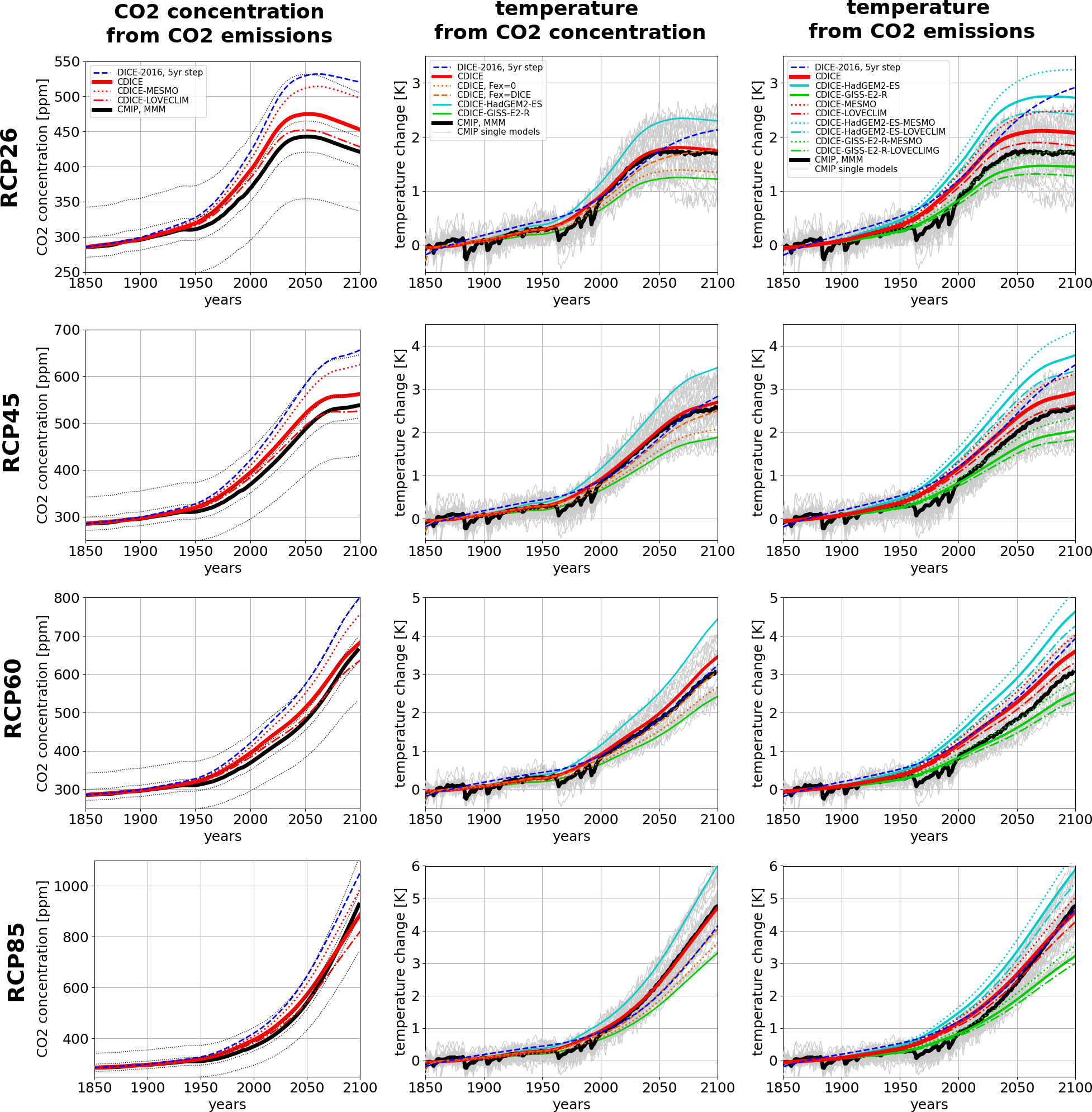}
  \captionof{figure}{Comparison of DICE with historical and future (years 1850 to 2100, scenarios RCP26, RCP45, RCP60, and RCP85, from top to bottom) CMIP5 data. Shown are atmospheric CO2 concentrations (left) as prescribed in CMIP5 (solid black, with dotted lines indicating $\pm 5$\% and $\pm 20$\% ranges) and as computed from CO2 emissions with DICE-2016 (dashed blue), CDICE (solid red), as well as CDICE-MESMO and CDICE-LOVECLIM (dotted and dash-dotted red). Also shown is temperature evolution based on prescribed CO2 concentrations (middle) and based on emissions (right) for DICE-2016, CDICE, and variants of CDICE.}
  \label{fig:CMIPBM}
\end{figure}

Our fourth test was chosen to be as close as possible to real applications, that is, a necessary gold standard test any CE should pass: the CMIP5 simulations for historical and future representative concentration pathways (RCP26, RCP45, RCP60, and RCP85, from top to bottom in Figure~\ref{fig:CMIPBM}). Compared to the other three tests, this test features some additional hurdles. First, the entire CE is examined, that is, the carbon cycle combined with the temperature equations. Second, strong mitigation scenarios are considered that test a CE's ability to capture a regime transition from, basically, emission-dominated to response-dominated. Third, non-CO2 forcings must be taken into account. They form part of the CMIP5 simulations and the real world alike and thus must be incorporated and examined in the context of any CE. To facilitate the interpretation of results, we 
separately examine in Figure~\ref{fig:CMIPBM} CO2 concentrations due to changing carbon emissions (left column, performance of carbon cycle), the temperature in response to changing CO2 concentrations (middle column, performance of temperature equations), and finally, temperature as a function of carbon emissions (right column, performance of full CE).

The CMIP5 data (gray and black lines in Figure~\ref{fig:CMIPBM}) provide a benchmark range: depending on the model, warming by 2100 is between 1 K and 2.5 K for RCP26 and between 3 K and over 6 K for RCP85. The range exists, although the models feed on prescribed concentrations, not emissions. A subset of CMIP5 models recomputed the RCP85 scenario based on prescribed emissions. By 2100, their CO2 concentrations lie in a range from 800 ppm to 1150ppm, or within about 20\% of the CMIP5 prescribed value of 935 ppm, and warming ranges roughly from 2.7 K to 6.5K~\citep[Figure 12.36 in the][]{IntergovernmentalPanelonClimateChange2015}. The episodic cooling apparent in Figure~\ref{fig:CMIPBM}, which is due to significant volcanic eruptions such as Mt. Pinatubo in 1991, is a reminder that the temperature evolution in CMIP5 encapsulates a whole range of forcings. The decreasing CO2 concentrations and temperatures in RCP26 are of particular interest as they allow us to benchmark the CE in the context of mitigation scenarios. 

Looking at the carbon cycle (left column in Figure~\ref{fig:CMIPBM}), we find that CO2 concentrations modeled by the DICE family are mostly within 20\% of those prescribed by CMIP5 (black dotted lines), thus within the range of emission-based CMIP5 simulations (see above). Concentrations are always highest for DICE-2016, followed by CDICE-MESMO, CDICE, and CDICE-LOVECLIM, in line with calibration (CDICE family) and evaluation (DICE-2016) against the 100 GtC pulse test (see section~\ref{sec:test_100GtC}). Looking at the different scenarios, the strong mitigation scenario RCP26 has concentrations prescribed in CMIP5 (black solid) that are lower than in any model of the DICE family. The situation changes as one moves to RCP45 and RCP85, where towards the end of the century (the year 2100), concentrations in CDICE are even lower than those prescribed in CMIP5. The behavior may indicate a dependence of the carbon cycle on the scenario (strong mitigation versus continued strong emission) either on the side of the DICE model family or on the side of CMIP5. A dedicated analysis is beyond the scope of the present paper.

Turning to the temperature response to CO2 concentrations as prescribed in CMIP5 (middle column), the curves for the different versions of CDICE echo corresponding calibrations (see section~\ref{sec:test_4xco2}). In particular, CDICE (red solid) falls on top of the CMIP5 multi-model-mean (black solid) independent of the RCP scenario considered. This is remarkable in several respects. The temperature equations in CDICE were calibrated against one perturbation (instantaneous quadrupling of CO2), whereas CO2 keeps constantly changing here. The temperature evolution is well captured even in RCP26 after 2050, when CO2 concentrations decrease again. Finally, the data by~\citet{Geoffroy2013}, used to calibrate CDICE, incorporate only a subset of CMIP5 models.\footnote{Only a subset of all CMIP5 data was available when~\citet{Geoffroy2013} performed their analysis.} Turning to DICE-2016 (blue dashed line), it features relatively high temperatures under strong mitigation (RCP26) and rather low temperatures in the strong forcing RCP85 scenario. One reason could be that the slow temperature response of DICE-2016 to an instantaneous increase in CO2 concentrations turns into a lagging response upon gradually changing CO2. Another reason could be the non-CO2 forcing (see Section~\ref{sec:meth_temperature}), which may be too strong in the RCP26 scenario, but too weak in the RCP85 scenario. For CDICE, the (assumed) non-CO2 forcing (dotted orange line) contributes around 0.5 to 1-degree warming by 2100. 

The performance of the entire CE (Figure~\ref{fig:CMIPBM}, right column) encapsulates the features just described. Some combinations of the independently calibrated carbon cycle and temperature equations fall outside the range of CMIP5 models (thin gray lines). For example, CDICE-HadGEM2-ES-MESMO and CDICE-HadGEM2-ES are too warm in RCP26, whereas CDICE-GISS-E2-R-LOVECLIME tends to be on the cool side in RCP85. The relative importance of the carbon cycle and temperature response depends on the forcing scenario. Upon strong forcing, like RCP85, temperatures differ mainly because of the different ECS calibrations (cyan, red, and green curves). In contrast, the calibration of the carbon cycle plays a minor role (dotted, solid, and dash-dotted curves). For RCP26, CEs that combine high ECS with efficient removal of carbon emissions (CDICE-HadGEM2-ES-LOVECLIM, cyan dash-dotted) have a similar temperature evolution as CEs combining inefficient carbon removal with an intermediate ECS (CDICE-MESMO, red dotted). In section~\ref{sec:simulation}, we examine the degree to which this degeneracy persists within an economic context.

In summary, CDICE falls within CMIP5 models for all future scenarios. This is not the case for some variants of CDICE that combine extreme calibrations of the carbon cycle and temperature equations. DICE-2016 performs better for the strong forcing scenario RCP85 than for the strong mitigation scenario RCP26. Concrete assumptions about the non-CO2 forcing term, notably its strength, are demonstrated to matter for global mean temperature change.
%

\subsection{Synthesis and lessons learned from the climate test cases}
\label{sec:test_synthesis}
%
%
The results presented in this section show that it is important to independently calibrate the two basic building blocks of any CE, that is, the carbon cycle and the temperature equations. In this paper, we provide two corresponding, highly idealized test cases that are based on published data from CMIP5. We advocate that the calibrated CE should, subsequently, be exposed to two additional tests: one to quantify the CE's transient climate response and the other to evaluate the CE's performance for realistic future scenarios, under strong forcing as well as under strong mitigation. Associated tests are described that rely only on publicly accessible data from CMIP5. We illustrated that DICE-2016, although failing both calibration tests, passes the two additional tests, possibly due to compensating errors. Additionally, we demonstrated that the functional form of the DICE family is fit for purpose: a re-calibrated version, CDICE, passes all four tests. 

The CMIP5 test cases highlight the existence of a target range for both the carbon cycle and the temperature equations. Full-fledged ESMs do not agree on global mean temperature change, even if this change is only due to prescribed gradual changes in atmospheric CO2 concentration, as in the one percent per year CO2 increase test case. Variants of CDICE using calibrations for the temperature response that are extreme yet in line with CMIP5 have been shown to bracket the range of CMIP5 models for all four RCPs examined. We advocate exploiting this bracketing behavior in the context of economic studies as a measure of the uncertainty arising from the climate part. The same applies with regard to the carbon cycle. We caution, however, that combining extreme calibrations for both the carbon cycle and the temperature equations may result in an overly extreme climate of the CE. 

Non-CO2 forcings were shown to play a prominent role and, as such, constitute another source of uncertainty for any CE. Their assumed linear increase with time in DICE-2016 has the advantage of being simple. Climate literature rather advocates that non-CO2 forcings amount to about one-third of the forcing from CO2 (cf. section~\ref{sec:description_of_bm}); the form adopted in CDICE. It is not obvious whether any of the two forms is clearly superior with regard to the envisaged applications; studying the interplay of climate, society, economy, and technology. Net-zero targets, mitigation scenarios, or scenarios presenting even more heavy use of fossil fuels may affect the mix of non-CO2 forcing agents and associated lifetimes. A more detailed description of such developments is not desirable as it would necessitate including non-CO2 agents, reservoirs, and processes, thereby violating the deliberately simple approach to climate taken in DICE. Retaining a simple form, but testing for associated sensitivities, seems a more promising avenue. 

Focusing again on CO2-forcings alone, the temperature part of CDICE was shown to perform very well with respect to any benchmark data for prescribed CO2 concentrations. Working with carbon emissions turned out to be more challenging. It is difficult for CDICE to reproduce equally well both high emission and mitigation scenarios, like RCP85 and RCP26 (cf. Figure~\ref{fig:CMIPBM}, left column). The point is relevant given that DICE should be applicable to a broad range of socio-economic scenarios, from fossil fuel dominated to mitigation and net-zero emissions. We speculate that the issue is rooted at least in part in the availability of only two time scales (of 5 and 102 years, see section~\ref{sec:meth_carbcycle}) related to the three reservoir carbon cycle. 

A reservoir-based carbon cycle model has the advantage that the total amount of carbon is preserved. In particular, carbon leaving the atmospheric reservoir is still present in another reservoir, notably the ocean, in the case of DICE. From there, it may continue to play a role for the climate system at some point in time. This capability of a reservoir-based approach is especially interesting with regard to strong mitigation scenarios, where the amount of carbon currently being stored in the (upper) ocean is likely to affect the carbon uptake capability of the ocean as emissions decline~\citep{ridge-mckinley:21}. For the same reason, it seems questionable whether yet simpler climate models without any reservoirs - an example of such a model may be the approximately linear relationship between cumulative carbon emissions and global mean temperature - are able to properly cope with strong mitigation scenarios.   

In summary, the emission-based global mean temperature evolution from 1850 to 2100 as modeled by either DICE-2016 or CDICE lies, despite the models' simple functional form, well within the range of CMIP5 results for most RCPs. For DICE-2016, this finding is noteworthy as the model, in contrast to CDICE, clearly fails more idealized tests (instantaneous quadrupling of CO2 and 100 GtC pulse). Moreover, it is deceiving, as compensating errors play an important role in DICE-2016. Noteworthy also is the challenge posed to either DICE-2016 or CDICE by the strong mitigation scenario RCP2.6: CDICE struggles, and DICE-2016 clearly fails and clearly warms far too much towards 2100. As already noted by~\citet{Traeger2014}, such strong mitigation scenarios excessively challenge the models' functional form.


\section{The social cost of carbon in partial equilibrium}
\label{sec:peSCC}

In standard economic models, climate change is treated as an externality, and, going back all the way to the work of~\cite{pigou1920wealth}, it is well known that Pareto efficiency can be restored by adding the marginal value of the externality to the market price of carbon. To reach this goal, IAMs compute the net present value of all future damages caused by the emission of an additional unit of carbon at some date $t$. The effect of these extra emissions on damages depends on the calibration of the damage function and the path of emissions.  Computing the net present value of the damages requires knowledge of all future interest rates.  A change in the market price of carbon will change the path of emissions and of interest rates. Consequently, one has to impose an economic model with many assumptions to obtain an accurate estimate of the SCC. However, one can gain important insights by examining the future path of damages caused by a pulse of emissions for a given RCP emissions scenario. One can then assume constant interest and growth rates and make general statements about the SCC. As we will show in section~\ref{SCC:CDICE} below, in the BAU case of the DICE-2016 model, the effects of the climate model on emissions and the interest rates are indeed tiny (of course, the effects on optimal abatement and mitigated emissions are very large). The social cost of carbon is defined for each date $t$ - as it turns out, if one assumes constant interest rates, the SCC relative to output is relatively constant across the next 50 years, and we, therefore, focus on the somewhat arbitrary date 2020 for reporting the SCC in this section.

In  this, and the subsequent section~\ref{sec:simulation}, we assume that exogenous forcing, $F^{\text{EX}}_{t},$ changes linearly with time from 0.5 in the year 2015 to 1.0 in 2100, that is, as specified in DICE-2016 (cf. equation~\eqref{eq:forcing}). This measure allows us to compare the impact of our CDICE calibration with DICE-2016 consistently. However, to be fully in line with state-of-the-art climate science, it might be better to assume $F^{\text{EX}}_{t} = 0.3 \cdot F^{\text{CO2}}_{t}$~\citep[see, e.g.,][]{ClimateChange2014SynthesisReportIPCC2014, gambhir-et-al:17}. In Appendix~\ref{sec:fex}, we show how such change in the full model affects our quantitative results.

 Following the majority of the economic literature on climate change, we assume that 
the temperature influences the economic activity via a damage function $\Omega(\cdot)$ that multiplies total output $ Y^{\text{Gross}}_t(\cdot) $ at time $t$. Given a constant interest rate $ r $, and a constant growth rate of output, $g$, the present value of damages at some future date $t$ is given by 
$ \left(\frac{1+g}{1+r} \right) ^t \Omega(T_{AT,t}) $. In this section, we take $\frac{1+r}{1+g}$ as given, and refer to it simply as the ``g-adjusted interest rate''.
Note that output growth can come from population growth as well as total factor productivity (TFP) growth, and it is likely not to be constant over the next 300 years. Real risk-free rates and returns to capital certainly vary over the business cycle, but also might change in the long term (see, e.g.,~\cite{bauer2021rising}).

The damage functions used in DICE-2016, but also in~\cite{Hansel2020} and~\cite{Howard2017} among others, all assume the following quadratic functional form:
\begin{equation}
   \Omega(T_{AT,t}) =   \psi_2 \cdot (T_{AT,t})^2 .
   \label{eq:qu_damage}
\end{equation}
Since, in this section, we only consider the SCC relative to the CDICE calibration (both the temperature equations as well as the carbon cycle calibrated to the MMM), the value of the  parameter $ \psi_2 $ is irrelevant. Of course, $\psi_2 $ matters for the absolute value of the SCC and for optimal abatement, but not for ratios of the SCC for different calibrations. However, as we show, the functional form is very important, and higher-order terms (as, for instance, discussed in~\cite{weitzman2012ghg}) can make a significant difference in the effect of climate model uncertainty on the SCC.

\subsection{Different climate calibrations and the SCC}
\label{SCC:CDICE}

\begin{figure}[ht] 
    \centering
    \includegraphics[width=0.7\linewidth]{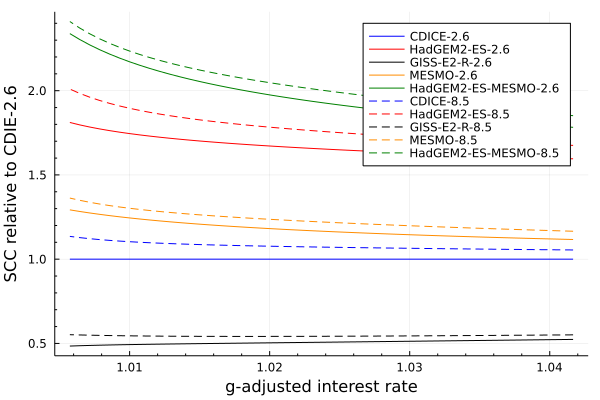} 
    \caption{ SCC for different CDICE calibrations, emission scenarios, and a quadratic damage function.} 
    \label{fig:scc_cdice} 
\end{figure}

\begin{figure}[ht] 
\begin{subfigure}[b]{0.5\linewidth}
    \centering
    \includegraphics[width=1.0\linewidth]{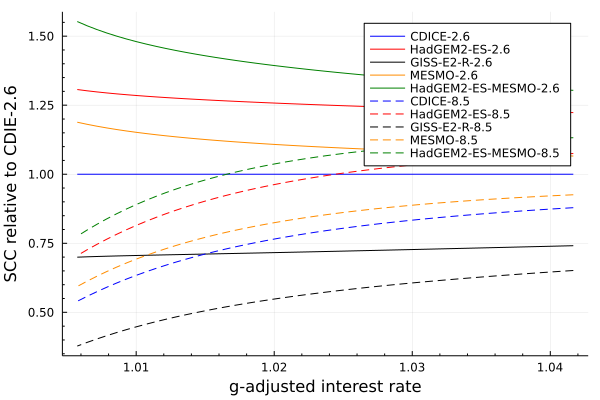} 
    \caption{Linear Damages} 
    \label{fig:scc_lindam} 
  \end{subfigure} 
   \begin{subfigure}[b]{0.5\linewidth}
    \centering
    \includegraphics[width=1.0\linewidth]{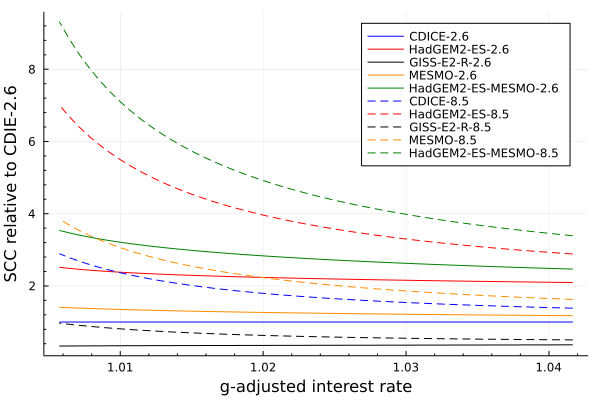} 
    \caption{Cubic Damages} 
    \label{fig:scc_cudam} 
  \end{subfigure}
  \caption{SCC for alternative damage functions.}
  \label{fig:scc:altdam}
\end{figure}

Figure~\ref{fig:scc_cdice} displays the SCC under two different RCP emission scenarios, that is, RCP8.5 with very high emissions and RCP2.6 with very low emissions, and four different calibrations of our CE. 
We consider the MMM for both the temperature equations and the carbon cycle (denoted by CDICE). Furthermore, we also look at HadGEM2-ES and GISS-E2-R as alternative temperature calibrations, as well as MESMO as an alternative carbon-cycle calibration.
As we vary the growth-adjusted interest rate from  0.5 percent to 4 percent, the SCC for the RCP 2.6 scenario with the basic CDICE  model (both temperature equation and carbon cycle calibrated to the MMM) is always normalized to one. All models considered in this Figure are a calibration of CDICE, for simplicity, the term CDICE is often omitted in the legends in Figures~\ref{fig:scc_cdice} and~\ref{fig:scc:altdam}.

There are four key takeaways from this stylized exercise. First, model uncertainty for the temperature equation can lead to three to five times higher SCC for very pessimistic scenarios as opposed to the most optimistic scenario. Second, the effect of an extreme carbon cycle is much smaller compared to the temperature block (it increases the SCC by about 20 percent relative to the MMM), but the combined effect of a pessimistic carbon cycle and a pessimistic temperature scenario can be very large. Third, the relative uncertainty introduced by climate model uncertainty does not change very much with discounting, at least for g-adjusted interest rates above 2 percent. Fourth, the SCC does not change very much with an alternative to the emissions scenario. It is uniformly higher when the emissions are very large, but relative to the effect of the climate model uncertainty and the discount factor, the changes are small.

It is important to note that the last two observations depend crucially on the assumption that damages are quadratic (cf. equation~\eqref{eq:qu_damage}). Typically, for any strictly convex cost function, one would expect that marginal costs increase in quantity and that, therefore, the SCC is substantially higher for RCP8.5 than for RCP2.6. However, the marginal costs of emissions also include the non-linear effects of emissions on the temperature. The concavity in the forcing equation~\eqref{eq:forcing} counters the convexity of damages, and thus, the overall shape of the cost function is an unknown a priori.

This point is further illustrated in Figure~\ref{fig:scc:altdam}, where we show the analogue of Figure~\ref{fig:scc_cdice} for a linear damage function (left panel) as well as a cubic damage function of the form $ \Omega(T_{AT,t}) =   \psi_3 \cdot(T_{AT,t})^3 $ (right panel).

%

For linear damages, the marginal costs decrease in emissions. For high emission scenarios, the SCC is lower than for low emission scenarios. The overall effect of climate model uncertainty is significantly smaller than in the quadratic cost case. For the cubic damages, on the other hand, marginal costs clearly increase. For high emission scenarios, the SCC is much higher. The effect of model uncertainty, in this case, is huge. Depending on the assumed interest rate, the SCC for the most pessimistic climate scenario can be almost eight times the SCC for the most optimistic scenario.

\subsection{The effect of miscalibration}
\label{sec:mis-calibration}

\begin{figure}[t!] 
    \centering 
    \includegraphics[width=0.7\linewidth]{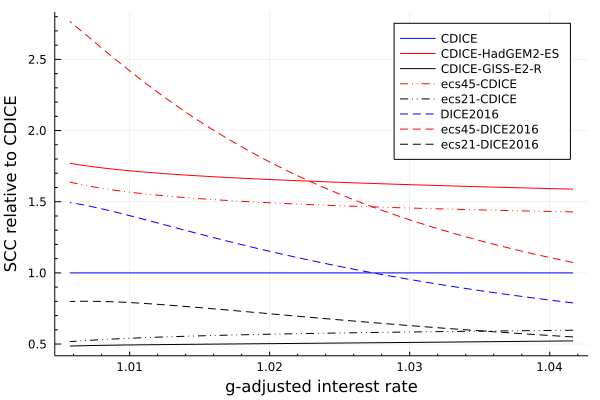} 
    \caption{SCC for different calibrations. RCP8.5 and quadratic damage function.} 
    \label{fig:scc_dice} 
\end{figure}

How important are the exact details of our calibration for the SCC? What happens, for instance, if, instead of using the CDICE-HadGem2-ES calibration, we just change the ECS-parameter in Equation~\eqref{eq:temp1} to 2.15 or to 4.55 without changing any other parameters? How does the SCC for the original DICE-2016 calibration differ from the SCC in our calibration?
Figure~\ref{fig:scc_dice} shows how the SCC changes with various calibrations of the CE. We focus here on the RCP 8.5 scenario with quadratic damages and compute the SCC at t=5 (the year 2020)  relative to the CDICE calibration.

Figure~\ref{fig:scc_dice} shows the SCC for CDICE-HadGEM2-ES   and CDICE-GISS-E2-R as alternative temperature calibrations and compares them to simply setting the ECS to the associated values (4.55 in CDICE-HadGEM2-ES and 2.15 in CDICE-GISS-E2-R), but leaving the CDICE calibration otherwise unchanged otherwise (cf. the dot-dashed lines labeled 'ecs45-CDICE' and 'ecs21-CDICE'). This is how uncertainty about the ECS is typically treated in economic models of climate change (see, e.g., \cite{Nordhaus2018b} or \cite{Hassler2018}).
We observe that even within the CDICE calibration, just changing the ECS is not a very good proxy for the extreme warming scenarios in CMIP5. In particular, the SCC for an ECS of 2.15 is about 15 percent too large relative to GISS-E2-R, and the SCC for an ECS of 4.55 is about 10 percent too small relative to CDICE-HadGEM2-ES. This indicates that for a quantitative treatment of climate model uncertainty, it is not enough to simply vary the ECS parameter in the CE. As explained above, the ECS is not a free parameter in Earth System models, but emerges as one characteristic of simulated climate change describing the very long-run behavior of temperature. For the social cost of carbon, this very long run is largely irrelevant, and the speed of warming is often more important. Modeling the climate uncertainty by simply varying the ECS parameters in the climate emulator incorrectly reduces the uncertainty about the SCC by a significant amount.

Figure~\ref{fig:scc_dice} also shows the ECS for the DICE-2016 calibration, the MMM as well as calibrations with an ECS of 2.15 and with 4.55. The differences with respect to CDICE are apparent: For intermediate values of the discount fact (say 0.97 or so), the SCC for CDICE and DICE-2016 are very similar, both for the MMM and for the calibrations with ECS of 2.15 and 4.55, respectively. In the case of a discount factor of 0.96, DICE-2016 underestimates the SCC substantially and is about 20 percent lower than for CDICE. For high values of this parameter, DICE-2016 produces much higher values for the SCC, for a value of 0.99, about 30 percent larger than CDICE.

\section{The social cost of carbon and optimal abatement in the DICE economy}
\label{sec:simulation}

In this section, we present the optimal solutions for the adjusted DICE-2016 model, which features both the calibration of the economic part as presented in DICE-2016~\citep{Nordhaus2018} merged with the re-calibrated CDICE climate part (cf. section~\ref{sec:climate}).
We will, in the following, also refer to this complete IAM as CDICE. A comprehensive summary of the complete calibration of CDICE is available in Appendix~\ref{sec:generic}. We will refer to the original calibration of the DICE model as DICE-2016.\footnote{Note that the calibration for the original DICE-2016 stems from the publicly available GAMS code specification. For details, see~\url{http://www.econ.yale.edu/~nordhaus/homepage/homepage/DICE2016R-091916ap.gms.}}

The key non-climate part of the DICE-2016 model consists of a single, infinitely lived, representative consumer and a single firm. As it is standard in economics the equilibrium allocation can be described as the solution
to a planner's problem (see, e.g.,~\cite{Golosov2014} for more details, and~\cite{kotlikoff2021making} for a critique of this approach). The planner / representative agent maximizes a time-separable utility function over (per capita) consumption $ (\frac{C_t}{L_t})_{t=0}^{\infty} $ with a constant intertemporal elasticity of substitution (IES),  $ \psi >0 $, and a time preference parameter, $ 0< \beta<1 $.\footnote{In Appendix~\ref{sec:generic}, we list all exogenous variables, equations, and parameters of DICE-2016 and CDICE.} The optimal value, $ V_0 $ is given by the following expression:

\begin{align}
\label{eq:obj_2016}
V_0=  & \max_{\left\{K_{t+1}, \mu_t\right\}_{t=0}^{\infty}} \sum_{t=0}^{\infty} \beta^{t}
  \frac{\left(\frac{C_t}{L_t}\right)^{1-1/\psi} -1}{1-1/\psi} L_t\\
  \text{s.t.} \quad \label{eq:kplus_2016}
  & K_{t+1} = \left(1  -\Theta\left(\mu_{t}\right) - \Omega\left(T_{\text{AT},t}\right) \right)
    A_{t} K_t^{\alpha} L_t^{1-\alpha} + (1-\delta) K_t - C_{t}  \\
    \nonumber & \text{\cref{eq:cc_mass}, \; \cref{eq:temp1}, \; \cref{eq:temp2}}\\
  & \label{eq:Kplus_nonnegative} 0 \leq K_{t+1} \\
  & \label{eq:mu_range} 0 \leq \mu_t \leq 1 \\
  & \label{eq:emissions} \text{where}\;  \; E_t=\sigma_t Y^{\text{Gross}}_{t} (1-\mu_t) + E^{\text{Land}}_t.
\end{align}
The CO2 emissions, denoted by $E_t$, consist of non-industrial emissions, $ E_t^{\text{Land}} $, as well as industrial emissions that are modeled as a fraction of output, $ \sigma_t Y_t^{\text{Gross}}$ , with $\sigma_{t}$ being emission intensity and $ \mu_t \ge 0 $ being mitigation. Output is produced in a Cobb-Douglas technology with capital, $K_t$, and labor, $L_t$. Mitigation is costly and decreases output at a rate $ \Theta(\mu_t) $. Higher temperatures decrease output at a damage $ \Omega(T^{\text{AT}}_{t})$. For a detailed specification and parametrization of all the equations including exogenous variables and damages, please refer to Appendix~\ref{sec:generic}.

First, the planner solves the maximization problem stated above without understanding that higher mitigation $ \mu_t $ leads to lower damages from a temperature increase. In the BAU scenario, she only chooses an investment path, and mitigation is set to zero. Then, she solves the problem optimally by choosing mitigation as well as an investment path. In both cases, the SCC is the marginal cost of atmospheric
carbon in terms of the numeraire good. Following the literature (see, e.g.,~\citet{Traeger2014} and~\citet{Cai2019}), we can write the SCC as the planner's marginal rate of substitution between the atmospheric carbon concentration and the capital stock. Thus, we have:\footnote{We mention here for the sake of accuracy that in DICE-2016, the cost of backstop is given in 2010 thousand USD per ton of CO2 in 2015, whereas the mass of carbon in the atmosphere is measured in tons of carbon. Thus to make the social cost of carbon formula numerically correct with respect to units of measurement, one needs to adjust it by a factor of 3.66, which is a conversion rate between carbon and CO2. The SCC is measured in 2010-US dollars per ton of carbon.}
\begin{align}
  \label{eq:SCCdef}
  SCC_{t} = -\frac{\partial V_{t}/\partial
  M_{\text{AT},t}}{\partial V_{t}/\partial
  K_{t}}.
\end{align}

The optimal carbon tax ($CT_{t}$) is the tax that equates the private and the social cost of carbon.~\citet{Nordhaus2018}, among others, defines the optimal carbon tax as a function of mitigation $\mu_{t}$. The social planner chooses the mitigation $\mu_{t}$, which is equivalent to choosing the carbon tax in units [USD/tC]\footnote{One needs to multiply the carbon intensity $\sigma_{t}$ in the denominator by 1000 because we define the carbon intensity in the units of 1000 GtC.}, that is,
\begin{align}
  \label{eq:CTdef}
  CT_{t} = \frac{\theta_{1,t}\theta_{2}\mu_{t}^{\theta_{2}-1}}{\sigma_{t}},
\end{align}
where $\theta_{1,t}$ is the abatement cost, and $\sigma_{t}$ is emission intensity.\footnote{Detailed description of all parameters and exogenous parameters of the model can be found in Appendix~\ref{sec:generic}.}
By definition, the SCC is equal to the optimal carbon tax if $\mu_{t}<1$. 

In the following, we first consider the CDICE model under different CMIP5 calibrations for the carbon cycle and the temperature equation. This is to illustrate that different reasonable calibrations for the CE can have large effects on the optimal abatement.
Second, we compare CDICE to DICE-2016 to show some economic consequences of the incorrect calibration in DICE-2016. We do so in two steps. First, we compare the predictions of the two calibrations with the exact economic calibration from DICE-2016. Second, we also investigate the effects of differences in the IES-parameter $ \psi $, which can have large effects on future g-adjusted interest rates.

\subsection{CDICE - Economic consequences of climate model uncertainty}
\label{sec:Cdice}

We consider the economic model of DICE-2016 combined with our nine different reasonable calibrations for carbon-cycle and temperature equations. We explore the economic consequences of the large uncertainty in our climate model.\footnote{All results we report in the following were obtained by using \lq\lq Deep Equilibrium Nets\rq\rq \ by~\cite{azinovic2019Deep}, a method for computing global solutions to high-dimensional dynamic economic models.}
We take as a starting date (t=0) the year 2015. In addition, the initial conditions are listed in Table~\ref{table:IniVals2015ClimDICE2}. 

\begin{figure}[t!]
 \begin{subfigure}[b]{0.5\linewidth}
     \centering
     \includegraphics[width=1.0\linewidth]{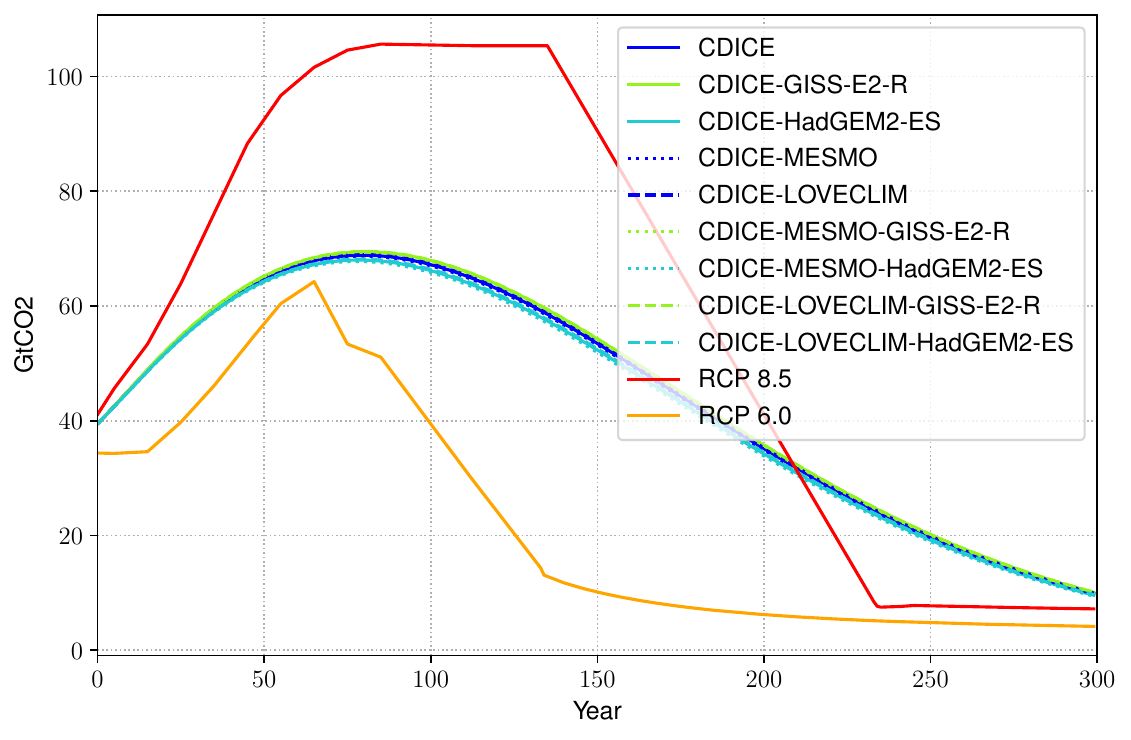}
   \caption{Emissions} 
     \label{fig:bau_emission}
  \end{subfigure}
  \begin{subfigure}[b]{0.5\linewidth}
     \centering
     \includegraphics[width=1.0\linewidth]{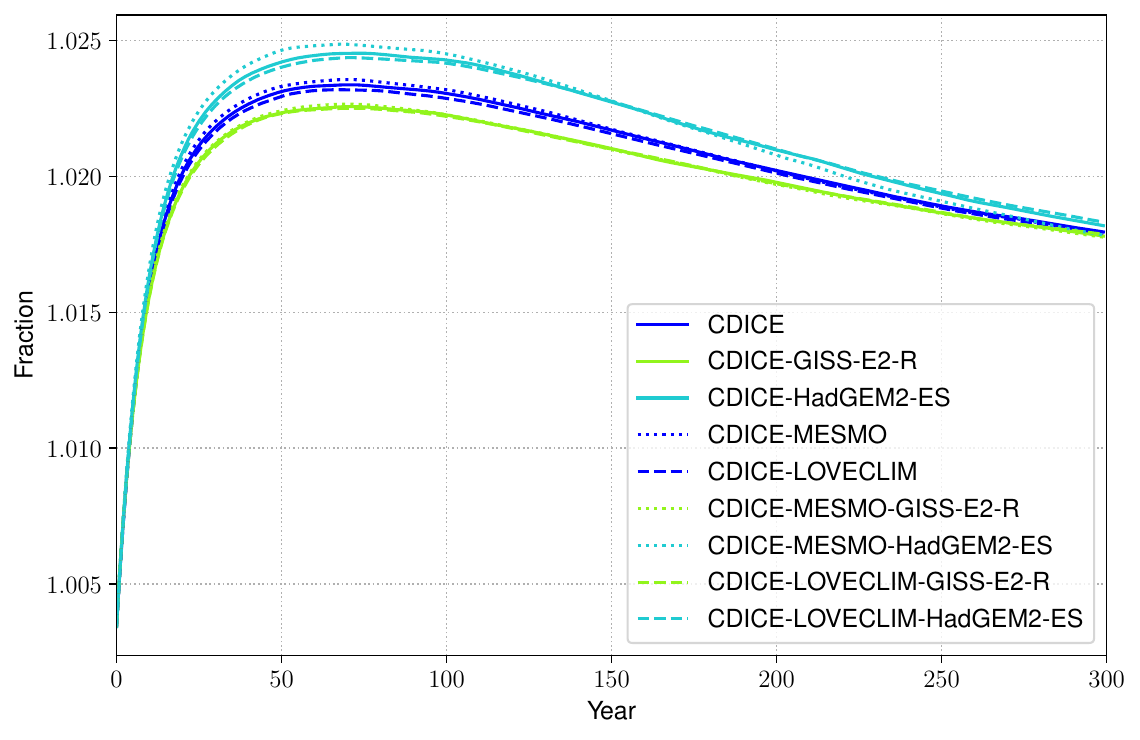}
    \caption{g-adjusted interest rate}
     \label{fig:bau_interest}
     \end{subfigure}
     \caption{BAU emissions (left panel) as well as growth adjusted interest rate (right panel) for different climate calibrations.}
    \label{fig:bau_CDICE1}
 \end{figure}

We start our analysis by examining the predictions of the model in a scenario without mitigation. The first observation is that in this framework, the emissions are not sensitive to the climate calibration used. Figure~\ref{fig:bau_emission} shows the path of emissions for our nine versions of CDICE and compares it to the emissions in RCP 8.5 and RCP 6. The assumed climate parameters have an effect on BAU emissions via differences in damages (higher temperatures imply higher damages that in turn imply less carbon emissions in this model), but the effect is quantitatively tiny. One reason for this behavior is that the damages only become large once emissions have started to decline substantially.  The fraction by which  emissions are reduced by these damages might be significant, but the absolute value of the reduction is small.
Figure~\ref{fig:bau_emission} also shows that the emissions from DICE-16 lie somewhere between RCP6 and RCP8.5. The lessons from our detailed analysis of these emission scenarios above should also carry over to the DICE-2016 BAU emissions. 

Figure \ref{fig:bau_interest} shows that the path of the growth adjusted interest rate is also more or less the same across different climate calibrations. There is some variation that can be explained by differences in future damages that translate into differences in the interest rate. Higher expected future damages imply a lower capital stock and a (slightly) higher interest rate today.
The variation over time, which is much more substantial, is mostly explained by the assumption in DICE-2016 that TFP growth, as well as population growth, flattens over time. The DICE-2016 calibration starts with an initial population growth of 1.4 percent and an initial TFP growth of 2.25 percent, with population growth declining fast over the next 50-100 years and TFP growth declining substantially slower. The way the time-varying growth rates translate into the time-varying interest rate depends crucially on the preferences of the representative agent. We will return to this issue in section~\ref{sec:discount} below.
Note for now that the initial g-adjusted interest rate is close to zero, whereas after 50 years, it is around 2 percent, and then finally converges to 1.5 percent in the very long run.

\begin{figure}[t!]
 \begin{subfigure}[b]{0.5\linewidth}
     \centering
     \includegraphics[width=1.0\linewidth]{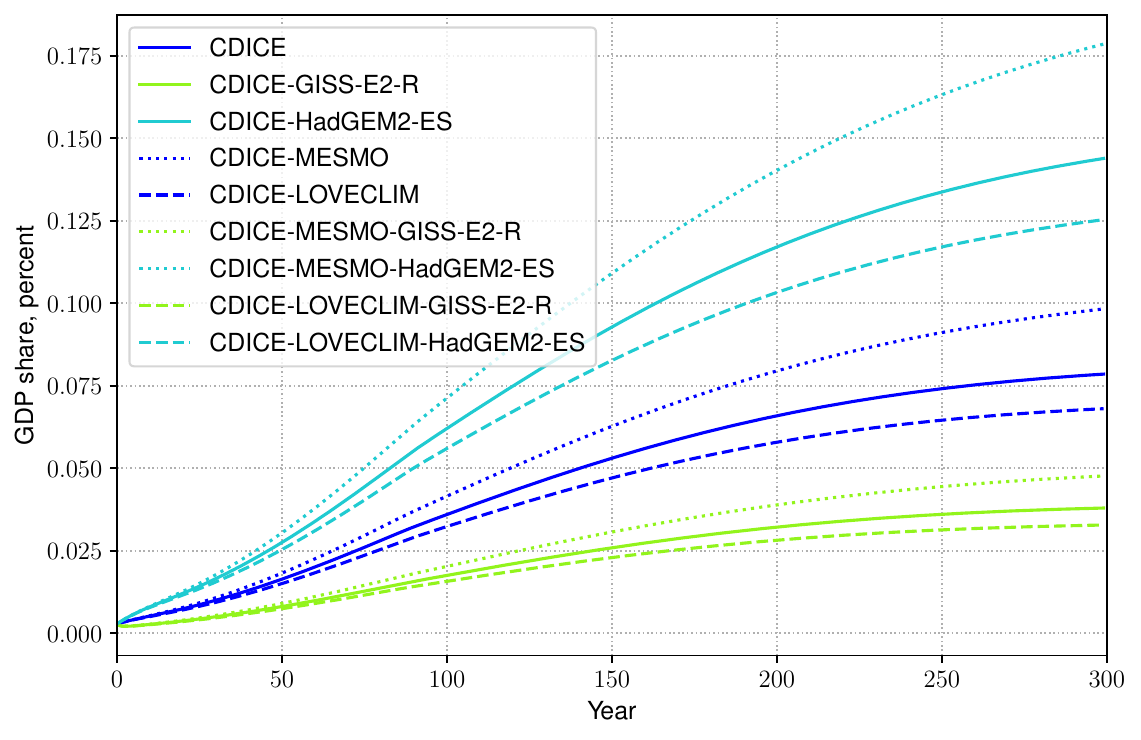}
    \caption{Damages}
     \label{fig:bau_damages}
  \end{subfigure}
  \begin{subfigure}[b]{0.5\linewidth}
     \centering
     \includegraphics[width=1.0\linewidth]{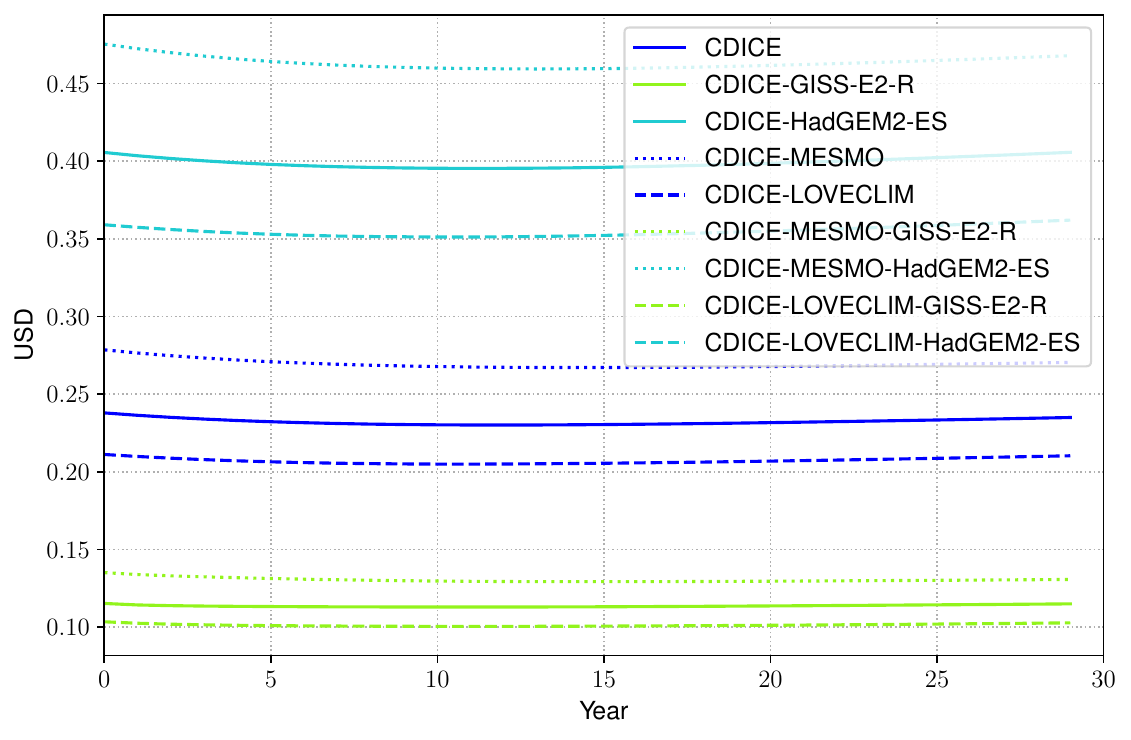}
    \caption{Relative SCC}
     \label{fig:bau_scc}
     \end{subfigure}
     \caption{This figure shows the evolution damages (left panel) and the relative SCC (right panel) over time for nine CDICE-calibrations in the business as usual case. Year zero on the graph corresponds to the starting year 2015.}
    \label{fig:bau_CDICE}
 \end{figure}

Figure~\ref{fig:bau_CDICE} shows the effect of different climate calibrations on damages as a percentage of GDP and the SCC as a percentage of GDP. As is to be expected from our discussions in section \ref{SCC:CDICE} above, our three different calibrations for the carbon cycle imply three vastly different predictions for CO2 concentrations.
Across different temperature equations, for a fixed carbon cycle, this then results in 9 vastly different paths for temperature.
 As already pointed out in section \ref{SCC:CDICE} above, the effects of the carbon cycle calibration on temperatures and damages are not nearly as large as the effect of the calibrations of the temperature equation, and the combination of an extreme carbon cycle and an extreme temperature equations yields extreme damages. This is true both for CDICE-MESMO-HadGEM2-ES as well as for CDICE-LOVECLIM-GISS-E2-R.
The range of plausible damages in 300 years goes from around three percent of GDP to over 17.5  percent of GDP, almost a factor of six.
As expected this has a very large effect on optimal abatement.
The relative SCC depicted over the next 50 years in Figure \ref{fig:bau_scc} is fully consistent with our results in section \ref{SCC:CDICE} above. In this section, we discussed only the SCC at a fixed date $ t=5 $, whereas we examine it here over the next 50 periods. The SCC initially falls over time, and this effect is caused by the stark rise in the g-adjusted interest rate discussed above. After that, it stays relatively flat, and the relative SCC between the different temperature calibrations is similar to that described in section \ref{SCC:CDICE} above.

\begin{figure}[htb]
 \begin{subfigure}[b]{0.5\linewidth}
     \centering
     \includegraphics[width=1.0\linewidth]{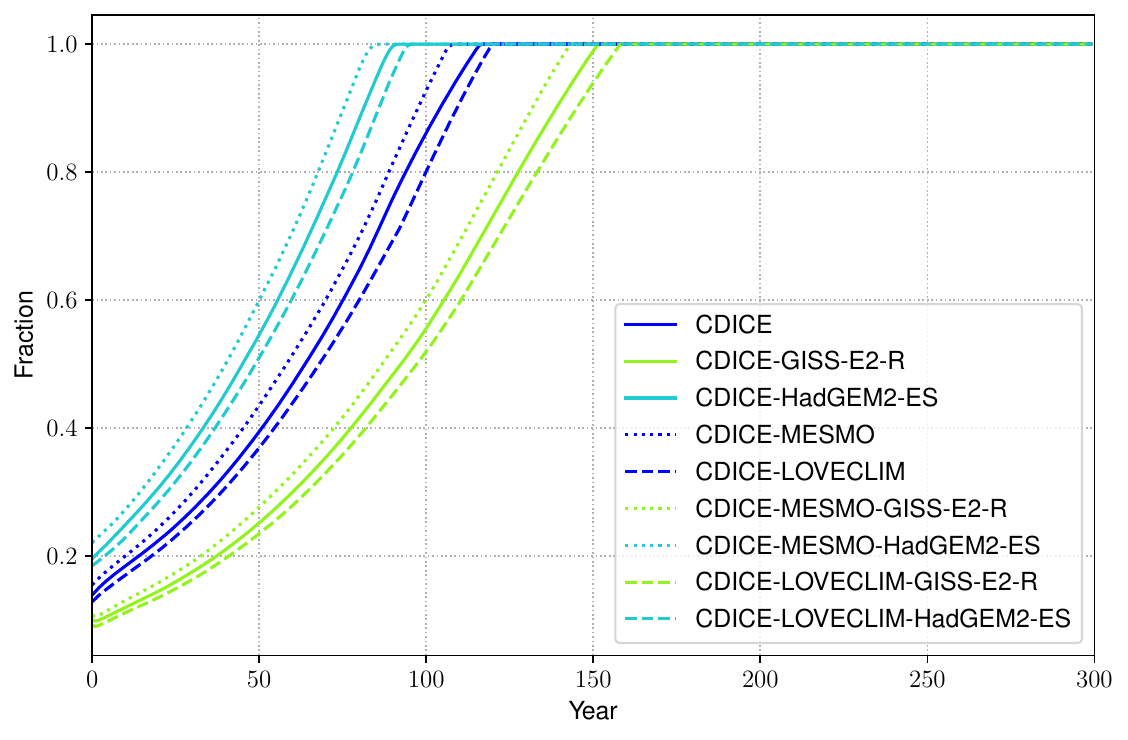}
    \caption{Opt. Abatement}
     \label{fig:opt_abate}
  \end{subfigure}
  \begin{subfigure}[b]{0.5\linewidth}
     \centering
     \includegraphics[width=1.0\linewidth]{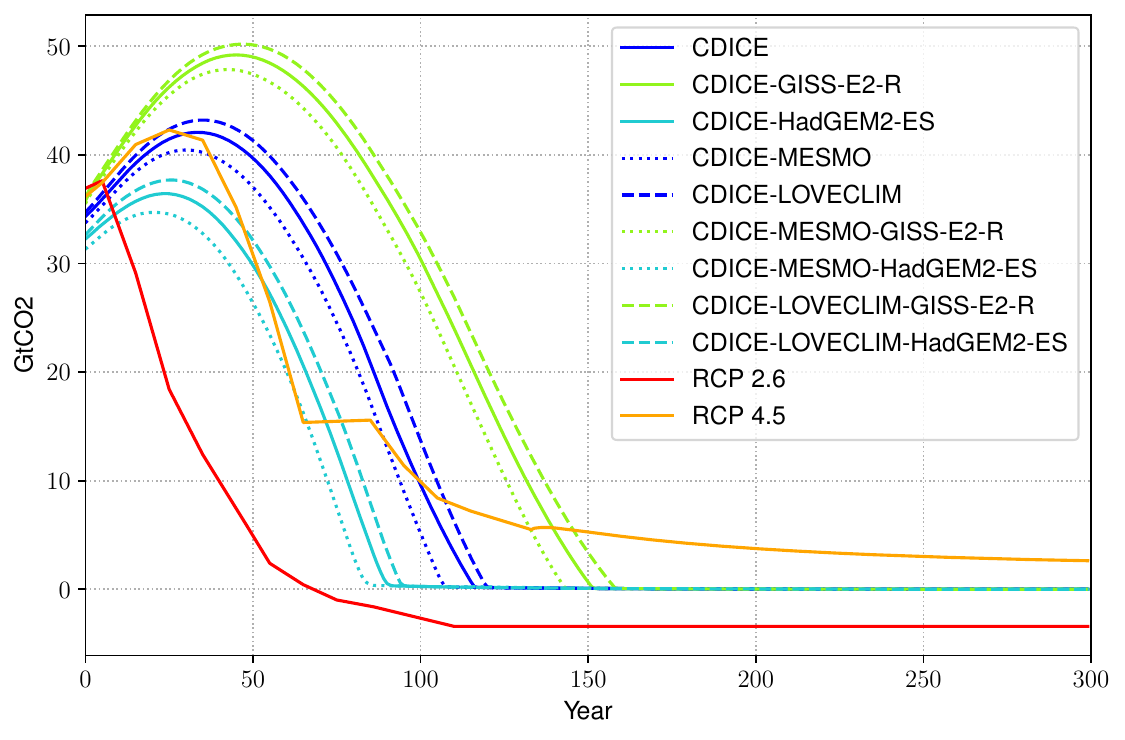}
    \caption{Opt. Emissions}
     \label{fig:optCO2}
     \end{subfigure}
     \caption{This figure shows optimal abatement (left panel) and emissions (right panel) over time.}
    \label{fig:opt1}
 \end{figure}

The large differences in damages and the SCC across calibrations imply large differences in the optimal solution. We now assume that the social planner optimally chooses mitigation, $ \mu_t, $ in the optimization-problem (\ref{eq:obj_2016}).
Figure \ref{fig:opt_abate} shows the optimal abatement choices for the nine different climate calibrations.
Differences in the social cost of carbon translate mechanically into differences in optimal abatement via Equation (\ref{eq:CTdef}) above. Since the exogenous parameters in (\ref{eq:CTdef}) are time-varying, optimal mitigation is time-varying. The differences in optimal mitigation between the different climate calibrations are huge over the next 80 years; about a factor of three.
By definition, differences in abatement lead to differences in emissions. Figure 
\ref{fig:optCO2} shows BAU emissions for our 9 cases compared to the emission scenarios in RCP 2.6 and in RCP 4.5. It is interesting to note that in all cases, optimal emissions stay far above RCP 2.6 emissions. This is a consequence of the specification of the damage function in DICE-2016, which we use here, and the fact that the model does not allow for carbon removal, which becomes an active part of future policy in RCP 2.6. Emissions in the MMM case closely mirror optimal emissions in RCP 4.5.

\begin{figure}[htb]
 \begin{subfigure}[b]{0.5\linewidth}
     \centering
     \includegraphics[width=1.0\linewidth]{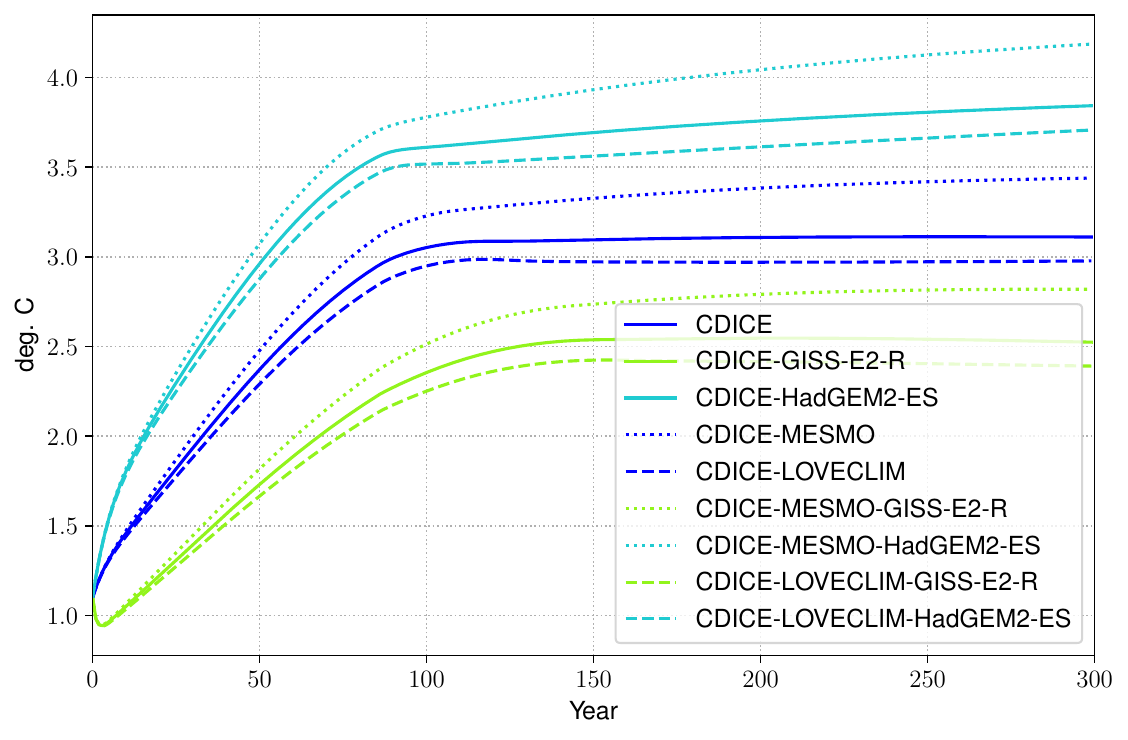}
    \caption{Opt. Temperature}
     \label{fig:opt_temp}
  \end{subfigure}
  \begin{subfigure}[b]{0.5\linewidth}
     \centering
     \includegraphics[width=1.0\linewidth]{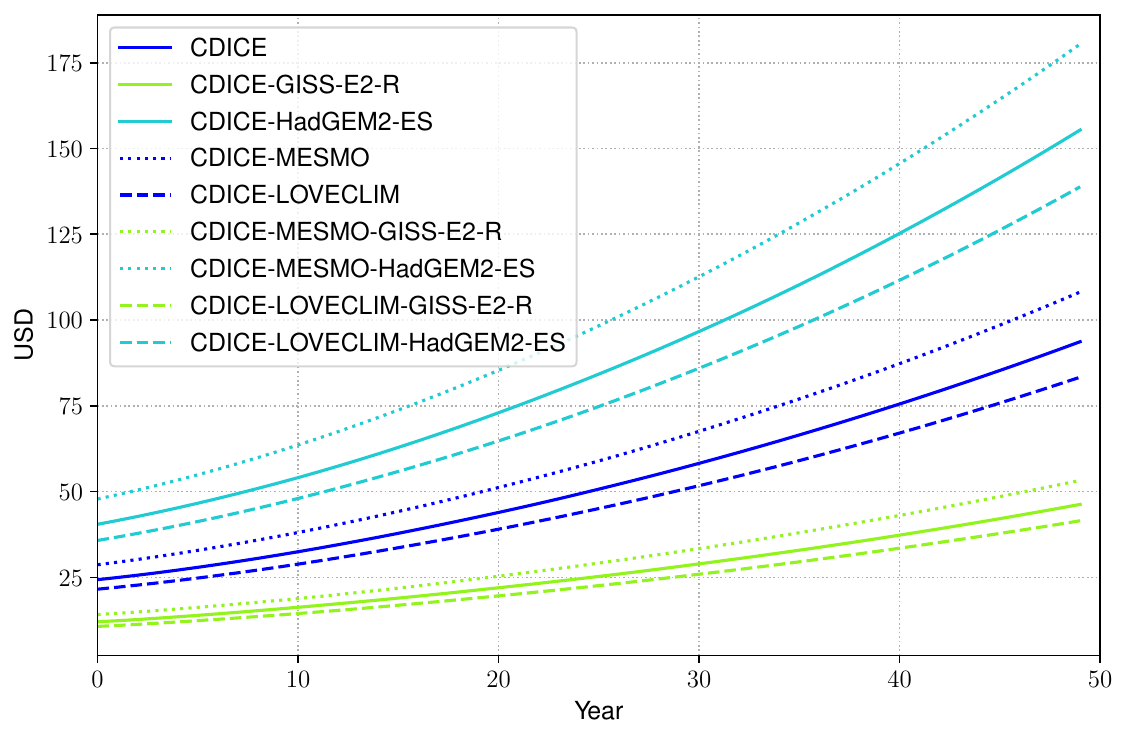}
    \caption{SCC}
     \label{fig:opt_scc}
     \end{subfigure}
     \caption{Evolution of temperature (left panel) and SCC (right panel) with optimal abatement.}
    \label{fig:opt2}
 \end{figure}

The fact that optimal emissions stay far above RCP2.6, which is the strong-mitigation scenario from IPCC, readily implies that the optimal temperature rises far above the Paris limit.  Figure \ref{fig:opt_temp} shows that in the MMM calibration, temperature rises (relative to preindustrial levels) by 3 degrees Kelvin and stays at that level for hundreds of years. In our most pessimistic climate calibration, the optimal temperature rises by over 4 degrees. As pointed out before, this depends crucially on the choice of the damage function, which we take directly from DICE-2016 for comparison purposes.
 As~\cite{nordhaus2008question} points out, \lq\lq \textit{the economic impact of climate change ... is the thorniest issue in climate-change economics}\rq\rq. See \cite{Hansel2020} or \cite{carleton2020valuing} for more realistic treatments of damages.

Naturally, the choice of the damage function leads to relatively low SCC  in levels. Figure \ref{fig:opt_scc} depicts the social cost of carbon in dollars over the next 50 years. Note that it increases only because of GDP growth. The relative SCC is almost identical to that in the BAU case depicted in Figure~\ref{fig:bau_scc} above. Recall that the optimal tax on carbon is equal to the SCC. Thus, the model's optimal tax lies, for the MMM case, slightly below what is typically discussed for economic policy purposes.

\subsection{The economic consequences of miscalibrated climate}

\begin{figure}[t!] 
  \begin{subfigure}[b]{0.5\linewidth}
    \centering
    \includegraphics[width=1.0\linewidth]{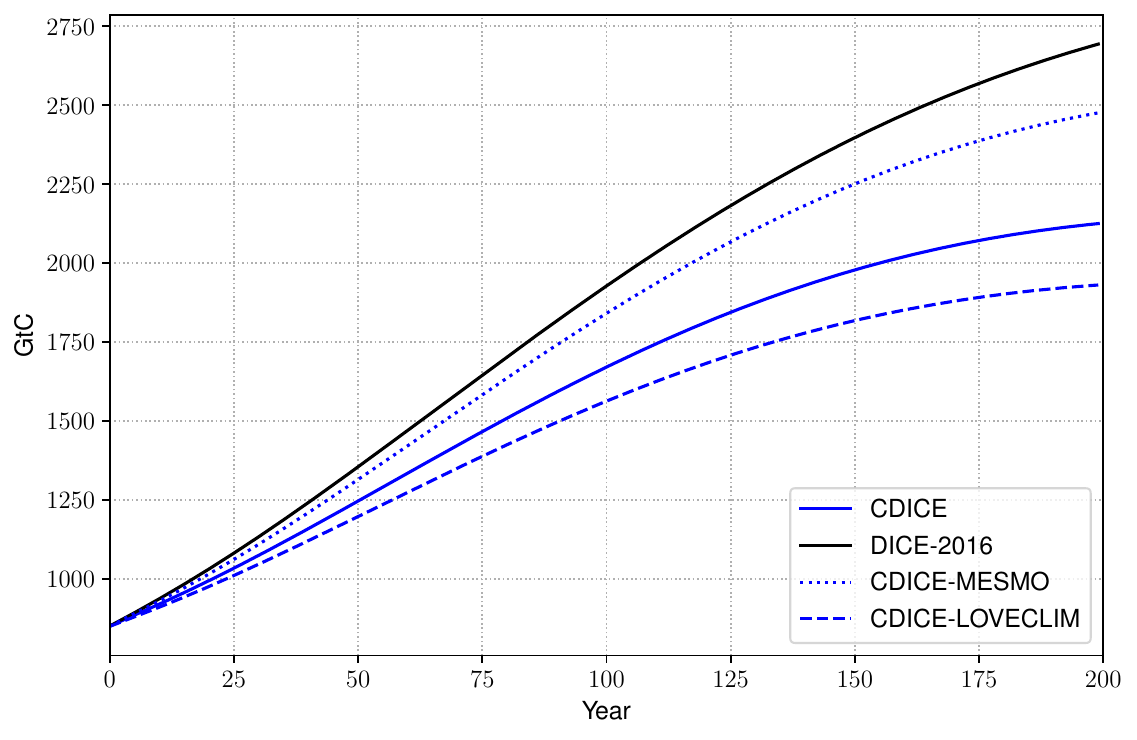} 
    \caption{Mass of carbon in the atmosphere, BAU case} 
    \label{fig:bau_mass:a} 
  \end{subfigure}
  \begin{subfigure}[b]{0.5\linewidth}
    \centering
    \includegraphics[width=1.0\linewidth]{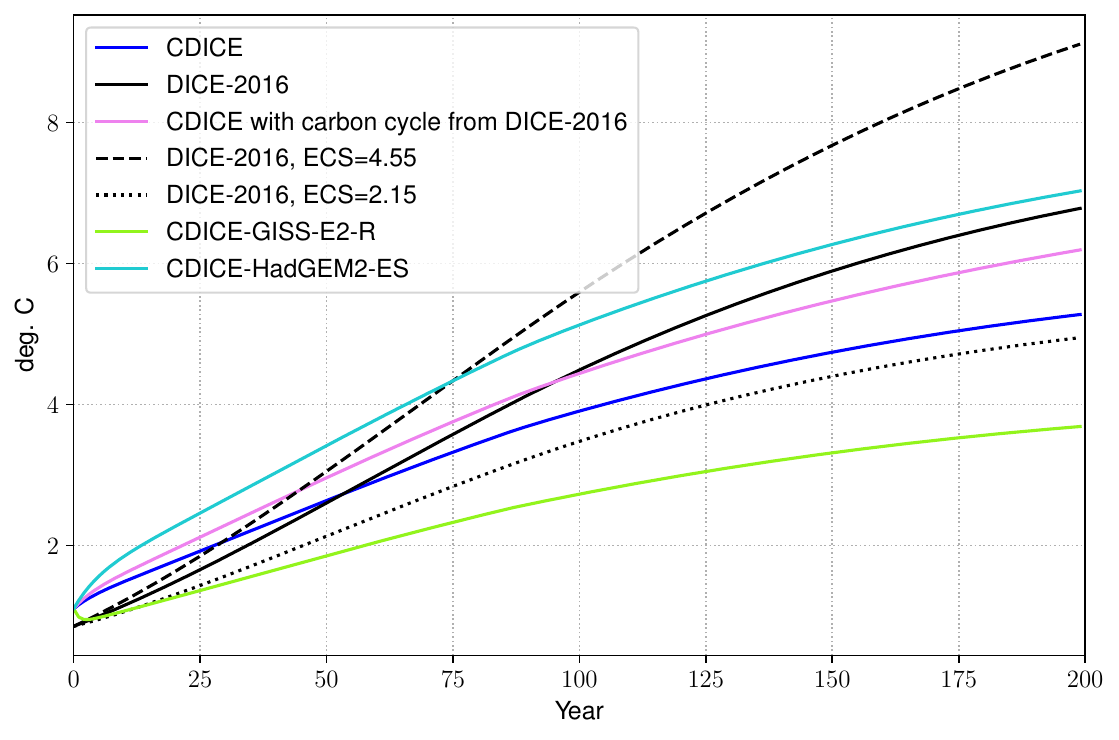}
    \caption{Temperature of the atmosphere, BAU case } 
    \label{fig:bau_temps} 
  \end{subfigure} 
  \caption{Mass of carbon (left panel) and atmospheric temperature (right panel) for the CDICE and DICE-2016 calibrations.}
  \label{fig:bau-misc}
 \end{figure}

As pointed out above, the calibration of the climate part of DICE-2016 has two serious flaws. Both the temperature equations and the carbon cycle are miscalibrated.
As mentioned in~\cref{sec:test_100GtC}, the carbon cycle in DICE-2016 overstates the fraction of CO2 emissions that end up in the atmosphere relative to the MMM in~\cite{joos2013carbon} and the extreme CDICE-MESMO case. We can confirm this by inspecting the BAU results. Compared to all three carbon calibrations in CDICE, DICE-2016 predicts way too much carbon in the atmosphere (cf. Figure~\ref{fig:bau_mass:a}). In fact, the amount of carbon in the atmosphere is larger than for the CDICE-MESMO calibration (which corresponds to the extreme case in \cite{joos2013carbon}) at any point in time.

 %

The effect is less pronounced if one considers the overall temperature response. In addition to a mistake in the carbon cycle, the DICE-2016 calibration also makes a mistake in the temperature equation.
As Figure \ref{fig:bau_temps} illustrates, in the short run (for the first 50 years or so), CDICE predicts more warming than the DICE-2016, despite the fact that there is far less CO2 in the atmosphere. Even for an ECS of 4.5, DICE-2016 initially warms slower than the CDICE MMM calibration.
The combined effect of temperature- and carbon-cycle miscalibration is crucial here; if one takes the CDICE MMM temperature equations together with the DICE-2016 carbon cycle, the temperature evolution is comparable to DICE-2016. The most notable difference is the initially overly slow warming in the DICE-2016 temperature equations.

Note that in the very long run, the dynamic system, defined by equations (\ref{eq:cc_mass}), (\ref{eq:temp1}) and (\ref{eq:temp2}), that governs climate in our simple model will converge to a new steady-state after all emissions are zero. In this steady-state, relative masses of carbon are given by the vector $ M_{EQ} / \sum_{l=1}^3 M_{EQ,l}$ and it can be easily seen that the fraction of emitted CO2 that remains in the atmosphere is about 0.26 for our CDICE calibration and about 0.22 for DICE-2016. The long-run effect on temperature is determined by the ECS, which is chosen to be 3.25 in CDICE and 3.1 in DICE-2016. Overall, the very long-run behavior of the two calibrations will be similar, with more warming in CDICE than in DICE-2016. Nevertheless, the miscalibration in DICE-2016 implies a consistently higher mass of carbon in the atmosphere over the next 500 years and a higher temperature after about 100 years. The important lesson is that parameters that determine the very long run of the climate system are more or less irrelevant to temperature over the next 300 years and, therefore, for the SCC that determines optimal policy today.

\begin{figure}[t!] 
  \begin{subfigure}[b]{0.5\linewidth}
    \centering
    \includegraphics[width=1.0\linewidth]{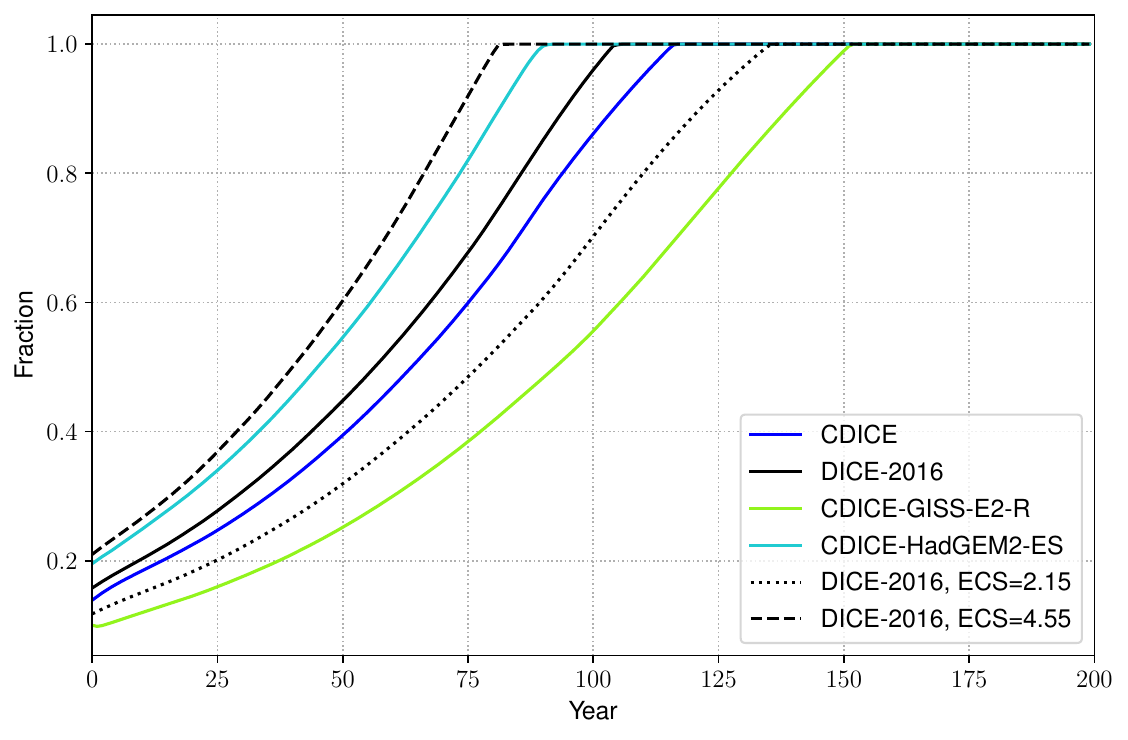} 
    \caption{Optimal abatement over time} 
    \label{fig:opt_abate_misc} 
  \end{subfigure}
  \begin{subfigure}[b]{0.5\linewidth}
    \centering
    \includegraphics[width=1.0\linewidth]{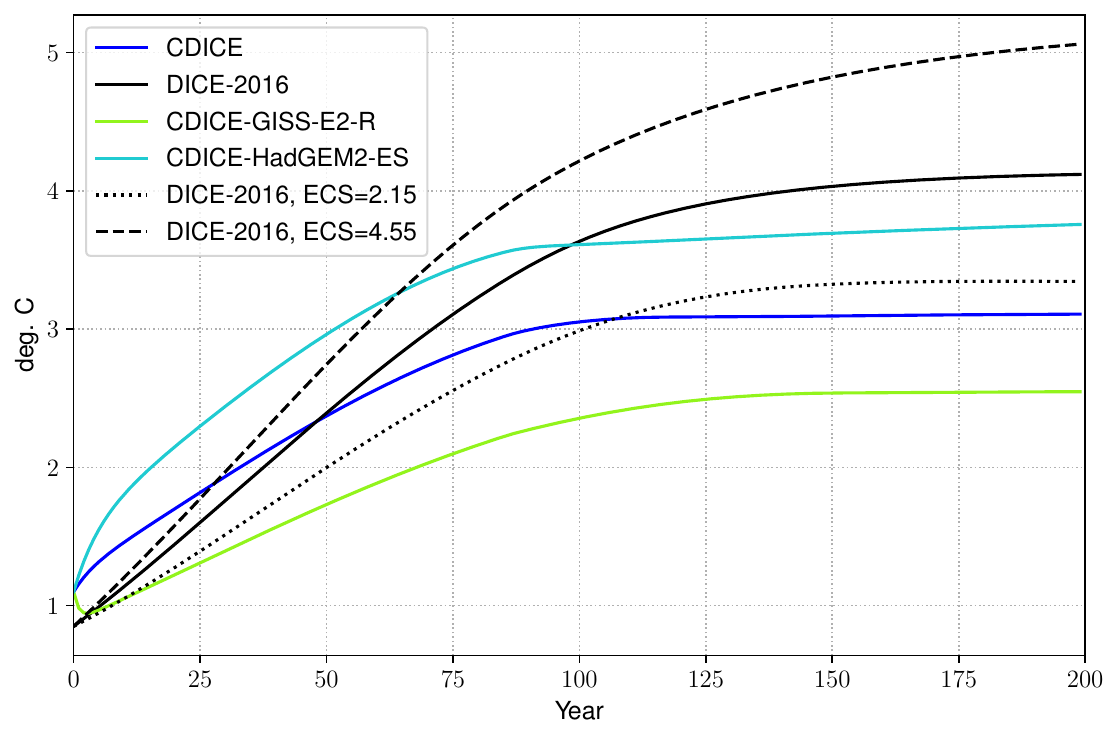}
    \caption{Temperature in atmosphere, optimal abatement case} 
    \label{fig:opt_tem_misc} 
  \end{subfigure} 
  \caption{Optimal abatement in CDICE and DICE-2016.}
  \label{fig:opt_misc1}
\end{figure}

So far, we are repeating statements that we already made in section \ref{sec:climate} above. The crucial question is what the implications of this miscalibration are for the optimal policy.
Figure \ref{fig:opt_abate_misc} depicts the optimal levels of abatement for the different calibrations.
Abatement is presented as a share of industrial emissions that are mitigated. Hence, its numerical values are bound to the interval $[0,1]$, where $0$ corresponds to the absence of any abatement, whereas $1$ implies using the full mitigation capacities.
The figure shows that the DICE-2016 calibrations imply a little more optimal abatement than the respective CDICE calibrations. The differences are so small because initial warming lies below the CDICE case and, with discounting, this is important for abatement over the next decade. This finding again highlights the relevance of both parts of the climate model, of the carbon cycle and the temperature response, and of caution against the potential for compensating errors.

From this figure, one might be tempted to argue that for policy recommendations it does not make much of a difference whether one uses the CDICE calibration or the DICE-2016 calibration. However, it makes a large difference for optimal temperature as Figure \ref{fig:opt_tem_misc} demonstrates.
The DICE-2016 (with the ECS of  3.1) atmospheric temperature is now outside of the CMIP5 range. It can be seen that after about 90 years, the black line (corresponding to DICE-2016) crosses the predicted temperature under the optimal mitigation in DICE-2016, and it is above the temperature predicted by the extreme climate sensitivity scenario CDICE-HadGEM2-ES. In the BAU scenario, DICE-2016 falls well within the CMIP5 range, whereas in the economic model with mitigation, this is no longer the case. This issue arises because, in the BAU scenario, DICE-2016 is `helped' by offsetting errors in the climate equations and the carbon cycle. In the case of optimal mitigation, the flaw in the carbon cycle dominates and leads to so little mitigation that the temperature rises by more than 4 degrees. This is already implicit in the BAU SCC. Despite the fact that DICE-2016 leads to much larger damages and much higher temperatures than CDICE, the SCC is almost identical. This implies (to first order) the same optimal carbon tax for CDICE and DICE-2016, resulting in far higher temperatures in DICE-2016.

 %

Although long-run temperatures differ so much, the reason why abatement and the SCC are comparable between CDICE and DICE-2016 is simply that these temperature differences only become significant in 75 years and, as shown in Figure~\ref{fig:bau_interest}, the g-adjusted interest rate is relatively high at that point in time. The damages that occur are heavily discounted. If one takes a different perspective and, instead of maximizing the utility of a social planner, restricts long-run warming to lie below 2 degrees Kelvin, the differences between the two climate-model calibrations become starker.
In fact, for the DICE-2016 calibration, it is impossible to keep the temperature below 2 degrees. Full abatement that reduces industrial emissions to zero is possible (and in fact, in DICE-2016, this is not even extremely costly, as the output is reduced by about seven percent) does not suffice to keep global warming below 2 degrees. This is very different in CDICE, where putting an immediate stop to industrial emissions will keep warming below 1.7 degrees. Somewhat more realistically, if one considers the ECP 2.6 emissions profile (that includes carbon removal), DICE-2016 warms by more than 2.5 degrees (as can already be seen in Figure \ref{fig:CMIPBM} above) and stays above 2.3 degrees beyond 2300 while CDICE warms by 2.1 degrees, but warming is below 1.8 degrees by 2300. 

Clearly, discounting is crucial for the small differences in abatement between the two calibrations. The interest rate is endogenous, as it depends on consumption growth, but also on the preferences of the representative agent.

\subsection{The role of discounting}
\label{sec:discount}

There is a  large literature (see, e.g.,~\cite{Hansel2018,stern2007economics}) about the \lq\lq right\rq\rq discount rate for the social planner, that is, about the correct value of the parameter $
\beta = \frac{1}{1+\rho} $ in the planner's utility function. It is sometimes argued that this parameter can be pinned down by the observed average rates of return to capital (or interest rates). In a growth economy, this is not quite correct as the curvature of the planner's felicity function, $ \psi$, and the time preference $ \beta $ are not jointly identified from average interest rates. The fact that in DICE-2016, the SCC is so sensitive to the growth-adjusted interest rate (documented in section~\ref{sec:peSCC} above) implies that changing these two preference parameters simultaneously while keeping average interest rates fixed has large effects on the SCC and on optimal abatement in the DICE-2016 calibration.
For simplicity, we take the baseline per-capita consumption growth rate to be 2 percent, which gives an interest rate of 4 percent in the DICE-2016 calibration (the long-run average growth rate in DICE-2016 is assumed to be zero, so we pick the 2 percent somewhat arbitrary to be consistent with the historical average and common long-run assumptions).

\begin{figure}[t!] 
  \begin{subfigure}[b]{0.5\linewidth}
    \centering
    \includegraphics[width=1.0\linewidth]{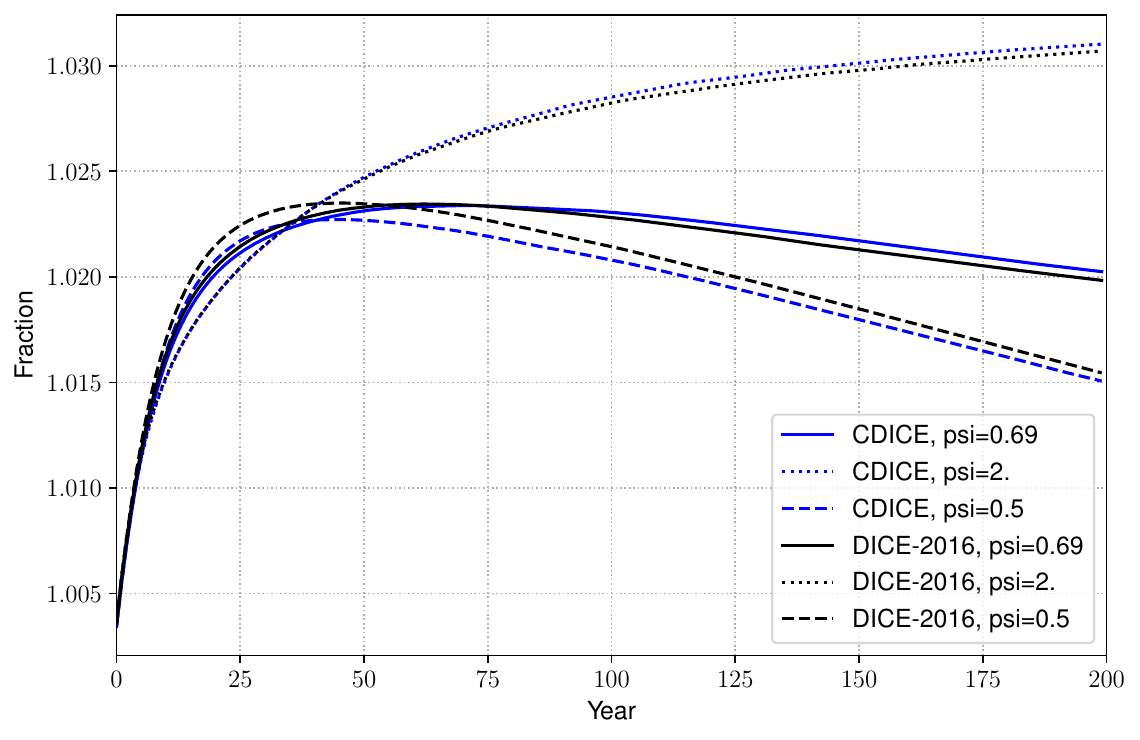} 
    \caption{g-adjusted interest rate} 
    \label{fig:bau_ies_interest} 
  \end{subfigure}
  \begin{subfigure}[b]{0.5\linewidth}
    \centering
    \includegraphics[width=1.0\linewidth]{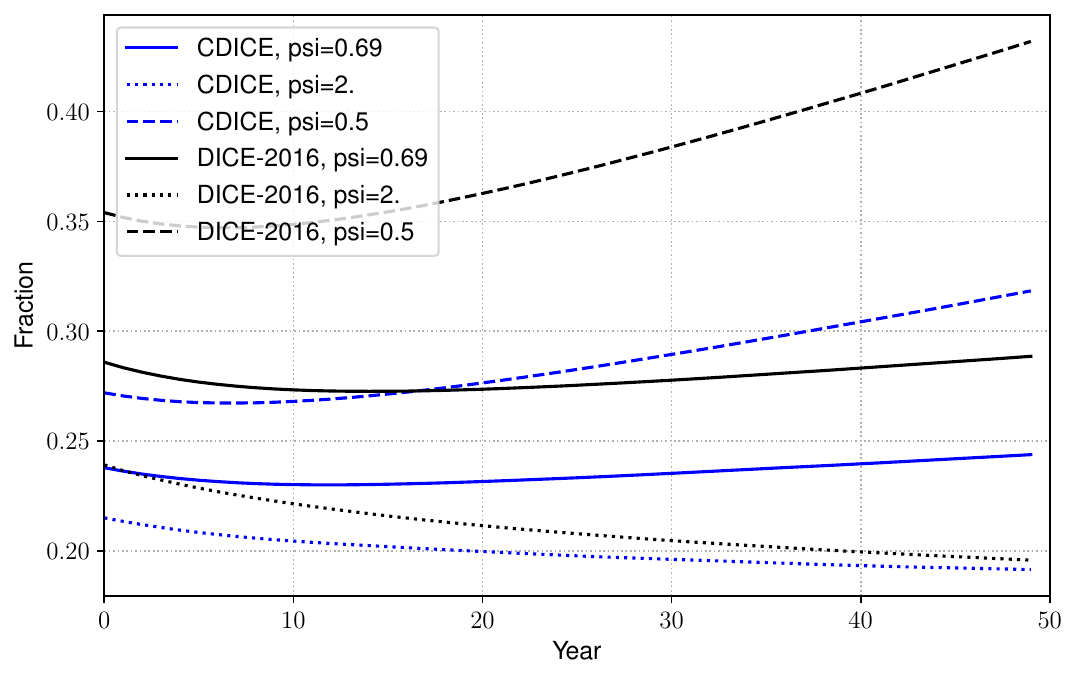}
    \caption{Relative SCC under BAU} 
    \label{fig:bau_ies_scc} 
  \end{subfigure} 
  \caption{Interest rate (left panel) and SCC (right panel) under different values for $ \psi $.}
  \label{fig:opt_misc1}
\end{figure}

Figure \ref{fig:bau_ies_interest} shows the paths of growth adjusted interest rate for different values of the IES, $ \psi $, in the business as usual case. G-adjusted interest rates are quite similar for the first 40 years, but there is significant divergence in the long run. After 200 years the difference in g-adjusted interest rates between $ \psi=0.5 $ and $ \psi=2 $ is almost a factor of 2. 
As Figure \ref{fig:bau_ies_scc} shows,
this naturally translates into a large difference in the social cost of carbon across different specifications of the IES. The figure shows the SCC relative to output. The absolute SCC increases over time in all cases because of output growth. Because of the incorrect timing in the climate calibration of DICE-2016, the effect of the parameter $ \psi $ on interest rates is amplified, and the differences in the SCC between low and high $ \psi $ are larger than 50 percent. While a value of $ \psi=2 $ might be considered an unrealistically high a value of the IES,  $ \psi=0.5 $ is certainly well within the range of the IES which is typically assumed.\footnote{There are large differences in the IES typically assumed in the macro-literature (one or larger) to those typically found in household studies. There it is found that the IES varies significantly across households, ranging from 0.3 to 1. See, e.g. \cite{calvet2021cross}.}

For the $ \psi=0.5 $ case, the relative social cost of carbon initially decreases slightly and then strongly increases over time. This is obviously caused by the fact that g-adjusted interest rates decrease over time. In contrast, for a high value of the IES, $ \psi=2 $, g-adjusted interest rates start decreasing significantly in about 40 years, leading to decreases in the SCC over time.

\begin{figure}[t!] 
    \centering 
    \includegraphics[width=0.9\linewidth]{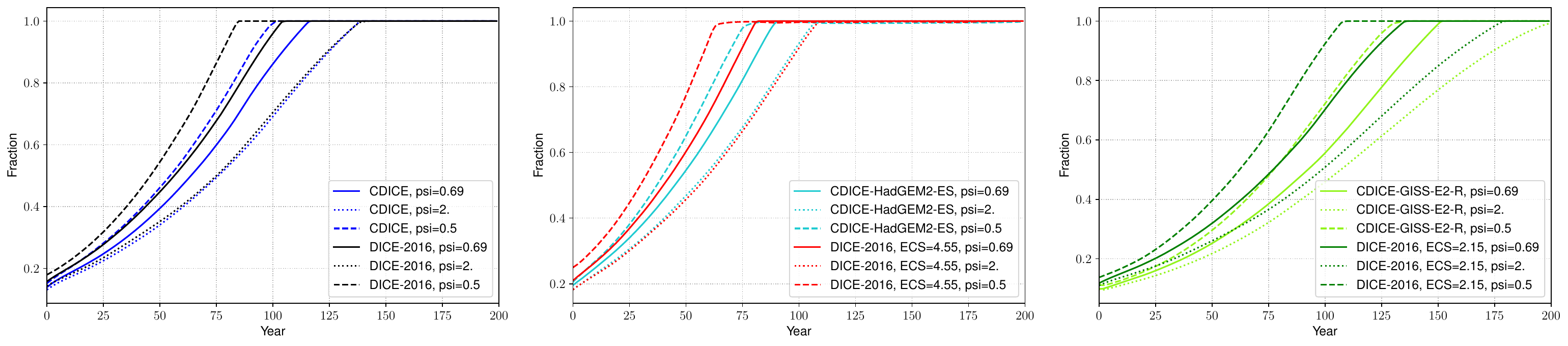} 
    \caption{Optimal abatement for different values of $ \psi$ and different climate calibrations. The left panel shows results for MMM, the middle panel for high ECS calibrations and the right panel for low ECS calibrations.} 
    \label{fig:opt:mu:ies} 
\end{figure}

These differences in the BAU SCC obviously imply large differences in optimal policies. Figure \ref{fig:opt:mu:ies} shows optimal abatement for the three values of $ \psi $, comparing the  DICE-2016  calibration to CDICE. The three panels in the figure show different ECS calibrations for DICE-2016 and HadGEM2-ES and GISS-E2-R for CDICE. Differences in abatement are very large between the different $ \psi $-cases. In particular, in DICE-2016, for large $ \psi $, the use of carbon is extended by about 50 years relative to the low IES case. This is true independently of the assumption on the ECS. Since the timing in the DICE-2016 calibration is wrong, the effects of future lower interest rates are much more pronounced than for CDICE. In fact, for a high value of $ \psi $ (higher future interest rates), the optimal policy in CDICE and DICE-2016 are quite similar; the differences become large as expected future interest rates become small. 
The range of plausible values for $ \psi $ adds another model uncertainty to economic models of climate change, and it turns out that the miscalibration in DICE-2016 amplifies the effect. In more interesting models with uncertainty and Epstein-Zin utility (see, e.g.,~\cite{Cai2019}), the use of the DICE-2016 climate calibration is likely to lead to serious problems.

\begin{figure}[th!] 
    \centering 
    \includegraphics[width=0.9\linewidth]{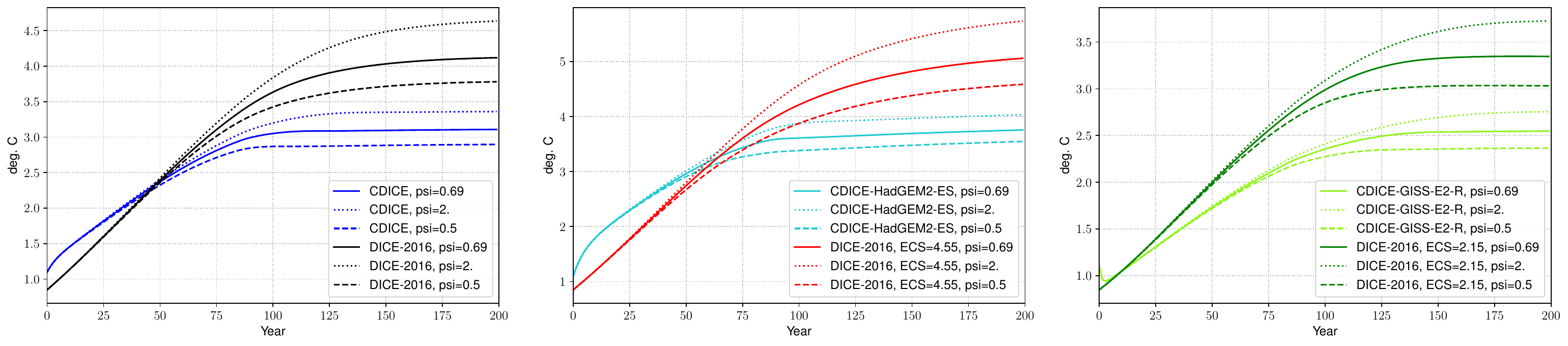} 
    \caption{Optimal temperatures for different values of $ \psi$ and different climate calibrations. The left panel shows results for MMM, the middle panel for high ECS calibrations and the right panel for low ECS calibrations.} 
    \label{fig:opt:tem:ies} 
\end{figure}

As above, optimal long-run temperatures in DICE-2016 are far above the values in CDICE. Now there is an even larger spread caused by different values for $ \psi $. Figure \ref{fig:opt:tem:ies} shows future temperature under optimal abatement for different values of $ \psi $ and for the three different specifications of ECS.
For the case of a high ECS, the optimal warming under DICE-2016 is now quite dramatic if one assumes a high value of $ \psi $. In 200 years, optimal warming is almost 6 degrees; far above anything considered acceptable by most climate scientists. Even for the MMM calibration, a high $ \psi $ implies warming by more than 4.5 degrees.  For this high IES case, warming prescribed by DICE-2016 falls significantly out of the CMIP5 range.

Future interest rates and growth rates matter a lot for optimal mitigation and optimal interest rates. Even if one settles on a path for TFP growth, different reasonable preference specifications lead to large differences in interest rates. The fact that the DICE-2016 calibration of the climate emulator contains two flaws (discussed in detail above) implies an unrealistically large sensitivity to discounting, and differences in future interest rates translate into an unrealistically large difference in the SCC.

The results presented here clearly highlight the need for reliable climate calibrations and show the advantages of our approach. The fact that DICE-2016 lies within the range of CMIP5 predictions in some cases does not ensure that it can be used as a reliable tool within a calibrated economic model.

%

\section{Conclusion}
\label{sec:conclusion}

IAMs provide a quantitative framework which takes into account economic and environmental stocks and flows as well as their interaction and which allows researchers to investigate climate change and possible mitigation pathways~\citep{Hassler2016}. IAMs differ very much in terms of structure, complexity, level of detail, and possible solution methods~\citep{Weyant2017,Hare2018,Flamos2019}. Models like the one by~\citet{Clarke2009} have very rich climate representation, whereas~\citet{Bosetti2007} provides a detailed economic model. Other models provide a detailed representation of the energy sector. Models like~ \citet{Nordhaus2018,Golosov2014,Traeger2019,Hambel2021,Hassler2018} have a more parsimonious structure and exhibit a higher level of aggregation but allow for monetary estimates both for the cost of curbing climate change and for the economic benefits that it brings, which is a necessary feature from a policy-making perspective. Having a very detailed model with the possibility of a cost-benefit analysis would be the preferable option. However, computational costs for such kinds of models are extremely high. Thus, there is a strong demand for IAMs, on the one hand, for capturing and reflecting relevant processes both in the economy and climate.
On the other hand, IAMs are necessary for quantitative estimation (uncertainty, risks, tipping points, endogenous growth), so they should be parsimonious enough to have reasonable computational complexity. The perfect candidate to satisfy both requirements could be the DICE model~\citep{Nordhaus2018}. However, it was strongly criticized by climate scientists and recently by economists for representing climate not in accordance with the recent climate science advancements~\citep{Dietz2020}. 

Our paper's three main contributions are as follows.
First, we develop a series of tests to evaluate the quality of CEs used in economic modeling. These tests include one test that is similar to~\cite{Dietz2020}, but overall they are more comprehensive in that they specifically test both, the temperature equations and the carbon cycle of CEs. 

Second, we update the parsimonious climate representation of climate in DICE  with a new calibration of coefficients that aligns it with the CMIP5 benchmark, making the DICE model both simple and transparent yet making it realistically represent climate. The functional form of the climate part in DICE is retained. This allows authors who have the climate module of DICE \lq\lq hard-wired\rq\rq \ in their economic model to obtain more reliable results by simply changing parameters.

Third, we explain that, although the DICE-2016 predictions in a BAU scenario fall within the range of CMIP5,  this is caused by two compensating errors in the calibration. We demonstrate that mitigation scenarios are highly sensitive to such compensating errors between the carbon cycle and the temperature response of the model. We advocate that this result is generic and not specific to DICE-2016 and that the proposed battery of tests enables the identification of such issues in climate models used in climate economics. Regarding DICE-2016, our conclusion is similar to that in~\cite{Dietz2020}; the calibration of DICE-2016 is not suited for reliable policy analysis.


\newpage
\begin{appendices}
\numberwithin{equation}{section}

\section{Summary table of the models and parameters}
\label{sec:models_summary}

\begin{table}[H]
\centering
\resizebox{\textwidth}{!}
{\begin{tabular}{ c c c c c c c c c c c  } 
 \toprule
 Name & $b_{12}$  & $b_{23}$ & $M^{EQ}_{AT,UO,LO}$ & $M^{INI}_{AT,UO,LO}$ & $c_1$ & $c_3$ & $c_4$ & $f_{2xco2}$ & $t_{2xco2}$ & $T^{INI}_{AT,OC}$ \\ 
 \midrule
 CDICE (MMM-MMM) & 0.054 & 0.0082 & (607, 489, 1281) & (851, 628, 1323) & 0.137 & 0.73 & 0.00689 & 3.45 & 3.25 & (1.1, 0.27)  \\ 
 CDICE-HadGEM2-ES & 0.054 & 0.0082 & (607, 489, 1281) & (851, 628, 1323) & 0.154 & 0.55 & 0.00671 & 2.95 & 4.55 & (1.1, 0.27) \\ 
 CDICE-GISS-E2-R & 0.054 & 0.0082 & (607, 489, 1281) & (851, 628, 1323) & 0.213 & 1.16& 0.00921 & 3.65 & 2.15 & (1.1, 0.27) \\ 
 CDICE-MESMO & 0.059 & 0.008 & (607, 305, 865) & (851, 403, 894) & 0.137 & 0.73 & 0.00689 & 3.45 & 3.25 & (1.1, 0.27) \\
 CDICE-LOVECLIM & 0.067 & 0.0095 & (607, 600, 1385) & (850, 770, 1444) & 0.137 & 0.73 & 0.00689 & 3.45 & 3.25 & (1.1, 0.27)  \\
 CDICE-MESMO-HadGEM2-ES & 0.059 & 0.008 & (607, 305, 865) & (851, 403, 894) & 0.154 & 0.55 & 0.00671 & 2.95 & 4.55 & (1.1, 0.27)  \\
 CDICE-MESMO-GISS-E2-R & 0.059 & 0.008 & (607, 305, 865) & (851, 403, 894) & 0.213 & 1.16& 0.00921 & 3.65 & 2.15 & (1.1, 0.27) \\
 CDICE-LOVECLIM-HadGEM2-ES & 0.067 & 0.0095 & (607, 600, 1385) & (850, 770, 1444) & 0.154 & 0.55 & 0.00671 & 2.95 & 4.55 & (1.1, 0.27)  \\
 CDICE-LOVECLIM-GISS-E2-R & 0.067 & 0.0095 & (607, 600, 1385) & (850, 770, 1444) & 0.213 & 1.16& 0.00921 & 3.65 & 2.15 & (1.1, 0.27) \\
 DICE-2016 & 0.12 & 0.007 & (588, 360, 1720)& (851, 460, 1740) & 0.1005 & 0.088 & 0.025 & 3.6813 & 3.1 & (0.85, 0.0068)\\
 \bottomrule
\end{tabular}}
\caption{Labelling of the models, and their respective parameterization.}
\label{table:model_calibration}
\end{table}

\section{Generic DICE formulation and calibration}
\label{sec:generic}

In section~\ref{sec:simulation}, we laid out the DICE model as a whole, that is, we specified the dynamic optimization problem that we solve with a calibration of parameters that are either given by DICE-2016 or by variants of our CDICE. We call this dynamic problem, specified  by~\cref{eq:obj_2016,eq:kplus_2016,eq:Kplus_nonnegative,eq:mu_range,eq:emissions}, the generic formulation of the DICE model in terms of its formal structure. The generic formulation allows for an easy switch in parameters between different versions of the DICE model (DICE-2007, DICE-2016) as well as a switch to the CDICE parametrization.\footnote{Note that in the Appendix \cref{sec:generic}, we also include DICE-2007, for completeness, as there is a vast number of research papers and policy studies that have used this formulation of the DICE model.} The functional forms of DICE-2007 and DICE-2016 are slightly different; however, they do not provide a critical difference in the results. This fact is especially relevant for the laws of motion of exogenous parameters in both models because of the differences in time steps (DICE-2007 operates with a ten-year time step and DICE-2016 operates with a five-year time step), differences in units of measurement for emissions (GtC in DICE-2007 versus GtCO2 in DICE-2016) and other minor divergences. We believe that it is important to have a unified specification for the DICE model, which in turn allows a broader group of researchers to consider updating their parametrization of the model with respect to our proposed CDICE.

Another feature of the generic specification of the DICE model is that we explicitly allow for the time step. The time step enters all relevant equations as the multiplication factor $\Delta_t$ with respect to the annual baseline calibration for all the parameters. This applies both for the main equations of the model specified in \cref{sec:climate,sec:economy} and for exogenously specified parameters of the model in \cref{sec:generic_exo}.

\cref{sec:generic} of this Appendix is organized as follows. We present the generic equations and calibration for all the exogenous parameters of the DICE-2007 (reads value 2007 in the tables with parameters), DICE-2016 (reads Value 2016), and CDICE first. Then we provide core equations of the model corresponding to the mass of carbon evolution, temperature evolution, and economic growth with calibrations corresponding to DICE-2007, DICE-2016, and CDICE. When necessary, we provide a detailed comment on how the functional forms of certain equations were changed to be the same in the generic formulation and at the same time equivalent to the initial formulation of DICE-2007 and DICE-2016 (for example, the law of motion for TFP).

\subsection{Exogenous variables}
\label{sec:generic_exo}
We start by introducing the law of motion for exogenous variables that are time dependent. The law of motion for labor is given in~\cref{eq:labor_1y} along with labor growth presented in~\cref{eq:gr_L}: 
\begin{align}
\label{eq:labor_1y}
    & L_{t} = L_{0} + \left(L_{\infty}-L_{0}\right) \left(1-\exp\left(-\Delta_t\delta^{L}t\right)\right), 
    \\
    \label{eq:gr_L}
    & g^{L}_{t}= \frac{\frac{dL_{t}}{dt}}{L_t}=\frac{\Delta_t\delta^{L}}{\frac{L_{\infty}}{L_{\infty}-L_{0}}\exp\left(\Delta_t\delta^{L}t\right)-1}.
\end{align}
The numerical values of the parameters for the world population and its growth rate given in~\cref{table:Labor}:
\begin{table}[H]
\centering
\begin{tabular}{ c c c c c  } 
 \toprule
 Calibrated parameter & Symbol  & Value 2007 & Value 2016/CDICE \\ 
 \midrule
 Annual rate of convergence & $\delta^L$  & 0.035 & 0.0268   \\ 
 World population at starting year [millions] & $L_0$  & 6514 & 7403 \\ 
 Asymptotic world population [millions] & $L_{\infty}$  & 8600 & 11500 \\ 
 Time step of a model & $\Delta_t$  & 10 & 5/1 \\
 \bottomrule
\end{tabular}
\caption{Generic parameterization for the evolution of labor.}
\label{table:Labor}
\end{table}

The total factor productivity evolves according to the equation~\cref{eq:TFP_1y} with its growth rate following~\cref{eq:gr_A}: 
\begin{align}
    \label{eq:TFP_1y}
    A_t &= A_0 \text{exp} \left( \frac{\Delta_t g^A_0 (1 - \text{exp}(-\Delta_t\delta^A t) )}{\Delta_t\delta^A}   \right) .\\
    \label{eq:gr_A}
    g^{A}_{t} & = \frac{\frac{dA_{t}}{dt}}{A_t} =
    \Delta_tg^A_0\exp\left(-\Delta_t\delta^A t\right).
\end{align}
 The respective parameters of the total factor productivity evolution and TFP growth rate are given in~\cref{table:TFP}:
\begin{table}[H]
\centering
\begin{tabular}{ c c c c c  } 
 \toprule
 Calibrated parameter & Symbol  & Value 2007 & Value 2016/CDICE \\ 
 \midrule
 Initial growth rate for TFP per year & $g^A_0$  & 0.0092 & 0.0152  \\ 
 Decline rate of TFP growth per year & $\delta^A$  & 0.001 & 0.005 \\ 
 Initial level of TFP & $A_0$  & 0.02722 & 5.115 \\ 
 Time step of a model & $\Delta_t$  & 10 & 5/1 \\ 
 \bottomrule
\end{tabular}
\caption{Generic parametrization for the evolution of TFP.}
\label{table:TFP}
\end{table}

The carbon intensity, defined in~\cref{eq:carbon_1y_2007} for DICE-2007 and in \cref{eq:carbon_1y_2016}, characterizes how much anthropogenic carbon inflows the climate system due to production activity:
\begin{align}
      \label{eq:carbon_1y_2007}
      \sigma_{t} =
      \sigma_{0}\exp \left( \frac{\Delta_tg^\sigma_0 (1- \text{exp}(-\Delta_t\delta^\sigma t) )}{\Delta_t\delta^\sigma}   \right). 
\end{align}
\noindent The carbon intensity in DICE-2016 is given by:\\
\begin{align}
\label{eq:carbon_1y_2016}
        \sigma_{t} =
      \sigma_{0}\exp \left( \frac{\Delta_tg^\sigma_0 }{\text{log}(1+ \Delta_t\delta^\sigma) } \left( (1 + \Delta_t\delta^\sigma)^t -1 \right) \right).
\end{align}
The parametrization for carbon intensity processes is given by~\cref{table:sigma}:
\begin{table}[H]
\centering
\begin{tabular}{ c c c c c  } 
 \toprule
 Calibrated parameter & Symbol  & Value 2007 & Value 2016/CDICE \\ 
 \midrule
 Initial growth of carbon intensity per year & $g^\sigma_0$  & -0.0073 & -0.0152  \\ 
 Decline rate of decarbonization per year & $\delta^{\sigma}$  & 0.003 & 0.001 \\ 
 Initial carbon intensity (1000GtC) & $\sigma_0$  & 0.00013418 & - \\ 
 Initial carbon intensity (1000GtC) & $\sigma_0$  & - & 0.00009556 \\ 
 Time step & $\Delta_t$  & 10 & 5/1 \\
 \bottomrule
\end{tabular}
\caption{Generic parameterization for the carbon intensity evolution.}
\label{table:sigma}
\end{table}

DICE-2016 uses a backstop technology capable of mitigating the full amount of industrial emissions that enter the atmosphere. The cost of backstop technology is assumed to be initially high but could be reduced over time, which is reflected in the definition of the coefficient of the abatement cost function $\theta_{1,t}$ as defined in \cref{eq:abat_1y_2007} for DICE-2007 and  \cref{eq:abat_1y_2016} for DICE-2016.\footnote{The scale parameter 1000 in the equations~\eqref{eq:abat_1y_2007} and~\eqref{eq:abat_1y_2016} reflects the fact that we use 1000 GtC unit of measurement; the parameter $\text{c2co2}$ transforms carbon intensity measured in GtC into GtCO2, as the backstop price in DICE-2016 is given for GtCO2 instead of GtC.}\\
The abatement cost in DICE-2007 is given by:\\
\begin{align}
        \label{eq:abat_1y_2007}
        &\theta_{1,t} =
        \frac{p^{\text{back}}_{0}(1+\exp\left(-g^{\text{back}}t \right))1000  \sigma_{t}}{\theta_{2}}.
\end{align}

\noindent The abatement cost in DICE-2016 is given by:\\
\begin{align}
        \label{eq:abat_1y_2016}
        &\theta_{1,t} =
        \frac{p^{\text{back}}_{0}\exp\left(-g^{\text{back}}t \right)1000 \cdot \text{c2co2} \cdot \sigma_{t}}{\theta_{2}}.
\end{align}
The parameters for the abatement cost are presented in~\cref{table:theta}:
\begin{table}[H]
\centering
\begin{tabular}{ c c c c c  } 
 \toprule
 Calibrated parameter & Symbol  & Value 2007 & Value 2016/CDICE \\ 
 \midrule
 Cost of backstop  2005 thUSD per tC 2005 & $p^{\text{back}}_{0}$  & 0.585 & -  \\ 
  Cost of backstop  2010 thUSD per tCO2 2015 & $p^{\text{back}}_{0}$  & - & 0.55  \\ 
 Initial cost decline backstop cost per year & $g^{\text{back}}$  & 0.005 & 0.005 \\ 
 Exponent of control cost function & $\theta_{2}$  & 2.8 & 2.6 \\
 Transformation coefficient from C to CO2 & $\text{c2co2}$  & - & 3.666 \\
 \bottomrule
\end{tabular}
\caption{Generic parametrization for the abatement cost.}
\label{table:theta}
\end{table}

The non-industrial emissions from land use and deforestation decline over time according to~\cref{eq:land_1y}, with parameters are presented in~\cref{table:eland}:
\begin{align}
        \label{eq:land_1y}
      E_{\text{Land},t} =
      E_{\text{Land},0}\exp\left(-\Delta_t\delta^{\text{Land}}t\right).
\end{align}

\begin{table}[H]
\centering
\begin{tabular}{ c c c c c  } 
 \toprule
 Calibrated parameter & Symbol  & Value 2007 & Value 2016/CDICE \\ 
 \midrule
 Emissions from land 2005 (1000GtC per year) & $E_{\text{Land},0}$  & 0.0011 & -  \\ 
  Emissions from land 2015 (1000GtC per year) & $E_{\text{Land},0}$  & - & 0.000709 \\ 
 Decline rate of land emissions (per year) & $\delta^{\text{Land}}$  & 0.01 & 0.023 \\ 
 Time step & $\Delta_t$  & 10 & 5/1 \\
 \bottomrule
\end{tabular}
\caption{Generic parametrization for the emissions from land.}
\label{table:eland}
\end{table}

The exogenous radiative forcings that result from non-CO2 GHG are described in~\cref{eq:forc_1y}:
\begin{align}
    \label{eq:forc_1y}
        F^{EX}_{t} = F^{EX}_{0} + \frac{1}{\text{T}/\Delta_t}(F^{EX}_{1} -F^{EX}_{0}) \min(t, \text{T}/\Delta_t).
  \end{align}
The parameters of the exogenous radiative forcings are given in~\cref{table:FEX}:
\begin{table}[H]
\centering
\begin{tabular}{ c c c c c  } 
 \toprule
 Calibrated parameter & Symbol  & Value 2007 & Value 2016/CDICE \\ 
 \midrule
 2000 forcings of non-CO2 GHG (Wm-2) & $F^{EX}_{0}$  & -0.06 & -  \\
 2015 forcings of non-CO2 GHG (Wm-2) & $F^{EX}_{0}$  & - & 0.5  \\ 
 2100 forcings of non-CO2 GHG (Wm-2) & $F^{EX}_{1}$  & 0.3 & 1.0  \\  
 Number of years before 2100 & $\text{T}$  & 100 & 85 \\ 
 Time step & $\Delta_t$ & 10 & 5/1 \\
 \bottomrule
\end{tabular}
\caption{Generic parametrization for the exogenous forcing.}
\label{table:FEX}
\end{table}

\subsection{The equations of the climate system}
\label{sec:climate2}

The laws of motion for the mass of carbon in the atmosphere are presented in the main body of the text with~\cref{eq:cc_mass}. The temperature equations are given in \cref{eq:temp1,eq:temp2}.  However, we believe it can be helpful to provide the full description of the equations above together with their parametrization that can be used to relate DICE-2007 to DICE-2016 and CDICE.

The evolution for the mass of carbon in all three reservoirs is given by the \cref{eq:MAT_appendix,eq:MUO_appendix,eq:MLO_appendix}, carbon emissions are determined with the \cref{eq:Emit_appendix}:

\begin{align}
& \label{eq:MAT_appendix}  M^{\text{AT}}_{t+1} = (1- \Delta_tb_{12}) M^{\text{AT}}_{t} + \Delta_tb_{12} \frac{M^{\text{AT}}_{\text{EQ}}}{M^{\text{UO}}_{\text{EQ}}}M^{\text{UO}}_{t} +  \Delta_t E_t ,\\
  & \label{eq:MUO_appendix}  M^{\text{UO}}_{t+1} = \Delta_tb_{12}M^{\text{AT}}_{t} + (1- \Delta_tb_{12} \frac{M^{\text{AT}}_{\text{EQ}}}{M^{\text{UO}}_{\text{EQ}}} - \Delta_tb_{23})M^{\text{UO}}_{t} + \Delta_tb_{23} \frac{M^{\text{UO}}_{\text{EQ}}}{M^{\text{LO}}_{\text{EQ}}}M^{\text{LO}}_{t} ,\\
  & \label{eq:MLO_appendix}  M^{\text{LO}}_{t+1} = \Delta_tb_{23}M^{\text{UO}}_{t} + (1-\Delta_tb_{23} \frac{M^{\text{UO}}_{\text{EQ}}}{M^{\text{LO}}_{\text{EQ}}})M^{\text{LO}}_{t},\\
  & \label{eq:Emit_appendix} E_t = \sigma_t Y^{\text{Gross}}_{t} (1-\mu_t) + E^{\text{Land}}_t .\\
\end{align}

The parameters for the laws of motion for the masses of carbon, as well as the starting values and equilibrium values, are given in~\cref{table:masses}.
\begin{table}[H]
\centering
\begin{tabular}{ c c c c c c } 
 \toprule
 Calibrated parameter & Symbol  & 2007 & 2016 & CDICE \\ 
 \midrule
 Carbon cycle, annual value & $b_{12}$  & 0.0189288 & 0.024 & 0.054 \\ 
 
 Carbon cycle, annual value & $b_{23}$  & 0.005 & 0.0014 & 0.0082\\ 

 Time step & $\Delta_t$  & 10 & 5 & 1 \\
 Equilibrium concentration in atmosphere (1000GtC) & $M^{\text{AT}}_{\text{EQ}}$  & 0.587473 & 0.588 &  0.607 \\
 Equilibrium concentration in upper strata (1000GtC) & $M^{\text{UO}}_{\text{EQ}}$  & 1.143894 & 0.360 & 0.489 \\
 Equilibrium concentration in lower strata (1000GtC) & $M^{\text{LO}}_{\text{EQ}}$  & 18.340 & 1.720 & 1.281 \\
 Concentration in atmosphere 2015 (1000GtC) & $M^\text{AT}_\text{INI}$  & 0.8089 & 0.851 & 0.851\\
 Concentration in upper strata 2015 (1000GtC) & $M^\text{UO}_\text{INI}$  & 1.255& 0.460 & 0.628\\
 Concentration in lower strata 2015 (1000GtC) & $M^\text{LO}_\text{INI}$  & 18.365 & 1.740 & 1.323\\
 \bottomrule
\end{tabular}
\caption{Generic parametrization for the mass of carbon.}
\label{table:masses}
\end{table}

The temperature evolution is determined by \cref{eq:temp1_appendix,eq:temp2_appendix,eq:forcing_appendix} with parameters and starting values given in~\cref{table:temps}:
\begin{align}
& \label{eq:temp1_appendix} T^{\text{AT}}_{t+1} = T^{\text{AT}}_{t} + \Delta_tc_1 F_t  - \Delta_tc_1 \frac{F_{\text{2XCO2}}}{T_{\text{2XCO2}}}T^{\text{AT}}_{t}- \Delta_tc_1c_3 \left(T^{\text{AT}}_{t}-T^{\text{OC}}_{t}\right),    \\
   & \label{eq:temp2_appendix}  T^{\text{OC}}_{t+1} = T^{\text{OC}}_{t} + \Delta_tc_4 \left(T^{\text{AT}}_{t}  - T^{\text{OC}}_{t}\right),\\
   & \label{eq:forcing_appendix} F_{t} = F_{\text{2XCO2}} \frac{\log(M^{\text{AT}}_{t}/M^{\text{AT}}_{\text{base}})}{\log(2)} + F^{EX}_{t}.
\end{align}

\begin{table}[H]
\centering
\begin{tabular}{ c c c c c c } 
 \toprule
 Calibrated parameter & Symbol  & 2007 & 2016 & CDICE \\ 
 \midrule
 Temperature coefficient, annual value & $c_{1}$  & 0.022 & 0.0201 & 0.137\\
 Temperature coefficient, annual value & $c_{3}$  & 0.3 & 0.088 & 0.73\\
 Temperature coefficient, annual value & $c_{4}$  & 0.01 & 0.005 & 0.00689\\ 
 Forcings of equilibrium CO2 doubling (Wm-2) & $F_{\text{2XCO2}}$  & 3.8 & 3.6813 & 3.45\\
 Eq temperature impact (\textdegree{}C per doubling CO2) & $T_\text{2XCO2}$  & 3.0 & 3.1 & 3.25\\
 Eq concentration in atmosphere (1000GtC) & $M^{\text{AT}}_{\text{base}}$  & 0.5964 & 0.588 & 0.607\\
 Atmospheric temp change (\textdegree{}C) from 1850  & $T^{\text{AT}}_0$  & 0.7307 & 0.85 & 1.1\\
 Lower stratum temp change (\textdegree{}C) from 1850  & $T^{\text{OC}}_0$  & 0.0068 & 0.0068 & 0.27\\
 Time step & $\Delta_t$  & 10 & 5 & 1\\
 \bottomrule
\end{tabular}
\caption{Generic parametrization for the temperature.}
\label{table:temps}
\end{table}

\subsection{Economy equations}
\label{sec:economy}
The capital evolution and gross output in both models DICE-2007 and DICE-2016 are given by:
\begin{align}
\label{eq:capital_appendix}
    K_{t+1} &= (1-\delta^K)^{\Delta_t} K_t + \Delta_t I_t,\\
    \label{eq:grossoutput_appendix}
    Y^{\text{Gross}}_t &= A_t L_t^{1-\alpha} K_t^{\alpha}.
\end{align}
The functional forms of damages differ in DICE-2007 and DICE-2016. In DICE-2007, damages, abatement costs, output net of damages and output net of damages and abatement costs  are given by:
\begin{align}
 \label{eq:damages2007_appendix}   \Omega_t &= \frac{1}{1 + \psi_1 T^{\text{AT}}_{t} + \psi_2 (T^{\text{AT}}_{t})^2}, \\
  \label{eq:abatcost2007_appendix}  \Theta_{t} &= \theta_{1,t} \mu_t ^ {\theta_2},\\
   \label{eq:netout2007_appendix} Y^{\text{Net}}_t &= Y^{\text{Gross}}_t \cdot \Omega_t =  \frac{Y^{\text{Gross}}_t}{1 + \psi_1 T^{\text{AT}}_{t} + \psi_2 (T^{\text{AT}}_{t})^2} ,\\
   \label{eq:output2007_appendix} Y_t &= Y^{\text{Gross}}_t \cdot \Omega_t \cdot (1-\Lambda_t) = \frac{Y^{\text{Gross}}_t (1-  \theta_{1,t} \mu_t ^ {\theta_2})}{1 + \psi_1 T^{\text{AT}}_{t} + \psi_2 (T^{\text{AT}}_{t})^2}.\\
\end{align}
The same variables for DICE-2016 are given by the following equations:
\begin{align}
    \label{eq:damages2016_appendix} \Omega_t &=  \psi_1 T^{\text{AT}}_{t} + \psi_2 (T^{\text{AT}}_{t})^2 ,\\
    \label{eq:abatcost2016_appendix} \Theta_{t} &= \theta_{1,t} \mu_t ^ {\theta_2},\\
    \label{eq:netout2016_appendix} Y^{\text{Net}}_t &= Y^{\text{Gross}}_t \cdot (1- \Omega_t )=  Y^{\text{Gross}}_t (1 - \psi_1 T^{\text{AT}}_{t} - \psi_2 (T^{\text{AT}}_{t})^2),  \\
    \label{eq:output2016_appendix} Y_t &= Y^{\text{Gross}}_t  \cdot (1-\Lambda_t - \Omega_t) = Y^{\text{Gross}}_t (1-  \theta_{1,t} \mu_t ^ {\theta_2} - \psi_1 T^{\text{AT}}_{t} - \psi_2 (T^{\text{AT}}_{t})^2).\\
\end{align}
Consumption in both DICE-2007 and DICE-2016 models is given by:
\begin{align}
    C_t &= Y_t - I_t.
\end{align}
However, the utility function in DICE-2007 differs slightly from the utility used in DICE-2016. Utility in DICE-2007 is the following:
\begin{align}
    U_t &= \sum_{t=0}^{T}{\beta^{t} \cdot \Delta_t \cdot \frac{\left(\frac{C_t}{L_t} \right)^{1-1/\psi} - 1 }{1-1/\psi} L_t }.
\end{align}
And utility in DICE-2016 is given by:
\begin{align}
    U_t &= \sum_{t=0}^{T}{\beta^{t} \cdot \Delta_t \cdot \frac{\left(\frac{1000C_t}{L_t} \right)^{1-1/\psi} - 1 }{1-1/\psi} L_t }.
\end{align}
The discount rate in both DICE-2007 and DICE-2016 models is the same and determined as:
\begin{align}
    \beta &= \frac{1}{(1+\rho)^{\Delta_t}}.
\end{align}
The parameters for the economic part of DICE-2007 and DICE-2016 are given in~\cref{table:econ}.
\begin{table}[H]
\centering
\begin{tabular}{ c c c c c  } 
 \toprule
 Calibrated parameter & Symbol  & Value 2007 & Value 2016 \\ 
 \midrule
 Capital annual depreciation rate & $\delta^K$  & 0.1 & 0.1  \\ 
 Elasticity of capital  & $\alpha$  & 0.3 & 0.3  \\
 Damage parameter  & $\psi_1$  & 0.0 & 0.0  \\
 Damage quadratic parameter  & $\psi_2$  & 0.0028388 & 0.00236  \\
 Exponent of control cost function  & $\theta_2$  & 2.8 & 2.6  \\
 Risk aversion  & $\psi$  & 0.5 & 0.69  \\
 Time preferences   & $\rho$  & 0.015 & 0.015  \\
 Time step & $\Delta_t$  & 10 & 5 \\
 \bottomrule
\end{tabular}
\caption{Generic parameterization for the parameters of economy.}
\label{table:econ}
\end{table}

\section{Sensitivity exercise: change in the exogenous forcing}
\label{sec:fex}

In section~\ref{sec:test_cmip5_rcp} we used an alternative evolution equation for exogenous forcings, namely 
\begin{equation}
    F^{\text{EX}} = 0.3 \cdot F^{\text{CO2}}_t.
\end{equation}
This specification was adopted instead of \cref{eq:forc_1y} that was used in DICE-2016. As was shown in Figure~\ref{fig:CMIPBM}, this switch from the original DICE-2016 exogenous forcings can make quite a difference in the temperature evolution. However, when computing the simulation results in section~\ref{sec:simulation}, we used the original exogenous forcings equation \cref{eq:forc_1y} to remain consistent with the DICE-2016 model and to focus on analyzing the influence of the re-calibrated climate part on policies, which was the goal of the section. Therefore, in this section of the Appendix, we want to investigate the consequences of implementing time-dependent exogenous forcings that  differ for the DICE-2016 and CDICE models.

Below we present some solution results for the optimization problem \cref{eq:obj_2016} when the agent chooses optimal investment and mitigation paths. Furthermore, we compare DICE-2016 and CDICE with original and alternative exogenous forcings. Overall changing exogenous forcings does not make much of a difference for the modeling results. However, there are a couple of interesting observations that are worth mentioning.

\begin{figure}[!htb]
\centering
  \begin{subfigure}[b]{0.5\linewidth}
    \centering
    \includegraphics[width=1.0\linewidth]{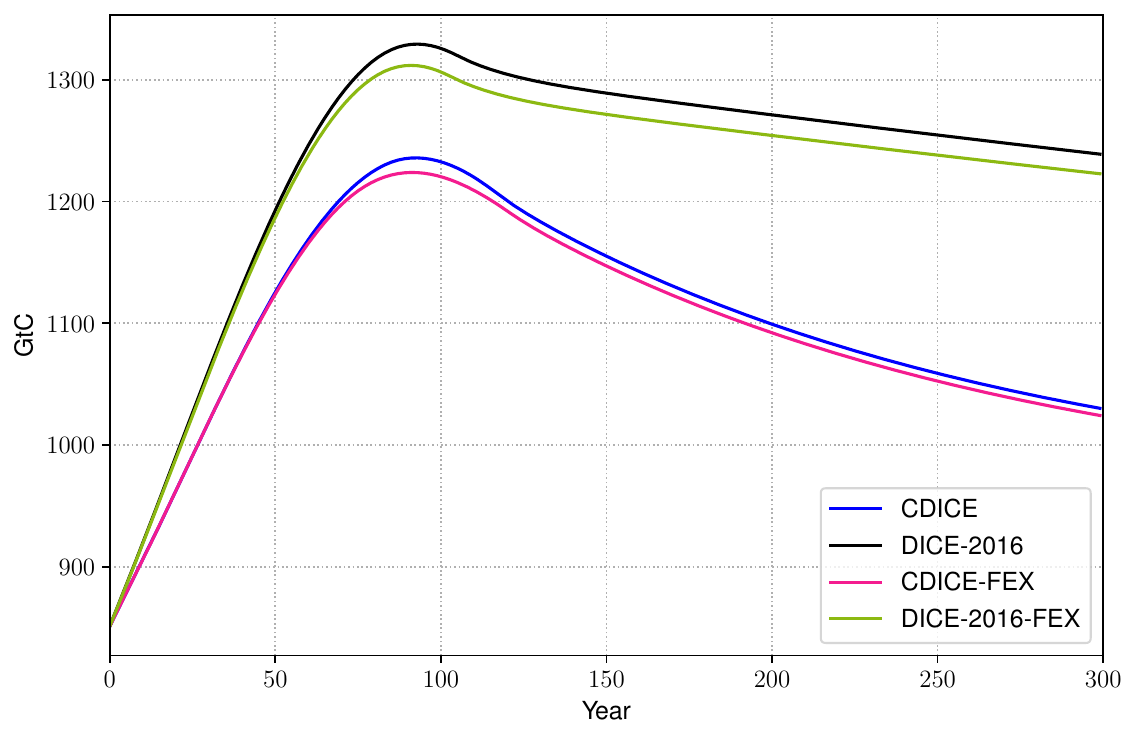}
    \caption{Mass of carbon in the atmosphere} 
    \label{fig:fex_mass_temp:a} 
  \end{subfigure}
  \begin{subfigure}[b]{0.5\linewidth}
    \centering
    \includegraphics[width=1.0\linewidth]{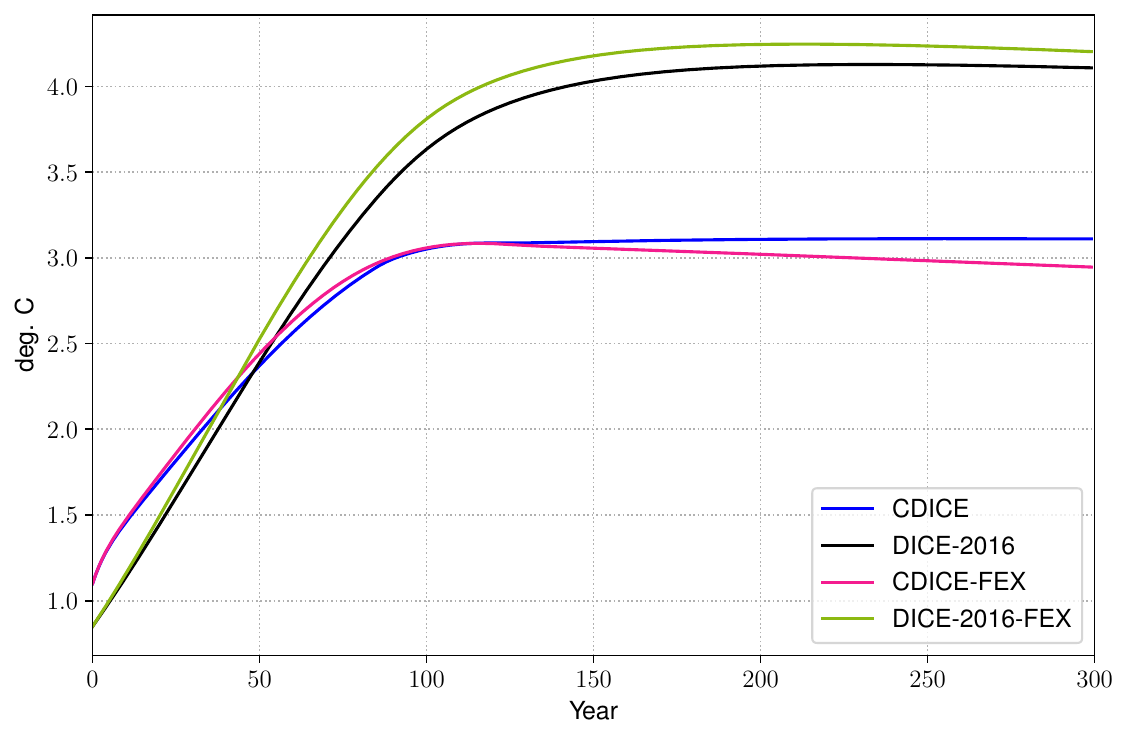}
    \caption{Temperature of the atmosphere} 
    \label{fig:fex_mass_temp:b} 
  \end{subfigure} 
  \caption{Mass of carbon in the atmosphere (left) and temperature of the atmosphere (right) for DICE-2016, and CDICE models with different exogenous forcing evolution (DICE-2016-FEX, CDICE-FEX show the model variables under the assumption of time dependent exogenous forcings) for an optimal abatement case. Year zero on the graphs corresponds to a starting year 2015.}
  \label{fig:fex_mass_temp}
\end{figure}

From Figure~\ref{fig:fex_mass_temp} we see that change in the exogenous forcings does not affect the CDICE model much but has some effect on DICE-2016. In DICE-2016 exogenous forcings being time-dependent results in a lower mass of carbon in the atmosphere than in DICE-2016 with the original forcings. However, it leads to an even higher temperature increase. The reason for this increased sensitivity of DICE-2016 is the same as for its excessive sensitivity to the discount rate. The carbon cycle and temperature equations are not in balance. Thus they overreact in transmitting changes in the inputs of the model towards outputs. 

\begin{figure}[!htb] 
  \begin{subfigure}[b]{0.5\linewidth}
    \centering
    \includegraphics[width=1.0\linewidth]{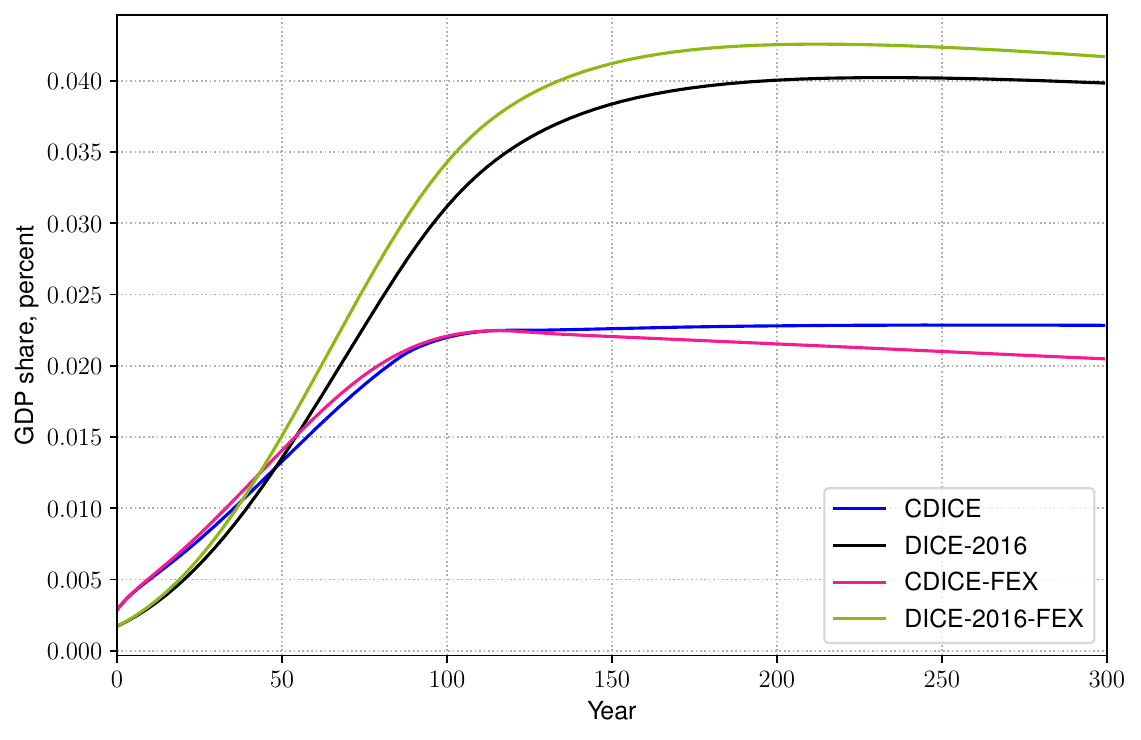} 
    \caption{Damages} 
    \label{fig:fex_dam_SCC:a} 
  \end{subfigure} 
  \begin{subfigure}[b]{0.5\linewidth}
    \centering
    \includegraphics[width=1.0\linewidth]{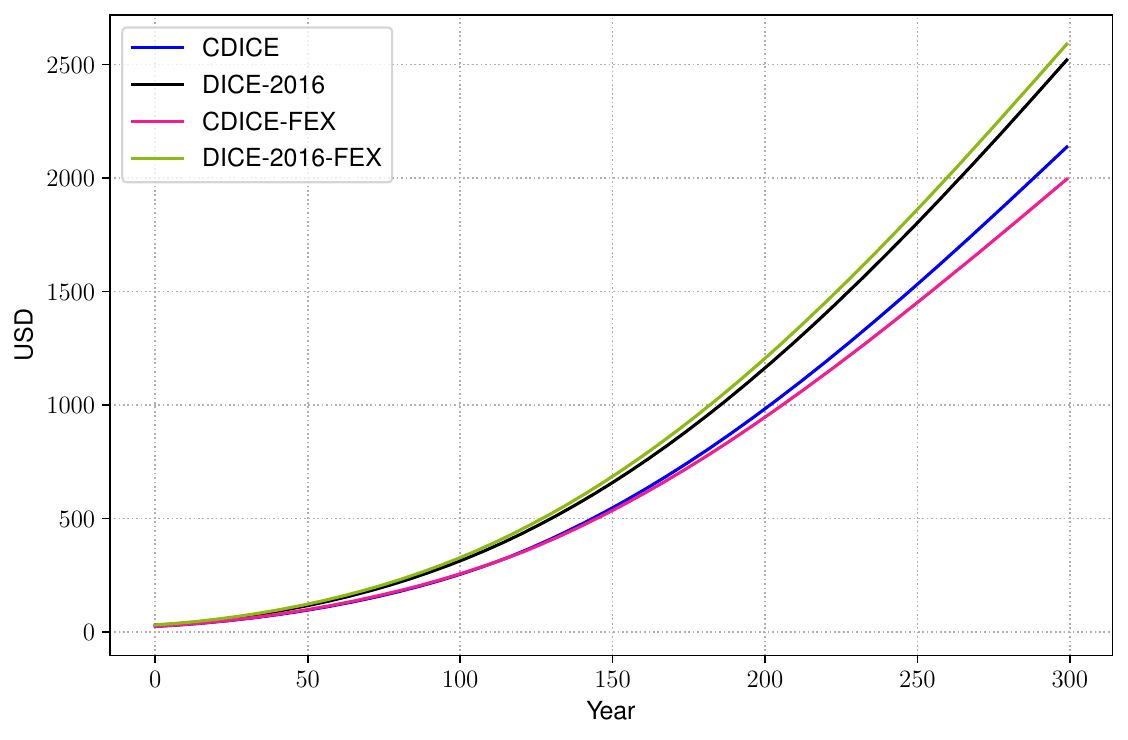}
    \caption{SCC} 
    \label{fig:fex_dam_SCC:b} 
  \end{subfigure} 
  \caption{Damages (left) and the social cost of carbon (right) for DICE-2016, and CDICE models with different exogenous forcings evolution (DICE-2016-FEX, CDICE-FEX show the model variables under the assumption of time dependent exogenous forcings) for an optimal abatement case. Year zero on the graphs corresponds to a starting year 2015.}
  \label{fig:fex_dam_SCC}
\end{figure}

Figure~\ref{fig:fex_dam_SCC} reports damages and the social cost of carbon for models under consideration. Both variables are coherently in line with the expectations that one might get analyzing temperature equations. DICE-2016 with time-dependent forcings predicts more damages and thus a higher social cost of carbon. For CDICE, both damages and the social cost of carbon remain roughly the same, with CDICE variables being slightly lower for alternative exogenous forcing formulation.


\end{appendices}

\newpage
\bibliography{references.bib}

\end{document}